\documentstyle[aps,prb,eqsecnum,twocolumn,floats,epsf]{revtex}
\begin{document}
\draft

\twocolumn[\hsize\textwidth\columnwidth\hsize\csname @twocolumnfalse\endcsname

\title{Renormalization-group study of Anderson and Kondo
impurities in gapless Fermi systems}
\bigskip
\author{Carlos Gonzalez-Buxton
and Kevin Ingersent\cite{byline}}
\address{Department of Physics, University of Florida,
P.\ O.\ Box 118440, Gainesville, Florida 32611}
\date{January 6, 1998}
\maketitle

\begin{abstract}
Thermodynamic properties are presented for four magnetic impurity models
describing delocalized fermions scattering from a localized orbital
at an energy-dependent rate~$\Gamma(\epsilon)$ which vanishes precisely
at the Fermi level, $\epsilon = 0$.
Specifically, it is assumed that for small $|\epsilon|$,
$\Gamma(\epsilon)\propto|\epsilon|^r$ with~$r>0$.
The cases $r=1$ and $r=2$ describe dilute magnetic impurities in unconventional
($d$- and $p$-wave) superconductors, ``flux phases'' of the two-dimensional
electron gas, and certain zero-gap semiconductors.
For the nondegenerate Anderson model, the main effects of the depression of
the low-energy scattering rate are the suppression of mixed valence in
favor of local-moment behavior, and a marked reduction in the exchange coupling
on entry to the local-moment regime, with a consequent narrowing of the
range of parameters within which the impurity spin becomes Kondo-screened.
The precise relationship between the Anderson model and the exactly screened
Kondo model with power-law exchange is examined.
The intermediate-coupling fixed point identified in the latter model by Withoff
and Fradkin~(WF) is shown to have clear signatures both in the
thermodynamic properties and in the local magnetic response of the impurity.
The underscreened, impurity-spin-one Kondo model and the overscreened,
two-channel Kondo model both exhibit a conditionally stable
intermediate-coupling fixed point in addition to unstable fixed points of
the WF~type.
In all four models, the presence or absence of particle-hole symmetry plays
a crucial role in determining the physics both at strong coupling and in the
vicinity of the WF~transition.
These results are obtained using an extension of Wilson's numerical
renormalization-group technique to treat energy-dependent scattering.
The strong- and weak-coupling fixed points of each model are identified and
their stability is analyzed.
Algebraic expressions are derived for the fixed-point thermodynamic
properties, and for low-temperature corrections about the stable fixed points.
Numerical data are presented confirming the algebraic results, identifying
and characterizing intermediate-coupling (non-Fermi-liquid) fixed points,
and exploring temperature-driven crossovers between different physical regimes.
\end{abstract}


]
\narrowtext

\section{Introduction}
\label{sec:intro}

In conventional metallic systems, it is well understood how many-body
correlations induced by dilute magnetic impurities in an otherwise
noninteracting conduction band can at low temperatures effectively quench all
spin fluctuations on each impurity site.\cite{Hewson}
This, the Kondo effect, depends critically on the presence of fermionic
excitations down to arbitrarily small energy scales.
The impurity properties are sensitive to the density of electronic states
$\rho(\epsilon)$ only through its value at the Fermi level, $\epsilon=0$.
Other details of the band shape have negligible effect on the
low-temperature physics.

A growing body \cite{Withoff,Borkowski:92,Cassanello,Chen,Ingersent:scaling,%
Ingersent:NRG,Buxton,Borkowski:96,Bulla} of theoretical work
shows that the standard picture of the Kondo effect must be fundamentally
revised in order to treat ``gapless'' systems, in which the effective density
of states vanishes precisely at $\epsilon=0$ but is nonzero everywhere
else in the vicinity of the Fermi energy.
The goal of the present paper is to extend the understanding of this issue
through a comprehensive account of the different physical regimes exhibited
by magnetic impurities in gapless host materials, including detailed
calculations of thermodynamic properties.

Gaplessness may be realized in a number of physical systems:
(1)~The quasiparticle density of states in an unconventional
superconductor can vary like~$|\epsilon|$ or~$|\epsilon|^2$ near line or point
nodes in the gap.\cite{Sigrist}
Heavy-fermion and cuprate superconductors are strong candidates for this
behavior.
(2)~The valence and conduction bands of certain semiconductors touch in such a
way that, for small~$|\epsilon|$, $\rho(\epsilon)$~is proportional
to~$|\epsilon|^{d-1}$ in $d$~spatial dimensions.
Examples include PbTe-SnTe heterojunctions,\cite{Volkov} and the ternary
compounds Pb$_{1-x}$Sn$_x$Se, Pb$_{1-x}$Sn$_x$Te, and Hg$_{1-x}$Cd$_x$Te,
each at a temperature-dependent critical composition.\cite{Hohler}
(Zero-gap mercury cadmium telluride has been proposed as the basis
for a giant magnetoresistance read-head for high-density storage.\cite{Solin})
(3)~Various two-dimensional electron systems --- including graphite
sheets,\cite{Semenoff} ``flux phases'' in a strong magnetic field,\cite{Fisher}
and exotic phases of the Hubbard model \cite{Baskaran} --- are predicted to
exhibit a linear pseudogap.
It is a matter of ongoing debate whether this pseudogap survives the presence
of disorder.\cite{disorder}
(4)~The single-particle density of states in the one-dimensional
Luttinger model varies like~$|\epsilon|^{2\alpha}$, where $\alpha$~changes
continuously with the strength of the bulk interactions.\cite{Voit}
In all these examples, the effective density of states can be approximated near
the Fermi level by a power law, $\rho(\epsilon)\propto|\epsilon|^r$ with $r>0$.

The first theoretical study of magnetic impurities in gapless Fermi systems
was carried out by Withoff and Fradkin,\cite{Withoff} who assumed a simplified
density of states in which the power-law variation extends across an entire
band of halfwidth $D$:
\begin{eqnarray}
\rho(\epsilon) = \left\{ \begin{array}{ll}
        \rho_0 |\epsilon/D|^r , \qquad & |\epsilon| \le D; \\[1ex]
        0, & \text{otherwise} . \end{array} \right.
                                                \label{rho_pure}
\end{eqnarray}
Using poor-man's scaling for the spin-$\frac{1}{2}$ (i.e., impurity degeneracy
$N=2$) Kondo model and a large-$N$ treatment of the Coqblin-Schrieffer model
--- both methods being valid for $0 \le r \alt \frac{1}{2}$ --- these authors
demonstrated that the Kondo effect takes place only if the dimensionless,
antiferromagnetic electron-impurity exchange~$\rho_0 J$ exceeds a critical
value, $\rho_0 J_c\approx r$; for $J<J_c$, the depletion of low-energy
conduction states causes the impurity to decouple from the band at low
temperatures.

Subsequent work has analyzed values of $r$ up to $2$ and beyond.
Large-$N$ methods have been applied\cite{Borkowski:92,Cassanello} to models
describing magnetic impurities in unconventional superconductors, in which the
power-law variation of $\rho(\epsilon)$ is restricted to a region
$|\epsilon|<\Delta$.
These studies, which may be directly relevant for Ni-doping
experiments\cite{YBCO} on YBa$_2$Cu$_3$O$_{7-\delta}$, have yielded results in
general agreement with Ref.~\onlinecite{Withoff}.
The logarithmic dependences on temperature~$T$ and frequency~$\omega$ which
characterize the standard ($r=0$) Kondo effect are replaced for $r>0$ by power
laws.
In the specific case $r=1$, these power laws acquire logarithmic
corrections.\cite{Cassanello}
For $r\le 1$ or $N=2$, any finite impurity concentration produces a small
infilling of the pseudogap which drives~$J_c$ to~zero.\cite{Borkowski:92}

Numerical renormalization-group (RG) calculations\cite{Chen,Ingersent:NRG}
for the case $N=2$ have revealed a number of additional features.
At particle-hole symmetry, the critical coupling~$J_c$ is infinite
for all $r>\frac{1}{2}$, while for $r<\frac{1}{2}$ the strong-coupling limit
exhibits anomalous properties, including values of the impurity entropy and
the effective impurity moment which are nonzero (even at $T=0$)
and which vary continuously with $r$.\cite{Chen}
Away from particle-hole symmetry, the picture is markedly
different.\cite{Ingersent:NRG}
Progressive introduction of band asymmetry or of impurity potential scattering
initially drives $J_c$ for $r>\frac{1}{2}$ back down towards the large-$N$
value $\rho_0 J_c\approx r$;
eventually, though, further increasing the asymmetry tends to freeze the motion
of conduction electrons near the impurity site, leading to an upturn in~$J_c$.
For $J>J_c$ and $r>0$, the impurity entropy and the effective impurity moment
both approach zero at $T=0$.
An electron phase shift of~$\pi$ suggests that the impurity contribution
to the resistivity also vanishes,\cite{Ingersent:NRG} instead of taking its
maximum possible value as it does in the standard Kondo effect.\cite{Hewson}

The spin-$s$ Kondo model presupposes the existence of a local moment at the
impurity site, i.e., an impurity level having an average occupancy
$\langle n_d\rangle \approx 2s$.
The more-fundamental Anderson model allows for real charge fluctuations on
the impurity site.
In the nondegenerate ($N=2$) version of this model, mixed-valence
($0<\langle n_d\rangle<1$) and empty-impurity ($\langle n_d\rangle\approx 0$)
regimes compete with $s=\frac{1}{2}$ local-moment behavior.
Poor-man's scaling has been applied \cite{Buxton} to an Anderson impurity
lying inside a power-law pseudogap.
The reduction in the density of states near the Fermi level has three main
effects, each of which grows more pronounced as $r$~increases:
the mixed-valence region of parameter space shrinks, and for $r\ge 1$
disappears altogether; there is a compensating expansion of the local-moment
regime and, to a lesser extent, of the empty-impurity regime;
and the value of the Kondo exchange~$J$ on entry to the local-moment regime
is reduced.
Since the threshold exchange for the Kondo effect ($J_c$~defined above) rises
with~$r$, these results imply --- at least in the cases of greatest interest,
$r=1$~and~$2$ --- that over a large region of phase space, the low-temperature
state possesses an uncompensated local moment.
This should be contrasted with systems having a regular density of states, in
which an Anderson impurity is always quenched at zero temperature.\cite{Hewson}

This paper contains a detailed study of four models describing a magnetic
impurity in a gapless host: the nondegenerate Anderson model, and
three variants of the Kondo model, representing (in the nomenclature
introduced by Nozi\`{e}res and Blandin \cite{Nozieres}) ``exactly screened,''
``underscreened,'' and ``overscreened'' impurity spins.
The nonperturbative RG formalism originally developed to describe magnetic
impurities in metals \cite{Wilson,KWW} is extended to provide algebraic results
for the stability of, and properties near, the weak- and strong-coupling fixed
points of each model.
Numerical implementation of the RG scheme enables characterization of the
thermodynamic properties at intermediate-coupling (non-Fermi-liquid) fixed
points, and allows the study of temperature-driven crossovers between
fixed-point regimes.

Three effects of the pseudogap are found to be common to all the models.
(1)~Over a finite fraction of parameter space, the impurity becomes
asymptotically free in the limit $T\rightarrow 0$.
The size of this weak-coupling region grows with increasing $r$.
(2)~The presence or absence of particle-hole symmetry plays a crucial role in
determining the low-energy behavior.
It turns out that each model can exhibit two distinct fixed points of the
Withoff-Fradkin type: one preserving, and the other violating, particle-hole
symmetry.
(These fixed points coexist only over a limited range of $r$~values.)
The strong-coupling physics is also very sensitive to particle-hole
(a)symmetry.
(3)~Power-law dependences of physical quantities on temperature and frequency
are generally different for sublinear and superlinear densities of states.
Specifically, in many places where the exponent $r$ enters physical quantities
for $r<1$, it is replaced for $r>1$ by either 1~or~$r-1$.
The case $r=1$, of particular interest in the contexts of high-$T_c$
superconductivity and of two-dimensional flux phases, exhibits logarithmic
corrections to simple power laws.

Extensive results are provided for the Anderson model.
We investigate in a systematic fashion the nature of the phase diagram at a
fixed, positive value of~$r$, and study the various trends produced by
increasing~$r$.
For $r\agt 0.5$, it proves impossible to observe Kondo screening of an
Anderson impurity in a system having a pure power-law density of states.
The suppression of the Kondo effect becomes less dramatic, however, if the
power law is restricted to a pseudogap of halfwidth $\Delta\ll D$.
(A preliminary version of these results was used to support the
perturbative scaling theory presented in Ref.~\onlinecite{Buxton}.)

Recently, Bulla {\it et al.}\cite{Bulla} have also applied the numerical RG
approach to the Anderson model with a pure power-law scattering rate, limited
to cases of strict particle-hole symmetry.
With one minor exception, the weak- and strong-coupling thermodynamics
reported for \mbox{$r<\frac{1}{2}$} are in agreement with
Ref.~\onlinecite{Buxton} and the present work.
The authors of Ref.~\onlinecite{Bulla} interpret these results, and certain
noninteger exponents describing their numerical data for the impurity spectral
function, as evidence for non-Fermi-liquid behavior.
We demonstrate, however, that the fixed-point properties are precisely those
expected for a noninteracting gapless system.
Hence, we argue that the weak- and strong-coupling limits can be
described within a generalized Fermi-liquid framework.

Previous studies of the exactly screened $s=\frac{1}{2}$ Kondo model with
power-law exchange have focused on the existence and position $J_c$ of the
intermediate-coupling fixed point, and on the thermodynamics in the weak-
and strong-coupling regimes.
Here we concentrate instead on the properties of the $J_c$ fixed point, which
is shown to have a clear signature in the impurity contribution to total
thermodynamic quantities and in a static response function which probes
the local behavior at the impurity site.
We also examine in some detail the relationship between the Kondo and Anderson
models in gapless hosts, and conclude that the models are independent to a
greater extent than in the standard case $r=0$.

Our investigation of the underscreened, $s=1$ Kondo model and the overscreened,
$s=\frac{1}{2}$, two-channel Kondo model focuses on the fixed-point physics.
Over a range of exponents $0<r\alt 1/4$, each problem exhibits an unstable
fixed point of the Withoff-Fradkin type at a critical coupling
\mbox{$\rho_0 J_c \approx r$}.
The novel feature, however, is the existence at some exchange $J^{\star}>J_c$
of a second intermediate-coupling fixed point which is locally stable with
respect to perturbations in $J$.
The $J^{\star}$ fixed point of the underscreened problem has no counterpart
in metals, but that of the overscreened model is the generalization to $r>0$
of the non-Fermi-liquid fixed point identified by Nozi\`{e}res and
Blandin.\cite{Nozieres}
For $r\agt 1/4$, the $J^{\star}$ fixed point disappears, and the $J_c$ fixed
point can be reached (and hence the Kondo effect observed) only under
conditions of strong particle-hole asymmetry.

Before proceeding further, we remark on a matter of terminology.
It will be shown in the next section that the conduction-band density of
states~$\rho(\epsilon)$ and the energy-dependent hybridization $t(\epsilon)$
(describing hopping between a magnetic level and the conduction band)
enter the Anderson impurity problem only in combination, through the
scattering rate
\begin{equation}
\Gamma(\epsilon) = \pi \rho (\epsilon) t^2 (\epsilon).
                                                \label{Gamma:def}
\end{equation}
The exchange and potential scattering in the Kondo model have the same
energy-dependence as $\Gamma(\epsilon)$.
In gapless systems is is natural to assume that $\rho(\epsilon)$ is given by
Eq.~(\ref{rho_pure}) while $t(\epsilon)$ is essentially constant.
However, the separate forms of the density of states and the hybridization are
unimportant provided that one is interested only in impurity properties.
In the remainder of the paper, we shall therefore refer to a power-law
{\em scattering rate\/} or {\em exchange}.
We shall focus mainly on the simplest case, that of pure power-law
scattering,
\begin{equation}
\Gamma(\epsilon) = \left\{ \begin{array}{ll}
        \Gamma_0 |\epsilon/D|^r , \qquad & |\epsilon| \le D; \\[1ex]
        0, & \text{otherwise} . \end{array} \right.
                                                \label{Gamma_pure}
\end{equation}
However, we will examine the effect of including more realistic features
such as band asymmetry and restriction of the power-law variation to a finite
pseudogap region.

The organization of this paper is as follows:
In Section~\ref{sec:general} we describe the generalization of Wilson's
numerical~RG method to handle magnetic impurity problems with an
energy-dependent impurity scattering rate.
The three sections that follow develop the analytical aspects of the technique
in the specific context of a pure power-law scattering rate, as given in
Eq.~(\ref{Gamma_pure}).
Section~\ref{sec:band} deals with the discretized conduction-band Hamiltonians
that lie at the heart of Wilson's method.
Section~\ref{sec:fixed} addresses the stability of the weak- and
strong-coupling fixed points of the four magnetic impurity models of interest,
while Section~\ref{sec:thermo} focuses on their thermodynamic properties.
The reader who is already familiar with the models we study and who wishes
to pass over the technical details of our treatment may wish to jump directly
to Section~\ref{sec:results}, where detailed numerical results are presented.
The results are summarized in Section~\ref{sec:summary}.
Two appendices contain mainly technical details.

\section{Generalized Formulation of the Numerical~RG Method}
\label{sec:general}

In this section, we describe a generalization of Wilson's nonperturbative
numerical~RG method \cite{Wilson,KWW} to treat impurity models in which the
scattering rate of conduction electrons from the impurity site is
energy-dependent.
The generalization, presented here in the context of the single-impurity Kondo
and Anderson Hamiltonians, was developed independently by several groups.
It has been applied to the two-impurity Anderson model,\cite{Sakai:90}
the two-impurity, two-channel Kondo model,\cite{Ingersent:94} the Anderson
lattice in infinite spatial dimensions,\cite{Sakai:94} and to the
single-impurity Kondo\cite{Ingersent:NRG} and Anderson\cite{Buxton,Bulla}
models with a power-law scattering rate.
Only the last of the papers cited reports any technical details.
Here we provide a comprehensive explanation of the method.

An alternative (but closely related) generalization of the numerical~RG
method, developed by Chen and Jayaprakash,\cite{Chen:95} has been used to
obtain equivalent physical results for the Kondo model with a pure power-law
scattering rate.\cite{Chen}
The relationship between the two formulations is discussed in
Ref.~\onlinecite{Bulla}.

\subsection{The Anderson impurity model}
\label{subsec:Anderson}

The nondegenerate Anderson Hamiltonian \cite{Anderson:model} for a single
magnetic impurity in a nonmagnetic host can be written as the sum of
conduction-band, impurity, and hybridization terms:
\begin{equation}
{\cal H}_A = {\cal H}_c+{\cal H}_d+{\cal H}_h,
                                                        \label{H_And:def}
\end{equation}
where
\begin{mathletters}
\begin{eqnarray}
{\cal H}_c &=& \sum_{{\bf k}} \epsilon_{\bf k}
        c^{\dag}_{{\bf k}\sigma} c_{{\bf k}\sigma} ,    \label{H_c:def} \\
{\cal H}_d &=& \epsilon_{d} n_d + U n_{d\uparrow} n_{d\downarrow},
                                                        \label{H_d:def} \\
{\cal H}_h &=& \sum_{\bf k} \frac{t_{\bf k}}{\sqrt{N_0}}
         (c^{\dag}_{\bf k \sigma} d_{\sigma} + \text{H.c.}).
                                                        \label{H_h:def}
\end{eqnarray}                                          \label{H_A:def}%
\end{mathletters}%
The energies $\epsilon_{\bf k}$ and~$\epsilon_d$ of electrons in the conduction
band and in the localized impurity state, respectively, are measured from the
Fermi energy; $N_0$~is the number of unit cells in the host;
$n_d = n_{d\uparrow} + n_{d\downarrow}$ is the total occupancy of the
impurity level;
and $U > 0$ is the Coulomb repulsion between a pair of localized electrons.
Without loss of generality, the hybridization matrix elements~$t_{\bf k}$
between localized and conduction states can be taken to be real and
non-negative.
Throughout the paper, summation over repeated spin indices ($\sigma$~in
the equations above) is implied.

For simplicity, we consider a spatially isotropic problem, i.e., one in which
$\epsilon_{\bf k} \equiv \epsilon_{|{\bf k}|}$ and
$t_{\bf k} \equiv t(\epsilon_{|{\bf k}|})$, so that the impurity interacts
only with $s$-wave conduction states centered on the impurity site.
The energies~$\epsilon$ of such $s$-wave states are assumed to be distributed
over the range $-(1\!+\!\mu)D \le \epsilon \le (1\!-\!\mu)D$.
It proves convenient to work with a dimensionless energy scale,
$\varepsilon = \epsilon/D$.
Then, dropping the kinetic energy of all non-$s$-wave conduction states,
Eqs.~(\ref{H_A:def}) can be transformed to the following one-dimensional form:
\begin{mathletters}
\begin{eqnarray}
{\cal H}_c &=& D \int^{1-\mu}_{-(1+\mu)} \!\! d \varepsilon \;
        \varepsilon \; c^{\dag}_{\varepsilon \sigma} c_{\varepsilon \sigma} ,
                                                        \label{H_c:1} \\
{\cal H}_d &=& \epsilon_d n_d + U n_{d\uparrow} n_{d\downarrow} ,
                                                        \label{H_d:1} \\
{\cal H}_h &=& \int^{1-\mu}_{-(1+\mu)} \!\! d \varepsilon \;
        \sqrt{\rho(\varepsilon D)D} \; t(\varepsilon D) \;
        (c^{\dag}_{\varepsilon \sigma} d_{\sigma} + \text{H.c.}) .
                                                        \label{H_h:1}
\end{eqnarray}                                          \label{H_A:1}%
\end{mathletters}%
The operator~$c_{\varepsilon \sigma}$, which annihilates an electron in an
$s$-wave state of energy $\varepsilon$, satisfies the anticommutation
relations $\{c^{\dag}_{\varepsilon \sigma},
c_{\varepsilon' \sigma'}\} = \delta (\varepsilon - \varepsilon') \,
\delta_{\sigma,\sigma'}$.

In this model, the impurity couples to a unique linear combination of $s$-wave
conduction states associated with an operator
\begin{equation}
    f_{0 \sigma} = F^{-1}
    \int^{1-\mu}_{-(1+\mu)} \! d \varepsilon \;
                     w(\varepsilon) \; c_{\varepsilon \sigma} ,
                                                \label{f_0}
\end{equation}
where
\begin{equation}
F^2 = \int^{1-\mu}_{-(1+\mu)} \! d \varepsilon \; w^2 (\varepsilon) .
                                                \label{F}
\end{equation}
The weighting function~$w(\varepsilon)$ entering Eqs.~(\ref{f_0}) and~(\ref{F})
is
\begin{equation}
w(\varepsilon) = \raisebox{1.3ex}{\scriptsize +}\!\!\!\!\!
                 \sqrt{\Gamma(\varepsilon D)/\Gamma_0},
                                                \label{w}
\end{equation}
where $\Gamma(\epsilon)$~is defined in Eq.~(\ref{Gamma:def}) and
$\Gamma_0$~is a reference value of the scattering rate (for example,
that at the Fermi level).
With these definitions, the hybridization term in the Hamiltonian can
be rewritten
\begin{equation}
{\cal H}_h = \sqrt{\Gamma_0 D/\pi} \; F\,
             (f^{\dag}_{0\sigma} d_{\sigma} + \text{H.c.}) .
                                                \label{H_h:f}
\end{equation}

\subsection{The Kondo impurity model}

The Kondo model \cite{Kondo} describes the interaction between a conduction
band and a localized impurity which has a spin~${\bf s}$ of
magnitude~$\sqrt{s(s+1)}$.
The Hamiltonian is
\begin{equation}
{\cal H}_K = {\cal H}_c + {\cal H}_s,           \label{H_Kondo:def}
\end{equation}
where ${\cal H}_c$ is the conduction-band Hamiltonian given in
Eq.~(\ref{H_A:def}a) and
\begin{equation}
{\cal H}_s = \sum_{\bf k, k'} \left[
          \frac{J_{\bf k, k'}}{N_0}\,c^{\dag}_{{\bf k}\sigma} \case{1}{2}
          \bbox{\sigma}_{\sigma\sigma'} c_{{\bf k}'\sigma'} \cdot{\bf s}
        + \frac{V_{\bf k, k'}}{N_0}\,c^{\dag}_{{\bf k}\sigma}
          c_{{\bf k}'\sigma} \right] .
                                                        \label{H_s:def}
\end{equation}
The first and second terms in ${\cal H}_s$ describe exchange and potential
scattering, respectively.

The impurity-spin-$\frac{1}{2}$ version of the Kondo model can be regarded as
a limiting case of the nondegenerate Anderson model [Eq.~(\ref{H_And:def})].
If $-\epsilon_d ,\, U+\epsilon_d \gg \Gamma ,\, k_B T$ (where $k_B T$ is
the thermal energy scale), then single occupancy of the Anderson impurity level
is overwhelmingly favored over zero or double occupancy, in effect localizing
a pure-spin degree of freedom at the impurity site.
The exchange and potential scattering coefficients can be determined using
the Schrieffer-Wolff transformation:\cite{Schrieffer}
\begin{mathletters}
\begin{eqnarray}
J_{\bf k,k'} &=& 2 \left(\frac{1}{|\epsilon_d|}+\frac{1}{|U+\epsilon_d|}\right)
               t_{\bf k} t_{\bf k'} ,                   \\
V_{\bf k,k'}& =& \frac{1}{2} \left( \frac{1}{|\epsilon_d|}-
                 \frac{1}{|U+\epsilon_d|} \right) t_{\bf k} t_{\bf k'} .
\end{eqnarray}                                          \label{SW}%
\end{mathletters}

Equations~(\ref{SW}) imply that the exchange and potential scattering both
exhibit the same dependence on~$\bf k$, and hence (in a spatially isotropic
problem) on $\varepsilon$.
Thus, just as for the Anderson model, the Kondo impurity interacts with a
single linear combination of conduction states.
Equation~(\ref{H_s:def}) can be rewritten
\begin{eqnarray}
{\cal H}_s
&=& \int^{1-\mu}_{-(1+\mu)} \!\! d \varepsilon \sqrt{\rho(\varepsilon D)D}
    \int^{1-\mu}_{-(1+\mu)} \!\! d \varepsilon' \sqrt{\rho(\varepsilon'D)D}
    \times \nonumber \\
& & \quad \left[J(\varepsilon D,\varepsilon'D)\,\bbox{\sigma}_{\sigma\sigma'}
    \cdot{\bf s}+V(\varepsilon D,\varepsilon'D)\,\delta_{\sigma,\sigma'}\right]
    c^{\dag}_{\varepsilon \sigma} c_{\varepsilon'\sigma'}
    \nonumber \\
&=& D \left[ \rho_0 J_0 \case{1}{2} \bbox{\sigma}_{\sigma\sigma'}
                  + \rho_0 V_0 \delta_{\sigma,\sigma'} \right]
      F^2 f^{\dag}_{0\sigma} f_{0\sigma'} ,
                                                        \label{H_s:f}
\end{eqnarray}
where~$f_{0\sigma}$ and~$F$ are defined in Eqs.~(\ref{f_0})--(\ref{w});
$\rho_0 J_0$~and~$\rho_0 V_0$ are reference values of
$\sqrt{\rho(\epsilon)\rho(\epsilon')} J(\epsilon,\epsilon')$ and
$\sqrt{\rho(\epsilon)\rho(\epsilon')} V(\epsilon,\epsilon')$, respectively.

While we shall primarily focus on the conventional Kondo model
[Eq.~(\ref{H_Kondo:def}) with $s=\frac{1}{2}$], we shall also present
results for the $s=1$ model and for the $s=\frac{1}{2}$, two-channel model.
Following Ref.~\onlinecite{Nozieres}, we refer to these three variants as the
``exactly screened,'' ``underscreened,'' and ``overscreened'' cases,
respectively.

The $N_c$-channel Kondo Hamiltonian,\cite{Nozieres} describing an impurity spin
degree of freedom interacting with $N_c>1$ degenerate bands (or ``channels'')
of conduction electrons, corresponds to Eq.(\ref{H_Kondo:def}) with
\begin{equation}
{\cal H}_c = \sum_{{\bf k},j} \epsilon_{\bf k}
        c^{\dag}_{{\bf k}j\sigma} c_{{\bf k}j\sigma} ,  \label{H_c2:def} \\
\end{equation}
and
\begin{equation}
{\cal H}_s = \! \sum_{{\bf k, k'}, j} \! \left[
          \frac{J_{\bf k, k'}^{(j)}}{N_0}\, c^{\dag}_{{\bf k}j\sigma}
          \case{1}{2} \bbox{\sigma}_{\sigma\sigma'} c_{{\bf k}'j\sigma'}
          \cdot{\bf s}
        \!+\! \frac{V_{\bf k, k'}^{(j)}}{N_0}\,c^{\dag}_{{\bf k}j\sigma}
          c_{{\bf k}'j\sigma} \right] ,
                                                        \label{H_s2:def}
\end{equation}
where $j = 1,$~$2$, \dots,~$N_c$ is the channel index.
In this paper we treat only the channel-symmetric version of the
two-channel problem, i.e., we take $N_c = 2$,
$J_{\bf k,k'}^{(1)} = J_{\bf k,k'}^{(2)}$, and
$V_{\bf k,k'}^{(1)} = V_{\bf k,k'}^{(2)}$.
(Multichannel variants of the Anderson model also exist, but they lie beyond
the scope of the present work.)

\subsection{Tridiagonalization of the conduction-band Hamiltonian}
\label{subsec:tridiag}

Given the form of the weighting function $w(\varepsilon)$ which defines
$f_{0\sigma}$~--- the particular linear combination of delocalized states
that interacts with the impurity degrees of freedom in the Anderson
or Kondo model --- the conduction-band Hamiltonian can be mapped exactly,
using the Lanczos procedure,\cite{Lanczos} onto a tight-binding Hamiltonian
describing a semi-infinite chain:
\begin{equation}
{\cal H}_c = D \sum^{\infty}_{n=0} \left[
   \varepsilon_n \, f^{\dag}_{n \sigma} f_{n \sigma}
  + \tau_n \, (f^{\dag}_{n \sigma} f_{n-1 , \sigma} + \text{H.c.}) \right] ,
                                                        \label{H_hop}
\end{equation}
where $\tau_0 \equiv 0$.
The operator~$f_{n\sigma}$ annihilates an electron in a spherical shell
centered on the impurity site; this shell may be reached, starting from
shell~$0$, by $n$~applications of the kinetic energy operator,
Eq.~(\ref{H_c:1}).
The $f_{n\sigma}$'s obey the anticommutation relations
$\{f^{\dag}_{n\sigma},f_{n'\sigma}\} = \delta_{n,n'}\delta_{\sigma,\sigma'}$.

The dimensionless coefficients $\varepsilon_n$~and~$\tau_n$ are determined
by the following recursion relations:\cite{Lanczos}
\begin{mathletters}
\begin{eqnarray}
\varepsilon_n &=& \langle f_{n\sigma} | {\cal H}_c/D | f_{n\sigma} \rangle , \\
\tau_{n+1} | f_{n+1 , \sigma} \! \rangle &=& ({\cal H}_c/D - \varepsilon_n)
        | f_{n \sigma} \rangle - \tau_n | f_{n-1, \sigma} \rangle ,\\
1 &=& \langle f_{n+1 , \sigma} | f_{n+1 , \sigma} \! \rangle .
\end{eqnarray}                                          \label{lanczos}%
\end{mathletters}%
Here, $| f_{n \sigma} \! \rangle = f^{\dag}_{n \sigma} | 0 \rangle$, where
$| 0 \rangle$ is the vacuum state.
There is no summation over $\sigma$ in Eqs.~(\ref{lanczos}).

The first two coefficients generated by the recursion relations are
\begin{equation}
\varepsilon_0 = F^{-2} \int^{1-\mu}_{-(1+\mu)} \! d\varepsilon
                \; \varepsilon \; w^2(\varepsilon)
                                                \label{eps_0}
\end{equation}
and
\begin{equation}
\tau_1 = F^{-2} \int^{1-\mu}_{-(1+\mu)} \! d\varepsilon
                \; (\varepsilon-\varepsilon_0)^2 \; w^2(\varepsilon).
\end{equation}
Beyond this point, the expressions for the coefficients entering
Eq.~(\ref{H_hop}) rapidly become complicated.
It is straightforward to show, however, that if the problem is symmetric about
the Fermi energy --- i.e., if $\mu = 0$ and $w(\varepsilon)=w(-\varepsilon)$
--- then $\varepsilon_n = 0$ for all~$n$.

For most functional forms of~$w(\varepsilon)$, the hopping
coefficients~$\tau_n$ rapidly converge with increasing $n$ to a constant value.
This prevents faithful approximation of the problem using any finite-length
chain, because terms in~${\cal H}_c$ involving sites $n$ which are remote from
the impurity are just as large as terms involving sites very close to the
impurity.

\subsection{Discretization of the conduction band}
\label{subsec:discrete}

Wilson showed \cite{Wilson} that, by replacing the continuum of conduction
band states by a discrete subset, one can introduce an artificial separation
of energy scales into the hopping coefficients $\tau_n$ entering
Eq.~(\ref{H_hop}).
This provides a convergent approximation to the infinite-chain
problem using finite-length chains, which correctly reproduces the impurity
contribution to system properties.
Wilson's procedure was developed for a flat conduction-band density
of states, and hence in the notation introduced above, for
$w(\varepsilon)=w_0$ (a constant).
Here, we present a generalization of the method to arbitrary $w(\varepsilon)$.

We divide the band into two sets of logarithmic energy bins, one each for
positive and negative values of $\varepsilon$.
The $m^{\text{th}}$ positive bin ($m = 0$,~1,~2,~\dots) extends over energies
\mbox{$\varepsilon^+_{m+1}<\varepsilon\le\varepsilon^+_m$}, where
\begin{equation}
\varepsilon^+_0 = 1\!-\!\mu; \quad
\varepsilon^+_m = (1\!-\!\mu) \Lambda^{1-z-m}, \; m>0.
                                                        \label{+bins}
\end{equation}
The corresponding negative bin covers the range
$\varepsilon^-_m\le \varepsilon<\varepsilon^-_{m+1}$, where
\begin{equation}
\varepsilon^-_0 = -(1\!+\!\mu); \quad
\varepsilon^-_m = -(1\!+\!\mu) \Lambda^{1-z-m}, \; m>0.
                                                        \label{-bins}
\end{equation}
Here $\Lambda$ parametrizes the discretization: numerical calculations
are typically performed with $\Lambda =$ 2--3, while the continuum
is recovered in the limit $\Lambda\rightarrow 1$.
Wilson's original treatment of the Kondo problem corresponds to setting
$\mu=0$ and $z=1$.
Values $z\not= 1$ are used in the direct calculation of
dynamical \cite{Yoshida} and thermodynamic \cite{Oliveira} quantities.
(Thermodynamic results are presented in Section~\ref{sec:results} below.)

Within the $m^{\text{th}}$ positive [negative] bin, we define a complete
set of destruction operators $a^{(q)}_{m\sigma}$ [$b^{(q)}_{m\sigma}$] and
an associated set of orthonormal functions $\psi^{(q)}_{a m} (\varepsilon)$
[$\psi^{(q)}_{b m} (\varepsilon)$], $q= 0, \pm 1, \pm 2, ...$, all of which
vanish for any $\varepsilon$ outside the $m^{\text{th}}$ positive [negative]
bin.
Given such a basis, one can write
\begin{equation}
c_{\varepsilon \sigma} = \sum_{m=0}^{\infty} \sum_{q= -\infty}^{\infty}
   [ \psi^{(q)}_{a m} (\varepsilon) a^{(q)}_{m \sigma} +
     \psi^{(q)}_{b m} (\varepsilon) b^{(q)}_{m \sigma} ]
\end{equation}
and
\begin{eqnarray}
{\cal H}_c = D \! \sum_{m,q,q'}
   \int^{1-\mu}_{-(1+\mu)} \!\!\!\! &d\varepsilon& \, \varepsilon \,
    \left[ \psi^{(q) *}_{a m} (\varepsilon) \psi_{a m}^{(q')} (\varepsilon) \,
              a^{(q) \dag}_{m \sigma} a_{m \sigma}^{(q')} \right. \nonumber \\
  &+& \left. \psi^{(q) *}_{b m} (\varepsilon) \psi_{b m}^{(q')} (\varepsilon) \,
              b^{(q) \dag}_{m \sigma} b_{m \sigma}^{(q')} \right] .
                                                          \label{H_c1}
\end{eqnarray}

The key step in generalizing the numerical~RG method to arbitrary
$w(\varepsilon)$ is the choice of a $q=0$ function within each bin
that has the same energy-dependence as $w(\varepsilon)$:
\begin{mathletters}
\begin{eqnarray}
\psi^{(0)}_{a m} (\varepsilon) &=& \left\{ \begin{array}{ll}
  w(\varepsilon) / F_{a m},  &
     \varepsilon^+_{m+1} < \varepsilon \le \varepsilon^+_m; \\[1ex]
  0,  &   \text{otherwise.}
                                 \end{array}  \right.   \\
\psi^{(0)}_{b m} (\varepsilon) &=& \left\{ \begin{array}{ll}
  w(\varepsilon) / F_{b m},  &
     \varepsilon^-_m \le \varepsilon < \varepsilon^-_{m+1}; \\[1ex]
  0,  &   \text{otherwise.}
                                 \end{array}  \right.
\end{eqnarray}
\end{mathletters}%
The orthonormality condition on these functions implies that
\begin{equation}
F^2_{a m} =
        \int^{\varepsilon^+_m}_{\varepsilon^+_{m+1}} \! d\varepsilon
              \; w^2 (\varepsilon),
\quad
F^2_{b m} =
        \int^{\varepsilon^-_{m+1}}_{\varepsilon^-_m} \! d\varepsilon
              \; w^2 (\varepsilon).
                                                \label{coeff}
\end{equation}
With this choice,
\begin{equation}
f_{0 \sigma} = F^{-1}
               \sum_{m=0}^{\infty} \left[ F_{a m} a^{(0)}_{m \sigma} +
                                          F_{b m} b^{(0)}_{m \sigma} \right],
\end{equation}
i.e., the impurity couples only to the $q=0$ mode within each bin.
Following an extension of the reasoning applied by Wilson,\cite{Wilson}
it can be shown that the coupling between modes $q \neq q'$ contained in the
kinetic energy [Eq.~(\ref{H_c1})] vanishes in the continuum
limit $\Lambda \! \rightarrow \! 1$.
To a good approximation this coupling can be neglected for $\Lambda>1$ as well.
(The ``discretization error'' arising from this approximation is estimated in
Section~\ref{sec:thermo}.)
We therefore assume that the $q \neq 0$ modes decouple completely from the
impurity, and contribute to the kinetic energy an uninteresting constant term
which is dropped henceforth.
Then,
\begin{equation}
{\cal H}_c \cong D \sum_{m=0}^{\infty}
     ( \varepsilon_{a m} \, a^{(0) \dag}_{m \sigma} a^{(0)}_{m \sigma}  +
       \varepsilon_{b m} \, b^{(0) \dag}_{m \sigma} b^{(0)}_{m \sigma} ),
                                                         \label{H_c}
\end{equation}
where
\begin{mathletters}
\begin{eqnarray}
\varepsilon_{a m} &=& F_{a m}^{-2}
        \int^{\varepsilon^+_m}_{\varepsilon^+_{m+1}} \! d\varepsilon
           \; \varepsilon \; w^2 (\varepsilon),         \\
\varepsilon_{b m} &=& F_{b m}^{-2}
        \int^{\varepsilon^-_{m+1}}_{\varepsilon^-_m} \! d\varepsilon
           \; \varepsilon \; w^2 (\varepsilon).
\end{eqnarray}
\end{mathletters}

Equation~(\ref{H_c}) can now be tridiagonalized using the Lanczos recursion
relations introduced in the previous section.
We define
\begin{equation}
f_{n \sigma} = \sum_{m=0}^{\infty} ( u_{n m} a^{(0)}_{m \sigma} +
              v_{n m} b^{(0)}_{m \sigma} ),
\end{equation}
where
\begin{equation}
u_{0m} = F_{am}/F, \qquad v_{0m} = F_{bm}/F.
\end{equation}
Then Eqs.~(\ref{lanczos}) imply that
\begin{mathletters}
\begin{eqnarray}
\varepsilon_n &=& \sum_{m} ( u^2_{n m} \varepsilon_{a m} + v^2_{n m}
               \varepsilon_{b m} ) ,            \\
\tau_{n+1} u_{n+1 , m} &=& ( \varepsilon_{a m} - \varepsilon_n ) u_{n m} -
           \tau_n u_{n-1 , m} ,                 \\
\tau_{n+1} v_{n+1 , m} &=& ( \varepsilon_{b m} - \varepsilon_n ) v_{n m} -
           \tau_n v_{n-1 , m} ,                 \\
1 &=& \sum_m (u_{n+1,m}^2 + v_{n+1,m}^2).
\end{eqnarray}                                  \label{lanczos:disc}%
\end{mathletters}%
These equations retain the feature of the undiscretized conduction band that
if $\mu = 0$ and $w(\varepsilon) = w(-\varepsilon)$, then $\varepsilon_n = 0$
for all $n$.

As will be discussed in greater detail below, the hopping coefficients
$\tau_n$ typically decrease like $\Lambda^{-n/2}$ for large $n$, while the
on-site energies $\varepsilon_n$ drop off at least this fast.
For this reason, it is convenient to work with scaled tight-binding
parameters
\begin{equation}
e_n = \alpha^{-1} \Lambda^{n/2} \varepsilon_n,
\qquad
t_n = \alpha^{-1} \Lambda^{n/2} \tau_n,
\end{equation}
where
\begin{equation}
\alpha = \case{1}{2} (1+\Lambda^{-1}) \Lambda^{\frac{3}{2}-z}
                                                \label{alpha}
\end{equation}
is a conventional factor \cite{Wilson,KWW,Yoshida} which approaches unity
in the continuum limit $\Lambda\rightarrow 1$.
With these definitions, the discretized conduction-band Hamiltonian becomes
\begin{eqnarray}
{\cal H}_c &=& \alpha D \sum_{n=0}^{\infty} \Lambda^{-n/2} \left[
        e_n f^{\dag}_{n\sigma} f_{n\sigma}
                                                \right. \nonumber \\
    & & \qquad + \; \left. t_n ( f^{\dag}_{n\sigma} f_{n-1,\sigma}
        + \text{H.c.} ) \right] .
                                                \label{H_c:disc}
\end{eqnarray}

In the special case $w(\varepsilon)=w_0$ with $\mu=1$ and $z=1$,
Wilson \cite{Wilson,xi_note} was able to derive a closed-form algebraic
expression for the hopping coefficients $t_n$.
Bulla {\it et al.}\cite{Bulla} have recently presented an ansatz for
$t_n$ when the scattering rate has the pure power-law form given in
Eq.~(\ref{Gamma_pure}).
In general, though, the algebraic expressions for the coefficients rapidly
become extremely cumbersome, and Eqs.~(\ref{lanczos:disc}) must be iterated
numerically.
A drawback of this approach is that the recursion relations prove
to be numerically unstable.\cite{Oliveira,Chen}
With double-precision arithmetic performed to roughly 16 decimal places,
it is typically possible to iterate Eqs.~(\ref{lanczos:disc}) only to $n=10$
for $\Lambda = 3$, and to $n=13$ for $\Lambda = 2$.

Chen and Jayaprakash have shown \cite{Chen} that it is possible to reorder the
calculation of the $t_n$'s in such a way as to circumvent the
instability.
In this work, however, we have adopted a brute-force approach, employing
a high-precision arithmetic package to compute the coefficients.
Typically, 120 decimal places suffice for the calculation of all coefficients
up to $n=30$, beyond which point the deviation of $t_n$ and $e_n$ from their
asymptotic values is insignificant (less than one part in $10^{15}$).

\subsection{The discretized impurity problem}
\label{subsec:discrete2}

After discretization of the conduction band, the one-impurity Anderson or
Kondo Hamiltonian can be written as the limit of a series of finite
Hamiltonians,\cite{N_note}
\begin{equation}
{\cal H} = \lim_{N \rightarrow \infty}
           \alpha \Lambda^{-N/2} D H_N, \label{H_limit}
\end{equation}
where $H_N$, describing an $(N\!+\!1)$-site chain,
is defined for all $N>0$ by the recursion relation
\begin{eqnarray}
H_N &=& \Lambda^{1/2} H_{N-1}
        + e_N f^{\dag}_{N\sigma} f_{N\sigma}    \nonumber \\
      & & +\; t_N (f^{\dag}_{N\sigma} f_{N-1,\sigma}
        + \text{H.c.}) - E_{G,N} .
                                                                \label{H_N}
\end{eqnarray}
Here, $E_{G,N}$ is chosen so that the ground-state energy of
$H_N$ is zero.
The Hamiltonian $H_0$ describes the atomic limit of the impurity
problem.
For the Anderson model,
\begin{eqnarray}
H_0 &=& e_0 f^{\dag}_{0\sigma} f_{0\sigma}
        + \tilde{\varepsilon}_d n_d + \tilde{U} n_{d\uparrow} n_{d\downarrow}
                                                        \nonumber \\
    & & + \; \tilde{\Gamma}^{1/2} (f^{\dag}_{0\sigma}d_{\sigma}+\text{H.c.})
        - E_{G,0},
                                                        \label{H_0A}
\end{eqnarray}
where
\begin{equation}
\tilde{\varepsilon}_d = \frac{\epsilon_d}{\alpha D},
\qquad
\tilde{U} = \frac{U}{\alpha D},
\qquad
\tilde{\Gamma} = \frac{F^2 \Gamma_0}{\pi \alpha^2 D} ;
                                                        \label{A_couplings}
\end{equation}
while for the Kondo models,
\begin{equation}
H_0 = (e_0 + \tilde{V})
        f^{\dag}_{0\sigma} f_{0\sigma} + \tilde{J} f^{\dag}_{0\sigma}
        \case{1}{2} \bbox{\sigma}_{\sigma,\sigma'} f_{0\sigma'} \cdot {\bf s}
        - E_{G,0},
                                                        \label{H_0K}
\end{equation}
with
\begin{equation}
\tilde{J} = F^2 \rho_0 J_0 / \alpha ,
\qquad
\tilde{V} = F^2 \rho_0 V_0 / \alpha .
                                                        \label{K_couplings}
\end{equation}
In the two-channel variant of the Kondo model, each $f$ and $f^{\dag}$
operator in Eqs.~(\ref{H_N}) and~(\ref{H_0K}) acquires a channel index $j$,
which is summed over.

An important feature of both the Anderson and Kondo Hamiltonians is their
behavior under the following particle-hole transformations:
\begin{equation}
\begin{array}{ll}
\text{Anderson} \quad &
        f_{n\sigma} \rightarrow (-1)^n f^{\dag}_{n,\sigma}, \quad
        d_{\sigma} \rightarrow -d^{\dag}_{\sigma}; \\[1ex]
\text{Kondo} &
        f_{n\sigma} \rightarrow (-1)^n f^{\dag}_{n,\sigma}, \quad
        {\bf s} \rightarrow -{\bf s}^{\dag}.
\end{array}
                                                \label{ph_trans}
\end{equation}
Examination of Eqs.~(\ref{H_N})--(\ref{K_couplings}) indicates that the
effective values of $t_n$, $\tilde{\Gamma}$, $\tilde{U}$ and $\tilde{J}$
remain unchanged under the transformations, but that $e_n \rightarrow -e_n$,
$\tilde{\varepsilon}_d \rightarrow -(\tilde{\varepsilon}_d+\tilde{U})$, and
$\tilde{V} \rightarrow -\tilde{V}$.

A symmetric weighting function such that $w(-\varepsilon)=w(\varepsilon)$
guarantees that $e_n=0$ for all~$n$.
In this case, one sees that the physical properties of the Anderson model
are identical for impurity energies $\epsilon_d$ and $-(\epsilon_d+U)$,
all other parameters being the same.
Thus, it is necessary to consider only $\epsilon_d \ge -U/2$
(or $\tilde{\varepsilon}_d \ge -\tilde{U}/2$) in order to fully explore the
physical properties of the model.\cite{KWW}

It should further be noted that, provided $w(-\varepsilon)=w(\varepsilon)$,
the symmetric Anderson model (defined by the condition $U+2\epsilon_d = 0$)
and the Kondo models with zero potential scattering ($V_0=0$) are
completely invariant under the transformations in Eqs.~(\ref{ph_trans}).
In cases where the impurity scattering rate is regular [$w(0) > 0$], the
presence or absence of particle-hole symmetry does little to affect the
physics.
By contrast, this symmetry turns out to play a crucial role in determining the
strong-coupling behavior of systems with power-law scattering.

\subsection{Iterative solution of the discretized problem}
\label{subsec:iterate}

The sequence of Hamiltonians $H_N$ defined by Eq.~(\ref{H_N}) can be solved
iteratively in the manner described in Refs.~\onlinecite{Wilson}
and~\onlinecite{KWW}.
The many-body eigenstates of iteration $N\!-\!1$ are used to construct the
basis for iteration $N$.
Before each step, the Hamiltonian is rescaled by a factor of $\Lambda^{1/2}$
so that the smallest scale in the energy spectrum remains of order unity,
and at the end of the iteration the ground-state energy is subtracted from
each eigenvalue.
This procedure is repeated until the eigensolution approaches a fixed point, at
which the low-lying eigenvalues of $H_N$ are identical to those of $H_{N+2}$.
(The spectra of $H_N$ and $H_{N+1}$ do not coincide because of a fundamental
inequivalence between the eigensolutions for chains containing odd and even
numbers of sites.
For example, particle-hole symmetry ensures the existence of a zero
eigenvalue for \mbox{$N\!+\!1$ odd}, whereas there is no such restriction for
\mbox{$N\!+\!1$ even}.)

All the Hamiltonians described in this paper commute with the total spin
operator
\begin{equation}
   {\bf S}_N = \sum_{n=0}^N f^{\dag}_{n\sigma} \case{1}{2}
                              \bbox{\sigma}_{\sigma\sigma'} f_{n\sigma'}
               + d^{\dag}_{\sigma} \case{1}{2}
                              \bbox{\sigma}_{\sigma\sigma'} d_{\sigma'},
\end{equation}
and with the charge operator
\begin{equation}
   Q_N = \sum_{n=0}^N (f^{\dag}_{n\sigma} f_{n\sigma} - 1) + (n_d - 1).
                                                        \label{charge:def}
\end{equation}
At particle-hole symmetry, $H_N$ commutes with all three components of
an ``axial charge'' operator ${\bf J}_N$:\cite{Jones}
\begin{mathletters}
\begin{eqnarray}
   J_{z,N} &=& \case{1}{2} Q_N,                           \\
   J_{+,N} &=& \sum_{n=0}^N (-1)^n f^{\dag}_{n\uparrow} f^{\dag}_{n\downarrow}
           - d^{\dag}_{\uparrow} d^{\dag}_{\downarrow}, \\
   J_{-,N} &=& (J_{+,N})^{\dag}.
\end{eqnarray}                                          \label{axial:def}%
\end{mathletters}%
Thus, the Hamiltonian $H_N$ can be diagonalized independently in subspaces
labeled by different values of the quantum numbers $S_z$, $S$, $Q$, and
(at particle-hole symmetry)~$J$.
Moreover, the energy eigenvalues are independent of~$S_z$, and when $J$~is a
good quantum number they are also independent of~$Q$.
It is therefore possible to perform the numerical~RG calculations using a
reduced basis consisting only of states with $S_z = S$ and, where appropriate,
$Q = -2J$.
(See Refs.~\onlinecite{KWW} and~\onlinecite{Jones} for further details.
Note that in the two-channel Kondo model, axial charge quantum numbers
$Q^{(j)}$ and $J^{(j)}$ can be defined separately for channel $j = 1$ and~$2$.)
By taking advantage of these symmetries of the Hamiltonian, the numerical
effort required to diagonalize $H_N$ can be considerably reduced.

Even with the optimizations described in the previous paragraph, after only
a few iterations the Hilbert space of $H_N$ becomes too large for a complete
solution to be feasible.
Instead, the basis is truncated according to one of two possible
strategies, which give essentially the same results: one either retains the
$M$ states of lowest energy, where $M$ is a predetermined number; or one
retains all eigenstates having an energy within some range $E_c$ of
the ground-state energy.
A compromise must be made between two conflicting goals: accurate reproduction
of results for the original undiscretized system, favored by choosing $\Lambda$
to be close to unity to minimize {\em discretization error}, and by making
$M$ or $E_c$ large to reduce {\em truncation error\/}; and a short
computational time, which points to large $\Lambda$ and small $M$ or $E_c$.
Unless otherwise noted, the thermodynamic quantities presented in this paper
were obtained using $\Lambda = 3$ and $E_c \ge 25$, for which choices the
primary source of error is the discretization.
(The magnitude of the error is estimated in Section~\ref{subsec:thermo_num}.)

In summary, the numerical~RG method can be generalized in a fairly
straightforward manner to treat energy-dependent scattering
of conduction electrons from the impurity site.
The conduction band is divided into logarithmic energy bins just as for the
case of a constant scattering rate.
However, the mode expansion within each energy bin has to be modified in order
that the impurity couples to a single mode ($q=0$).
This in turn alters the hopping coefficients $t_n$ which enter the
recursive definition of the Hamiltonians $H_N$ [see Eq.~(\ref{H_N})];
away from strict particle-hole symmetry, furthermore, the hopping terms are
complemented by on-site terms of the form
$e_n f^{\dag}_{n\sigma} f_{n\sigma}$.
Finally, a factor $F$, which depends on the overall normalization of
the weighting function $w(\varepsilon)$, is introduced into the starting
Hamiltonian~$H_0$ [see Eqs.~(\ref{H_0A})--(\ref{K_couplings})].

\section{Conduction Band Hamiltonians}
\label{sec:band}

For the remainder of the paper, we focus on systems in which the scattering
rate $\Gamma(\epsilon)$ vanishes in power-law fashion at the Fermi level.
Initially, we consider the simplest possible form of $\Gamma(\epsilon)$,
given by Eq.~(\ref{Gamma_pure}).
In the notation of Section~\ref{sec:general}, this corresponds to a
particle-hole-symmetric problem in which $\mu = 0$ and the weighting function
$w(\varepsilon)$ satisfies
\begin{eqnarray}
w(\varepsilon) = \left\{ \begin{array}{ll}
        |\varepsilon|^{r/2} , \qquad & |\varepsilon| \le 1; \\[1ex]
        0, & \text{otherwise} . \end{array} \right.
                                                     \label{w_pure}
\end{eqnarray}
Here $r$ can take any non-negative value; $r=0$ corresponds to a constant
density of states.
Subsequently, we shall generalize this form to restrict the power-law variation
in $w(\epsilon)$ to a finite region, and to allow for particle-hole asymmetry.
However, the essential physics is captured by the prototypical function
in Eq.~(\ref{w_pure}).

In this section, we apply the formalism described in Section~\ref{sec:general}
to construct and analyze a family of discretized conduction-band Hamiltonians
based on Eq.~(\ref{w_pure}),
\begin{equation}
H_N^{(L)} = \sum_{n=L}^N
        \Lambda^{(N-n)/2}  [ e_n f^{\dag}_n f_n
        + t_n ( f^{\dag}_n f_{n-1} + \text{H.c.} ) ].
                                                   \label{H_N^L}
\end{equation}
The parameter $L$ determines the innermost shell $f_L$ onto which conduction
electrons can hop, so $t_L$ is necessarily zero.
The case $L=0$ represents the free-electron problem, while $L=1$ and $L=2$
will turn out to describe electronic excitations at different strong-coupling
fixed points of the Anderson and Kondo models.
In each of these limits, spin-up and spin-down electrons decouple from
one another, so the index $\sigma$ can be dropped.

All information about the energy-dependence of scattering from the impurity
site enters the discretized Anderson and Kondo Hamiltonians through the
normalization factor $F$, and through the tight-binding parameters
$e_n$~and~$t_n$.
From Eq.~(\ref{F}), it is straightforward to see that $F^2=2/(1+r)$, while the
particle-hole symmetry of Eq.~(\ref{w_pure}) ensures that $e_n=0$ for all~$n$.
By contrast, the coefficients $t_n$ must be determined numerically,
as outlined in Section~\ref{subsec:discrete}.
(Bulla {\it et al.}\cite{Bulla} have recently deduced an algebraic formula
which appears to fit the numerical value of~$t_n$ for all~$n$.)

Values of $t_n$ for $r = 0$, $0.2$, and~$1$ are illustrated in
Table~\ref{table:coeffs}.
Whereas for a constant scattering rate ($r=0$), $t_n$~rapidly
approaches unity, in the power-law cases the asymptotic value
alternates with the parity of~$n$:
\begin{equation}
\lim_{n\rightarrow\infty} t_n = \left\{
   \begin{array}{ll}
      t^{\star}, & \text{$n$ odd;} \\
      t^{\star} \Lambda^{-r/2}, \qquad & \text{$n$ even;}
   \end{array} \right.
                                                \label{t_lim}
\end{equation}
where
\begin{equation}
t^{\star} =
        \frac{2}{1+\Lambda^{-1}} \frac{1\!+\!r}{2\!+\!r} \;
        \frac{1\!-\!\Lambda^{-(2+r)}}{1\!-\!\Lambda^{-(1+r)}}.
                                                \label{t*}
\end{equation}

\begin{table}[t]
\begin{tabular}{rccc}
  & \multicolumn{3}{c}{scaled hopping coefficient, $t_n$} \\[0.5ex]
\cline{2-4} \\[-2ex]
n & $r = 0$ & $r = 0.2$ & $r = 1$ \\[0.5ex] \hline \\[-2ex]
  1 & 0.8320502943 & 0.8839953062 & 1.0277402396 \\
  2 & 0.9076912302 & 0.8239937041 & 0.5598150205 \\
  3 & 0.9651711910 & 0.9890229755 & 1.0707750620 \\
  4 & 0.9879052953 & 0.9005285244 & 0.6177933421 \\
  5 & 0.9959128837 & 1.0141188271 & 1.0818574579 \\
  6 & 0.9986313897 & 0.9103510027 & 0.6246054709 \\
  7 & 0.9995431009 & 1.0170838067 & 1.0831683420 \\
  8 & 0.9998476229 & 0.9114602761 & 0.6253674693 \\
  9 & 0.9999491990 & 1.0174154894 & 1.0833149885 \\
 10 & 0.9999830654 & 0.9115837523 & 0.6254521995 \\
 11 & 0.9999943550 & 1.0174523707 & 1.0833312949 \\
 12 & 0.9999981183 & 0.9115974746 & 0.6254616147 \\
 13 & 0.9999993728 & 1.0174564690 & 1.0833331068 \\
 14 & 0.9999997909 & 0.9115989993 & 0.6254626609 \\
 15 & 0.9999999303 & 1.0174569244 & 1.0833333082 \\
 16 & 0.9999999768 & 0.9115991687 & 0.6254627771 \\
 17 & 0.9999999923 & 1.0174569750 & 1.0833333305 \\
 18 & 0.9999999974 & 0.9115991876 & 0.6254627900 \\
 19 & 0.9999999991 & 1.0174569806 & 1.0833333330 \\
 20 & 0.9999999997 & 0.9115991897 & 0.6254627914 \\
 21 & 0.9999999999 & 1.0174569812 & 1.0833333333 \\
 22 & 1.0000000000 & 0.9115991899 & 0.6254627916 \\
 23 & 1.0000000000 & 1.0174569813 & 1.0833333333 \\
 24 & 1.0000000000 & 0.9115991899 & 0.6254627916 \\
 25 & 1.0000000000 & 1.0174569813 & 1.0833333333
\end{tabular}
\vspace{1ex}
\caption{
Tight-binding parameters~$t_n$ for three different
powers~$r$ entering Eq.~(\protect\ref{w_pure}).
}
\label{table:coeffs}
\end{table}

The asymptotic behavior of~$t_n$ given by Eq.~(\ref{t_lim})
should be contrasted with that obtained for all weighting functions
in which $w(0)$ is finite and nonzero:
\begin{equation}
\lim_{n\rightarrow\infty} e_n = 0,
\qquad
\lim_{n\rightarrow\infty} t_n = 1.
\end{equation}
All such problems can be showed to exhibit essentially the same
physics,\cite{Wilson,KWW} and in particular to be described by the same set
of RG~fixed points.
The functional form of $w(\varepsilon)$ away from the Fermi energy determines
only the deviations of $e_n$ and $t_n$ from their
asymptotic values.
These deviations act as irrelevant perturbations in the RG~sense.
The only exception is $e_0$, which acts as a marginal
variable, equivalent to an additional potential scattering term of the type
appearing in Eq.~(\ref{H_s:def}).
The scenario of power-law scattering considered in this paper is
interesting precisely because $w(0) = 0$, which places the impurity problem
in a different universality class.

\subsection{Free-electron Hamiltonian ($L=0$)}
\label{subsec:L=0}

The free-electron Hamiltonian, describing the noninteracting conduction
band in the absence of any magnetic impurity level, corresponds to the
large-$N$ limit of $H_N^{(0)}$ as defined in Eq.~(\ref{H_N^L}).

{\em Eigenvalues:}
Numerical diagonalization indicates that for large $N$, the eigenvalues
approach limiting values, which we denote
\begin{eqnarray}
\text{$N$ odd:} &\quad&
        \eta^{\star}_j, \; j=\pm 1,\pm 2,\ldots,
                \pm\case{1}{2}(N\!+\!1);
                                                \nonumber \\[-1.5ex]
                                                \label{eta*:def} \\[-1.5ex]
\text{$N$ even:} & &
        \hat{\eta}^{\star}_j, \; j=0,\pm 1,\pm 2,\ldots,
                \pm\case{1}{2}N.
                                                \nonumber
\end{eqnarray}
Due to the particle-hole symmetry, $\hat{\eta}^{\star}_0 = 0$.
For $|j| \gg 1$ the eigenvalues are well-approximated by
\begin{equation}
\eta^{\star}_j, \; \hat{\eta}^{\star}_j =
        \text{sgn}(j) \; t^{\star} \Lambda^{|j|-\nu_N},            \label{eta*}
\end{equation}
where
\begin{equation}
\nu_N = \left\{
        \begin{array}{ll}
        1, \qquad & \text{$N$ odd;} \\[1ex]
        \case{1}{2}, & \text{$N$ even.}
        \end{array} \right.
\end{equation}

{\em Eigenvectors:}
We also consider the single-particle eigenstates of $H_N^{(0)}$,
associated with particle operators $g_j$ ($j \ge 0$) and
hole operators $h_{-j}$ ($j < 0$).
It will later prove useful to have expansions of the original operators $f_n$
in terms of these eigenoperators:
\begin{equation}
f_n = \left\{
    \begin{array}{ll}
        \sum^{(N+1)/2}_{j=1} A_{n j} [
        g_j + (-1)^n h^{\dag}_j ],
      & \text{$N$ odd;} \\[2ex]
        A_{n 0} g_0 + \sum_{j=1}^{N/2} A_{n j}
        [ g_j + (-1)^n h^{\dag}_j ],
      & \text{$N$ even.}
    \end{array}
\right.
                                                \label{f_n:L=0}
\end{equation}
It is found, again numerically, that for $j\gg 1$, the coefficients $A_{0j}$
and $A_{1j}$ separate into an $N$-dependent prefactor and a part that depends
only on the parity of $N$:
\begin{equation}
A_{n j} = \Lambda^{-(2n+1+r)N/4} \alpha_{nj}, \qquad n = 0,\; 1,
                                                \label{A_nj}
\end{equation}
where
\begin{eqnarray}
\alpha_{nj} &=& \alpha_n(z) \Lambda^{(2n+1+r)(j-\nu_N)/2},      \nonumber \\
{[\alpha_0(z)]^2} &=& \case{1}{2} [ 1 - \Lambda^{-(1 + r)} ]
        \Lambda^{(1+r)(z-1/2)},                         \label{alphas} \\
{[\alpha_1(1)]^2} &=& \case{1}{2} [ 1 - \Lambda^{-(3 + r)} ] \Lambda^{(3+r)/2}.
                                                                \nonumber
\end{eqnarray}
(The general $z$-dependence of $\alpha_1$ cannot be written so compactly
as that of $\alpha_0$.)

Just as in the standard case $r=0$, the expansion of all
other $f_n$'s becomes more complicated.\cite{Wilson}
For $n>1$, $f_n$ contains components which vary as
$\Lambda^{-(2m+1+r)N/4}$, for $m = n$, $n\!-\!2$, $n\!-\!4$,
\dots,~$n\,\text{mod}\,2$.

We emphasize that Eqs.~(\ref{eta*}), (\ref{A_nj}) and~(\ref{alphas}) are
very good approximations even for comparatively small values of $N$ and $j$.
For $\Lambda = 3$, for instance, these formulae hold to at least
seven decimal places for all $3 \le j \le N/2$.
The rate of convergence to the asymptotic forms with increasing $j$ seems
to be independent of $r$, at least in the range $0\le r \le 2$.

\subsection{Symmetric strong-coupling Hamiltonian (L=1)}
\label{subsec:L=1}

In Section~\ref{subsec:symm_strong}, we shall discuss the {\em symmetric
strong-coupling\/} fixed point of the Anderson and Kondo models.
At this fixed point, an infinite coupling between the localized level and
conduction electrons at the impurity site completely suppresses hopping of
conduction electrons onto or off shell~0, i.e., the $f_0$ degrees of freedom
are ``frozen out.''
In this situation, the conduction-band excitations of the system are described
by an effective Hamiltonian $H_N^{(1)}$ obtained by setting $L=1$ in
Eq.~(\ref{H_N^L}).

{\em Eigenvalues:}
For large $N$, the eigenvalues of $H_N^{(1)}$ are found to approach
limiting values which we denote
\begin{eqnarray}
\text{$N$ odd:} &\quad&
        \omega^{\star}_j, \; j=0, \pm 1,\pm 2,\ldots,
                \pm\case{1}{2}(N\!-\!1);
                                                \nonumber \\[-1.5ex]
                                                \\[-1.5ex]
\text{$N$ even:} & &
        \hat{\omega}^{\star}_j, \; j=\pm 1,\pm 2,\ldots,
                \pm\case{1}{2}N.
                                                \nonumber
\end{eqnarray}
Due to the particle-hole symmetry, $\omega^{\star}_0 = 0$, while
for \mbox{$N/2 \gg |j| \gg 1$}, the eigenvalues are well-approximated by
\begin{equation}
\omega^{\star}_j, \; \hat{\omega}^{\star}_j =
\text{sgn}(j) \; (t^{\star} \Lambda^{-r_1/2}) \Lambda^{|j|-\nu_{N\!-\!1}},
                                                                \label{omega*}
\end{equation}
where $r_1\equiv\min(r,1)$.
Alternatively, one can write
\begin{equation}
\omega^{\star}_j = \Lambda^{- r_1/2} \hat{\eta}^{\star}_j, \qquad
\hat{\omega}^{\star}_j = \Lambda^{- r_1/2} \eta^{\star}_j .
\end{equation}

These expressions point to a significant difference between the cases $r=0$
and $r>0$.
In the former instance, the strong-coupling energies given by
$H_N^{(1)}$ become identical to the weak-coupling energies of
$H_{N-1}^{(0)}$ in the limit of large $N$.
In other words, the transition from weak to strong coupling is equivalent
to an interchange between the \mbox{odd-$N$} and \mbox{even-$N$}
spectra.\cite{Wilson}
This relation between weak and strong coupling no longer holds true
in the presence of a power-law scattering rate.
For $r>1$, however, an even simpler pattern emerges:
$\omega^{\star}_j = \eta^{\star}_j$  ($j\not= 0$) and
$\hat{\omega}^{\star}_j = \hat{\eta}^{\star}_{j-\text{sgn}(j)}$.
Thus, for superlinear scattering rates, the strong-coupling spectrum differs
from that at weak coupling only by the insertion of one additional zero-energy
eigenstate.

{\em Phase shifts:}
The $L=1$ eigenvalues can also be expressed in terms of $s$-wave phase shifts
$\delta_0(\epsilon)$ applied to the $L=0$ spectrum: for $N$~odd,
\begin{equation}
   \omega^{\star}_j = \Lambda^{-\text{sgn}(\eta^{\star}_j)
        \delta_0(\alpha\Lambda^{-N/2}\eta^{\star}_j)/\pi} \eta^{\star}_j,
\end{equation}
and similarly for $N$~even.
The values of $\delta_0(\epsilon)$ in the limits
$\epsilon\rightarrow 0^{\pm}$ can be deduced from Eqs.~(\ref{eta*})
and~(\ref{omega*}), while the form of the leading corrections away from the
Fermi energy can be inferred by studying either pure potential-scattering from
the impurity site\cite{Chen} or power-law mixing between conduction electrons
and a noninteracting resonant level (see Appendix~\ref{app:reslevel}).
The result is
\begin{equation}
\delta_0(\epsilon) = (1-r_1)\,\frac{\pi}{2}\,\text{sgn}(-\epsilon)
        + {\cal O}(|\epsilon/D|^{|1-r|}).                \label{delta_0}
\end{equation}
The main feature of Eq.~(\ref{delta_0}) is a jump of $(1-r_1)\pi$ in the phase
shift on crossing the Fermi energy.
The interpretation of this jump will be deferred until
Section~\ref{subsubsec:thermo_SSC}.

{\em Eigenvectors:}
We can expand the annihilation operators $f_{n \sigma}$ in terms of
single-particle eigenoperators of $H_N^{(1)}$:
\begin{equation}
f_n = \left\{
    \begin{array}{ll}
        B_{n 0} g_0 + \sum^{(N-1)/2}_{j=1} B_{n j} [
        g_j - (-1)^n h^{\dag}_j ],
      & \text{$N$ odd;} \\[2ex]
        \sum_{j=1}^{N/2} B_{n j}
        [ g_j - (-1)^n h^{\dag}_j ],
      & \text{$N$ even.}
    \end{array}
\right.
                                                \label{f_n:L=1}
\end{equation}
For sufficiently large $N$ and $N/2 \gg j\gg 1$, the coefficients $B_{1j}$
and $B_{2j}$ separate into an $N$-dependent prefactor and a part that depends
on $N$ only through $\nu_{N-1}$:
\begin{equation}
B_{n j} = \Lambda^{-(2n-2+|1-r|)N/4} \beta_{nj}, \qquad n = 1,\; 2,
                                                \label{B_nj}
\end{equation}
where
\begin{equation}
\beta_{nj} = \beta_n \Lambda^{(2n-2+|1-r|)(j-\nu_{N\!-\!1})/2}.
                                                \label{beta_nj}
\end{equation}
The parameters $\beta_n$ can be determined numerically, but we have not
obtained algebraic expressions for their dependence on $\Lambda$ and $r$.
There is an important exception to Eq.~(\ref{B_nj}) for $r>1$:
in the limit $N\rightarrow\infty$, $B_{10}$ for $N$~odd [$B_{11}$ for $N$~even]
approaches a constant value which is independent of $N$.
Therefore, the effective scaling is
$f_1 \propto \Lambda^{-(1-r_1) N/4}$ and $f_2 \propto \Lambda^{(2+|1-r|)N/4}$.
These forms will turn out to have important implications for the stability of
the symmetric strong-coupling fixed point.

The approximate expressions for the eigenvalues and eigenvectors of $H_N^{(1)}$
do not apply so widely as those for $L=0$.
First, Eqs.~(\ref{omega*}), (\ref{B_nj}), and (\ref{beta_nj}) are restricted
not only to $j\gg 1$, but also to $j\ll N/2$.
Second, the rate of convergence with increasing $N$ is slower than for $L=0$
and depends explicitly on $r$: for instance, the deviation of each
eigenvalue ($\omega_j$ or $\hat{\omega}_j$) from its large-$N$ limit
($\omega^{\star}_j$ or $\hat{\omega}^{\star}_j$) is proportional to
$\Lambda^{-|1-r|N/2}$.
There is again an exception for $r>1$: the two smallest even-$N$ eigenvalues
obey the relation $\hat{\omega}_{\pm 1}\propto\Lambda^{-(r-1)N/4}$, and hence
converge to their asymptotes ($\hat{\omega}^{\star}_{\pm 1}=0$) even more
slowly than the other eigenvalues.

The results above provide clear evidence that a linear scattering rate
represents a singular case.
The expansion of $f_1$ acquires an $N$-independent component at $r=1$, and
for this value of $r$ alone several quantities (the eigenvalues and
eigenvectors as functions of $N$, the conduction-band phase shift as a
function of energy) converge in a logarithmic, rather than exponential,
fashion.
As pointed out by Cassanello and Fradkin,\cite{Cassanello} this case in
some sense represents the upper critical dimension of the theory.

\subsection{Asymmetric strong-coupling Hamiltonian ($L=2$)}
\label{subsec:L=2}

It will be shown below that, in most cases, the stable strong-coupling fixed
point of the Anderson and Kondo problems is not described by the
conduction-band Hamiltonian $H^{(1)}_N$ introduced in the previous subsection.
Instead, the system reaches either the {\em frozen-impurity\/} fixed point,
described by $H^{(0)}_N$, or the {\em asymmetric strong-coupling\/} fixed
point, at which both the $f_0$ and $f_1$ degrees of freedom are frozen out
and the low-lying excitations of the system are described by $H_N^{(2)}$.
Due to the asymptotic form of the hopping coefficients $t_n$
[see Eq.~(\ref{t_lim})], which only depend on whether $n$ is odd or even,
the low-energy properties of $H_N^{(2)}$ are equivalent
to those of $H_{N-2}^{(0)}$, provided that one relabels the
operators $f_n$ appropriately, i.e., $f_n$ at the $L=2$ fixed point has the
same expansion as $f_{n-2}$ at the $L=0$ fixed point.
The $L=2$ eigenstates are related to those for $L=0$ by a low-energy phase
shift $\delta_0(\epsilon)= \pi\text{sgn}(-\epsilon)$.

\section{Strong- and Weak-Coupling Fixed Points of the Impurity Models}
\label{sec:fixed}

In this section, we analyze RG~fixed points of the Anderson and Kondo models
with a power-law scattering rate described by Eq.~(\ref{w_pure}).
The focus is on hosts having a single conduction channel, although
reference will be made in passing to the two-channel Kondo model.

We consider only those fixed points that can be obtained by setting each
impurity parameter entering $H_0$ [see Eqs.~(\ref{H_0A}) and~(\ref{H_0K})]
either to zero or to infinity.
In such cases, a local Fermi-liquid description applies: the low-energy
many-body excitations of the system can be constructed as the product of
two independent sets of single-particle excitations, one set describing the
conduction band, the other arising from any active impurity degrees of freedom.
By studying small deviations from the fixed-point Hamiltonian, one can
determine the stability of the fixed point and the functional dependence of
certain physical properties.
These results can be obtained by largely algebraic means.
(The only numerical step is the derivation of the results presented in
Section~\ref{sec:band}, which involves diagonalization of simple quadratic
Hamiltonians.)

In Section~\ref{sec:results}, we shall also discuss a number of fixed points
which appear at intermediate (neither zero nor infinite) couplings.
Such fixed points are generally non-Fermi-liquid in nature, and at present can
be studied only via a full implementation of the numerical~RG scheme
outlined in Section~\ref{sec:general}.

\subsection{Stability of RG~fixed points}
\label{subsec:stability}

Within the nonperturbative RG approach, a fixed-point Hamiltonian
$H^{\star}$ satisfies
\begin{equation}
   H_{N+2} = H_N = H^{\star}.
\end{equation}
In this context, ``='' means that two Hamiltonians have identical low-energy
spectra, and that they share the same set of matrix elements of any physically
significant operator between their low-lying eigenstates.\cite{Wilson}

Any deviation from a fixed point Hamiltonian must be describable in the form
\begin{equation}
\delta H_N = H_N - H^{\star}
        = \sum_{\gamma} \tilde{\gamma} O_{\gamma},     \label{delH}
\end{equation}
where $\tilde{\gamma}$ is a dimensionless coupling,
and $O_{\gamma}$ is composed of operators associated with
those degrees of freedom (from among $f_{n\sigma}$, $d_{\sigma}$ or $\bf s$)
that remain active at the fixed point, multiplied by an overall factor
of $\Lambda^{N/2}$ which reproduces the scaling of $H_N$ implied
by Eq.~(\ref{H_N}).
The only constraint on the combination of operators entering
$O_{\gamma}$ is that the perturbation must preserve all symmetries
of the original model.

As explained in detail in Refs.~\onlinecite{Wilson} and~\onlinecite{KWW},
one can use the expansion of the operators $f_n$ developed in
Section~\ref{sec:band} to analyze the stability of the strong- and
weak-coupling fixed points points with respect to all possible perturbations.
Consider, for example, the weak-coupling limit in which the electronic degrees
of freedom are described by $H_N^{(0)}$.
Equations~(\ref{f_n:L=0}) and~(\ref{A_nj}) imply that the perturbation
$O_V = \Lambda^{N/2} (f^{\dag}_{0\sigma} f_{0\sigma} - \frac{1}{2})$
can be written as $\Lambda^{-r N/2}$ times an $N$-independent part composed
of the single-particle and single-hole operators, $g_j$ and $h_j$.
Making use of the effective temperature $T_N$ associated with iteration
$N$ (see Section~\ref{sec:thermo} for more details),
\begin{equation}
k_{\text{B}} T_N = \alpha \Lambda^{-N/2} D / \bar{\beta},
                                                        \label{T_N}
\end{equation}
where $\bar{\beta}$ is a small dimensionless parameter,
one sees that $O_V \propto T^r$, i.e., the perturbation is
{\em irrelevant\/} for all $r>0$.

In the remainder of this section, we identify the most relevant (or least
irrelevant) operators in the vicinity of the various Fermi-liquid fixed points.
At each fixed point, the expansion of $f_n$ ($n \ge L$) contains a piece which
varies like $T^{(n-L)+\frac{1}{2}|1\pm r|}$.
The dominant perturbations are therefore those operators $O_{\gamma}$ that
contain the fewest possible $f_n$'s, and in which the $f_n$'s that are present
have the smallest possible indices $n$.

We shall present our analysis in the context of the nondegenerate Anderson
model.
Features of the various Kondo models will be noted where they are different.

\subsection{Free-impurity fixed point}
\label{subsec:free}

The {\em free-impurity\/} or ``free-orbital'' \cite{KWW} fixed point of the
Anderson model corresponds to setting
$\tilde{\varepsilon}_d = \tilde{U} = \tilde{\Gamma} = 0$ in Eq.~(\ref{H_0A}).
This fixed point, which has no analogue in the Kondo models, is described by
an effective Hamiltonian
\begin{equation}
H_{\text{\it WC},N}^{\star} = H_N^{(0)} - E_{G,N}.
                                                \label{H_weak}
\end{equation}
Each many-body eigenstate is the product of an eigenstate of $H_N^{(0)}$
and a zero-energy eigenstate of a free impurity level.

By combining the reasoning outlined in the previous subsection with the
$N$-dependences given in Section~\ref{subsec:L=0}, one can identify four
operators which are, or may be, relevant in the vicinity of the
free-impurity fixed point:
\begin{equation}
\begin{array}{lllll}
O_{\epsilon_d} &=&
        \Lambda^{N/2} (n_d - 1)
        &\propto& T^{-1}, \\[1ex]
O_U &=&
        \Lambda^{N/2} (n_d - 1)^2
        &\propto& T^{-1}, \\[1ex]
O_{\Gamma} &=&
        \Lambda^{N/2} (f^{\dag}_{0\sigma} d_{\sigma}\!+\!\text{H.c.})
        &\propto& T^{-(1-r)/2}, \\[1ex]
O_{\Gamma_d} &=&
        \Lambda^{N/2} n_{d,-\sigma}
        (f^{\dag}_{0\sigma} d_{\sigma}\!+\!\text{H.c.})
        &\propto& T^{-(1-r)/2}.
\end{array}
                                                \label{weak_delH}
\end{equation}
Here, $O_{\epsilon_d}$, $O_U$, and $O_{\Gamma}$ are essentially equivalent to
the on-site energy, on-site Coulomb repulsion, and hybridization terms
(respectively) in the original Hamiltonian, while $O_{\Gamma_d}$~represents
correlated hybridization.
Of these operators, only $O_U$~and~$O_{\Gamma}$ respect particle-hole symmetry
and are allowed in the symmetric limit of the Anderson model.
Note that $O_{\epsilon_d}$~and~$O_U$ are always
relevant, whereas $O_{\Gamma}$~and~$O_{\Gamma_d}$ are
relevant for $r<1$ but are irrelevant for $r>1$.
Since there is at least one relevant operator for both the symmetric and
asymmetric cases, and also for all~$r$,
{\em the free-impurity fixed point is always unstable.}

\subsection{Valence-fluctuation fixed point}
\label{subsec:valence}

The {\em valence-fluctuation\/} fixed point \cite{KWW} of the Anderson model
corresponds to the original model with
$\tilde{\varepsilon}_d=\tilde{\Gamma}=0$, but $\tilde{U} = \infty$.
This is clearly not a fixed point of the symmetric Anderson model since
$\tilde{U}+2\tilde{\varepsilon}_d \not = 0$, and it has no analogue in the
Kondo models.
It is described by the same effective Hamiltonian as the free-impurity fixed
point, but doubly occupied impurity configurations are eliminated from the
Hilbert space.
Of the four dominant perturbation at the free-impurity fixed point
[see Eqs.~(\ref{weak_delH})], only $O_{\epsilon_d}$ and $O_{\Gamma}$ survive.
Since the former is a relevant operator for all $r$,
{\em the valence-fluctuation fixed point is always unstable.}

\subsection{Local-moment fixed point}
\label{subsec:local}

The {\em local-moment\/} fixed point corresponds to the original Anderson
model with $\tilde{\Gamma}=0$ and $\tilde{U}=-2\tilde{\varepsilon}_d=\infty$.
The effective Hamiltonian at the fixed point is still given by
Eq.~(\ref{H_weak}), but only singly occupied impurity states are allowed,
so there is a decoupled spin-one-half degree of freedom,
\begin{equation}
{\bf s} = d^{\dag}_{\mu} \; \case{1}{2}
        \bbox{\sigma}_{\mu\mu'} \; d_{0\mu'},
\end{equation}
localized at the impurity site.
This is the weak-coupling fixed point of the Kondo models, i.e.,
it corresponds to setting $\tilde{J}=\tilde{V}=0$ in Eq.~(\ref{H_0K}).

None of the perturbations in Eqs.~(\ref{weak_delH}) is allowed
at the local-moment fixed point.
Instead, the dominant perturbations are as follows:
\begin{equation}
\begin{array}{lllll}
O_J &=& \Lambda^{N/2} f^{\dag}_{0\sigma} \, \case{1}{2}
        \bbox{\sigma}_{\sigma\sigma'} \, f_{0\sigma'} \cdot {\bf s}
        &\propto& T^r, \\[1ex]
O_V &=& \Lambda^{N/2} (f^{\dag}_{0\sigma}f_{0\sigma} - 1)
        &\propto& T^r, \\[1ex]
O_{t_1} &=& \Lambda^{N/2} (f^{\dag}_{0\sigma}f_{1\sigma}\!+\!\text{H.c.})
        &\propto& T^{1+r}, \\[1ex]
O_{U_0} &=& \Lambda^{N/2} (f^{\dag}_{0\sigma}f_{0\sigma} - 1)^2
        &\propto& T^{1+2r}.
\end{array}
                                                \label{local_delH}
\end{equation}
Here, $O_J$ and $O_V$ describe exchange (Kondo) scattering and
pure potential scattering, respectively;
$O_{t_1}$ is a term from the kinetic energy; and
$O_{U_0}$ represents a Coulomb interaction between two
conduction electrons in shell~0.
Only $O_V$ breaks particle-hole symmetry.

In the standard case $r=0$, exchange and pure-potential scattering are
marginal; further analysis \cite{Anderson:scaling,Wilson} reveals that
antiferromagnetic [ferromagnetic] exchange is marginally relevant
[marginally irrelevant], hence the fixed point is unstable [stable].
For $r>0$, by contrast, all perturbations are irrelevant, so
{\em the local-moment fixed point is stable irrespective of the sign of $J$.}
This is the first of several important differences between the fixed-point
behaviors for $r=0$ and $r>0$.

We note that for the two-channel Kondo model, each of the operators
$O_{\gamma}$ listed in Eq.~(\ref{local_delH}) should be replaced by a pair
of operators, $O_{\gamma}^{(\pm)}=O_{\gamma}^{(1)}\pm O_{\gamma}^{(2)}$,
where $O_{\gamma}^{(j)}$ ($j = 1,$~2) is identical to $O_{\gamma}$ except
that all its $f$ operators carry a channel label~$j$.
Throughout this paper it is assumed that the two conduction channels couple to
the impurity spin with equal strength, in which case only the symmetric
operator $O_{\gamma}^{(+)}$ can enter $\delta H_N$.
In addition, one can construct allowed perturbations that contain $f$'s
belonging to both channels.
At the local-moment fixed point, the leading perturbation of this type is
\begin{equation}
O_{U'_0} = \Lambda^{N/2} (f^{\dag}_{01\sigma} f_{01\sigma} - 1)
                         (f^{\dag}_{02\sigma} f_{02\sigma} - 1)
           \propto T^{1+2r}.
\end{equation}
Similar remarks concerning the two-channel Kondo model apply at each of the
remaining fixed points described in this Section.

\subsection{Symmetric strong-coupling fixed point}
\label{subsec:symm_strong}

The {\em symmetric strong-coupling\/} fixed point is obtained by setting
$\tilde{\Gamma}=\infty$ in Eq.~(\ref{H_0A}) while keeping
$\tilde{\varepsilon}_d$ and $\tilde{U}$ finite, or by setting
$\tilde{J}=\infty$ in Eq.~(\ref{H_0K}).
We first consider the ground state of the atomic Hamiltonian $H_0$.
In the nondegenerate Anderson model and the conventional $s=\frac{1}{2}$ Kondo
model, any moment at the impurity site is competely screened by $f_0$
electrons.
(In the Anderson model, the ground-state impurity occupancy
$\langle n_d\rangle$ varies continuously with the parameters
$\tilde{\varepsilon}_d$ and $\tilde{U}$.
For any given set of couplings, however, there is a unique ground state,
and hence no residual impurity degree of freedom.)
The $s=1$ Kondo model and the two-channel Kondo model have spin-one-half ground
states;\cite{Nozieres} in the former instance, the impurity is
{\em underscreened}, in the latter, it is {\em overscreened}.

In all cases, an infinite gap separates the ground state(s) from all other
eigenstates of the atomic Hamiltonian.
As a consequence, the $f_{0\sigma}$ degrees of freedom are frozen out and
the effective Hamiltonian becomes
\begin{equation}
H_{\text{\it SSC},N}^{\star} = H_N^{(1)} - E_{G,N}.
\end{equation}

Based on the results of Section~\ref{subsec:L=1}, the dominant perturbations
at this fixed point are
\begin{equation}
\begin{array}{lllll}
O_{J_1} &=& \Lambda^{N/2} f^{\dag}_{1\sigma} \, \case{1}{2}
        \bbox{\sigma}_{\sigma\sigma'} \, f_{1\sigma'} \cdot \bbox{\tau}
        &\propto& T^{-r_1}, \\[1ex]
O_{V_1} &=&
        \Lambda^{N/2} (f^{\dag}_{1\sigma}f_{1\sigma} - 1)
        &\propto& T^{-r_1}, \\[1ex]
O_{U_1} &=&
        \Lambda^{N/2} (f^{\dag}_{1\sigma}f_{1\sigma} - 1)^2
        &\propto& T^{1-2r_1}, \\[1ex]
O_{t_2} &=&
        \Lambda^{N/2} (f^{\dag}_{1\sigma}f_{2\sigma}\!+\!\text{H.c.})
        &\propto& T^{(1-r_1+|1-r|)/2},
\end{array}
                                                \label{symm_delH}
\end{equation}
where, as before, $r_1=\min(r,1)$.
$O_{J_1}$, a Kondo-like operator involving the residual spin $\bbox{\tau}$,
is present only in the underscreened and overscreened models.
$O_{V_1}$ describes nonlocal potential scattering of electrons in shell~1
from the impurity site.
Both $O_{J_1}$ and $O_{V_1}$ are relevant perturbations for all $r>0$.
$O_{U_1}$, representing Coulomb repulsion between $f_1$ electrons,
is relevant for $r>\frac{1}{2}$.
Finally, the kinetic energy term $O_{t_2}$ is marginal for
$r=1$, but is irrelevant otherwise.

The fixed point is stable for $r>0$ if, and only if, three conditions are
satisfied: the impurity moment is exactly screened (to rule out $O_{J_1}$
as an allowed perturbation); the power $r$ is less than $\frac{1}{2}$ (to
ensure that $O_{U_1}$ is irrelevant); and the problem exhibits
particle-hole symmetry (so that $O_{V_1}$ is disallowed).
Thus, one sees that {\em the symmetric strong-coupling fixed point is
generically unstable.}
This represents another significant departure from the standard case $r=0$,
in which the fixed point is always marginally stable, except in overscreened
problems, where it is marginally unstable.\cite{Nozieres}

\subsection{Asymmetric strong-coupling fixed point}
\label{subsec:asymm_strong}

The {\em asymmetric strong-coupling\/} fixed point of the Anderson model and
of the single-channel Kondo models is obtained in the same
fashion as the symmetric strong-coupling fixed point considered above, with
a further condition: either the coefficient~$e_1$ entering
Eqs.~(\ref{H_c:disc}) and~(\ref{H_N}) is made infinite; or the model
Hamiltonian~$H_N$ is augmented by a term $\tilde{V}_1 O_{V_1}$,
where $O_{V_1}$ is defined in Eqs.~(\ref{symm_delH}) and
$|\tilde{V}_1|\rightarrow\infty$.
As a result, the $f_1$~degrees of freedom are frozen, in addition to
those associated with shell~$0$ and the impurity.
The effective Hamiltonian becomes
\begin{equation}
H_{\text{\it ASC},N}^{\star} = H_N^{(2)} - E_{G,N}.     \label{H_ASC}
\end{equation}
The two fixed points described by $e_1 = +\infty$ and \mbox{$e_1 = -\infty$}
(or by $\tilde{V}_1 = +\infty$ and $\tilde{V}_1 = -\infty$)
have different ground-state charges as defined in Eq.~(\ref{charge:def}), but
they are otherwise physically equivalent and will henceforth be treated as a
single fixed point.

The dominant perturbations are
\begin{equation}
\begin{array}{lllll}
O_{J_2} &=& \Lambda^{N/2} f^{\dag}_{2\sigma} \, \case{1}{2}
        \bbox{\sigma}_{\sigma\sigma'} \, f_{2\sigma'} \cdot \tilde{\bbox{\tau}}
        &\propto& T^r, \\[1ex]
O_{V_2} &=&
        \Lambda^{N/2} (f^{\dag}_{2\sigma}f_{2\sigma} - 1)
        &\propto& T^r, \\[1ex]
O_{t_3} &=&
        \Lambda^{N/2} (f^{\dag}_{2\sigma}f_{3\sigma}\!+\!\text{H.c.})
        &\propto& T^{1+r}, \\[1ex]
O_{U_2} &=&
        \Lambda^{N/2} (f^{\dag}_{2\sigma}f_{2\sigma} - 1)^2
        &\propto& T^{1+2r},
\end{array}
                                                \label{asymm_delH}
\end{equation}
where $\tilde{\bbox{\tau}}$ describes a residual spin-one-half degree of
freedom present only in the underscreened Kondo model.
Since these operators are all irrelevant,
{\em the asymmetric strong-coupling fixed point is stable for all $r>0$.}

In the standard case ($r=0$), the symmetric and asymmetric strong-coupling
fixed points represent two points on a continuous line of marginally stable
fixed points described by a family of effective Hamiltonians
\begin{equation}
H_N^{\star}(\tilde{V}_1) = H_N^{(1)} +
        \tilde{V}_1 \Lambda^{N/2} (f^{\dag}_{1\sigma}f_{1\sigma} - 1)
        - E_{G,N}.
                                                \label{H_V}
\end{equation}
These fixed points share essentially the same physical properties,
independent of the value of $\tilde{V}_1$.
The effect of a power-law scattering rate to destroy all but two of these
fixed points and to make the case $\tilde{V}_1 = 0$ unstable with
respect to the breaking of particle-hole symmetry.

The two-channel Kondo model also has a stable, strong-coupling fixed point
at \mbox{$\tilde{J}=\infty$}, \mbox{$e_1=\pm\infty$}.
However, we have not found any choice of the bare parameters $J_0$ and $V_0$
that produces flow to this limit, in which the ground state carries a residual
spin-one-half degree of freedom.
Instead, it is helpful to consider the Hamiltonian
$H_N+\tilde{J}_1 O^{(+)}_{J_1}+\tilde{V}_1 O^{(+)}_{V_1}$
[see Eqs.~(\ref{symm_delH}) and the last paragraph of
Section~\ref{subsec:local}].
The asymmetric strong-coupling fixed point of interest is reached by first
setting $\tilde{J}=\infty$ and $\tilde{V}=0$ to lock the impurity into an
overscreened spin doublet, and then taking the simultaneous limits
$\tilde{J}_1\rightarrow\infty$ and
$\tilde{V}_1\rightarrow\pm\infty$ in such a way that
$1/2 < |\tilde{V}_1| / \tilde{J}_1 < 3/4$.
Under this prescription, the impurity combines with shells 0~and~1 to produce
two degenerate ground-state configurations carrying quantum numbers
$(S,Q^{(1)},Q^{(2)}) = (0,\mp 1,0)$ and $(0,0,\mp 1)$.
This pair of spinless states represents a ``flavor-one-half'' degree of
freedom.
[The generators of electron flavor symmetry are obtained from the standard
spin generators by interchanging spin and channel indices:
$\uparrow\leftrightarrow 1$, $\downarrow\leftrightarrow 2$.
For example, the $z$-component of flavor measures the difference between the
number of electrons in channels 1~and~2, i.e.,
$J_z^{\text{flavor}} = \frac{1}{2}(Q^{(1)}-Q^{(2)})$.]
The fixed point is described by the effective Hamiltonian
$H_{\text{\it ASC},N}^{\star}$ defined in Eq.~(\ref{H_ASC});
the leading irrelevant perturbations are $O^{(+)}_{V_2}$ and
a flavor analogue of $O^{(+)}_{J_2}$.

\subsection{Frozen-impurity fixed point}
\label{subsec:frozen}

The {\em frozen-impurity\/} fixed point of the Anderson model is obtained by
setting $\tilde{\varepsilon}_d = +\infty$ in Eq.~(\ref{H_0A}).
Here, the impurity level becomes completely depopulated, and the
excitations of the system are just described by Eq.~(\ref{H_weak}).
The leading perturbations at this fixed point are
$O_V$, $O_{t_1}$ and
$O_{U_0}$ from Eqs.~(\ref{local_delH}),
so the fixed point is clearly stable for all $r>0$.

The electronic excitations at the frozen-impurity and asymmetric
strong-coupling fixed points are described by $H^{(0)}_N$ and $H^{(2)}_N$,
respectively.
As pointed out in Section~\ref{subsec:L=2}, these two Hamiltonians
can be made equivalent by a suitable relabeling of the operators~$f_n$.
The leading irrelevant perturbations about the two fixed points
become identical under this relabeling.
Thus, {\em the frozen-impurity and asymmetric strong-coupling fixed points
are physically equivalent,} up to a shift in the ground-state charge.
In treating the Anderson Hamiltonian, it will prove more convenient to refer
to the frozen-impurity fixed point (since $\langle n_d\rangle$ can be made
arbitrarily small in this model), whereas the asymmetric strong-coupling fixed
point more naturally describes the Kondo models (which correspond to the limit
\mbox{$n_d = 2s$}).

\section{Thermodynamic Properties}
\label{sec:thermo}

This section is primarily concerned with the numerical and analytical
calculation of the contribution made by magnetic impurities to various
thermodynamic properties.
First, though, we remark briefly on the thermodynamics of the
pure host Fermi systems.

\subsection{Host Thermodynamic Properties}
\label{subsec:thermo_host}

As we have emphasized in the Introduction, the impurity properties
of the models we consider depend on the conduction-band density of states
and the energy-dependent hybridization only in the particular combination
$\Gamma(\epsilon)=\pi\rho(\epsilon)t^2(\epsilon)$.
However, in order to compute the properties of the pure system in the
absence of magnetic impurities, it is necessary to specify $\rho(\epsilon)$
explicitly.
If the power-law energy-dependence of the scattering rate arises solely from
the hybridization, then the host properties will be those of a conventional
metal.
Here we focus on the opposite limit, more appropriate for describing the
gapless systems listed in the Introduction, in which the hybridization is
essentially constant and the density of states has the form given in
Eq.~(\ref{rho_pure}).
In this case, unit normalization of~$\rho(\epsilon)$ implies that
\begin{equation}
\rho_0 = \frac{1+r}{2D}.                        \label{rho0_pure}
\end{equation}

With these assumptions, it is straightforward to show that for $k_B T \ll D$,
the host entropy, specific heat capacity, and static susceptibility are
given by
\begin{mathletters}
\begin{eqnarray}
\frac{S^{(0)}}{k_B} &=& 2 N_0 \, (2\!+\!r) \, \phi(1\!+\!r)
                \left(\frac{k_B T}{D}\right)^{1+r} \!\!\!,      \\
\frac{C^{(0)}}{k_B} &=& 2 N_0 \, (1\!+\!r) \, (2\!+\!r) \, \phi(1\!+\!r)
                \left(\frac{k_B T}{D}\right)^{1+r} \!\!\!,      \\
\frac{k_B T \chi^{(0)}}{(g\mu_B)^2} &=& \frac{N_0}{2} \, \bar{\phi}(1\!+\!r)
                \left(\frac{k_B T}{D}\right)^{1+r}\!\!\!.
\end{eqnarray}                                                  \label{host}%
\end{mathletters}%
$N_0$ is the number of unit cells making up the solid, and
\begin{equation}
\phi(x) = \zeta_1(x+1), \quad
\bar{\phi}(x) = x \left[ \zeta_1(x) - \zeta_2(x) \right],
                                                        \label{phi's}
\end{equation}
where, for all $x>0$
and all positive integers $n$, we define\cite{Riemann}
\begin{equation}
\zeta_n(x) = \int_0^{\infty} \!\!\!\! du \; \frac{u^{x-1}}{(e^u + 1)^n}.
                                                        \label{zeta_n}
\end{equation}
(The functions $\phi$ and $\bar{\phi}$ will also enter the impurity
properties calculated in Section~\ref{subsec:thermo_pert}.)

One sees from Eqs.~(\ref{host}) that the exponent $r$ which determines the
density of states is directly reflected in the temperature-dependence of the
host properties.
For later reference, we define the host Wilson\cite{Wilson} (or Sommerfeld)
ratio,
\begin{equation}
R_W^{(0)}
= \lim_{T\rightarrow 0} \frac{4\pi^2}{3}
        \frac{k_B^2 T \chi^{(0)}}{(g\mu_B)^2 C^{(0)}}
= \frac{\pi^2\bar{\phi}(1\!+\!r)}{3(1\!+\!r)(2\!+\!r)\phi(1\!+\!r)},
                                                \label{Rw_host}
\end{equation}
Since $\phi(1)=\pi^2/12$ and $\bar{\phi}(1)=\frac{1}{2}$, Eq.~(\ref{Rw_host})
reduces in the limit $r\rightarrow 0$ to the standard result,
$R_W^{(0)}=1$.

\subsection{Impurity Thermodynamic Properties}
\label{subsec:thermo_imp}

The impurity contribution to a thermodynamic property $A$ is defined to be the
change in the total measured value of $A$ brought about by adding a single
impurity to the system.
Each such contribution can be computed from an expression of the form
\begin{eqnarray}
A_{\text{imp}} &=& \langle {\cal A} \rangle_{\text{imp}}
  = \langle {\cal A} \rangle - \langle {\cal A} \rangle_0       \nonumber \\
&=& \!\lim_{N\rightarrow\infty} \left[
    \text{Tr} \! \left( {\cal A} e^{-\beta_N \! H_N} \right) \!-\!
    \text{Tr}_0 \! \left(\!{\cal A} e^{-\beta_N\!H_N^{(0)}} \right)\right] \! ,
                                                        \label{A_imp}
\end{eqnarray}
where $\cal A$ is an operator which depends on the property of interest,
\begin{equation}
\beta_N = \alpha D \Lambda^{-N/2} / (k_B T)             \label{beta_N}
\end{equation}
is the natural energy scale of iteration $N$ divided by the thermal energy
scale, and ``$\text{Tr}_0$'' means a trace taken over an impurity-free system.

For example, the impurity contributions to the entropy and the specific heat
are obtained as
\begin{equation}
S_{\text{imp}} = - \frac{\partial F_{\text{imp}}}{\partial T}, \qquad
C_{\text{imp}} = - T \frac{\partial^2 F_{\text{imp}}}{\partial T^2}.
                                                        \label{S,C_imp}
\end{equation}
Here, $F_{\text{imp}}$ is the difference between the total Helmholtz free
energy of the system with and without the impurity:
\begin{eqnarray}
F_{\text{imp}}
&=& -k_B T \ln Z_{\text{imp}}           \nonumber \\
&=& \lim_{N \rightarrow \infty} k_B T
                        \ln\left(Z^{(0)}_N/Z_N\right),  \label{F_imp}
\end{eqnarray}
with
\begin{equation}
   Z_N = \text{Tr} e^{-\beta_N H_N}, \qquad
   Z^{(0)}_N = \text{Tr}_0 e^{-\beta_N H^{(0)}_N}.
\end{equation}

Another quantity of interest is the impurity contribution to the zero-field
magnetic susceptibility, given by
\begin{eqnarray}
\frac{k_B T \chi_{\text{imp}}}{(g\mu_B)^2}
&=& \langle S_z^2 / Z \rangle_{\text{imp}} \nonumber \\
&=& \lim_{N \rightarrow \infty} \left\{
    \frac{\text{Tr}\left( S_z^2 e^{-\beta_N H_N}
                \right)}{Z_N} \right.                   \label{chi_imp} \\
& & \qquad \; \left. - \,
     \frac{\text{Tr}_0 \left( S_z^2 e^{-\beta_N H^{(0)}_N}
                \right)}{Z^{(0)}_N} \right\} ,          \nonumber
\end{eqnarray}
where $\mu_B$ is the Bohr magneton, $g$ is the Land\'{e} $g$ factor
(assumed to be the same for conduction and localized electrons),
and $S_z$ is the $z$-component of the total spin of the system.
The quantity $3k_B T\chi_{\text{imp}}$ equals the square of the effective
moment contributed by the impurity to the system.

The numerical~RG formulation provides a controlled approximation for computing
the impurity contributions to each thermodynamic property according to
Eq.~(\ref{A_imp}).
The method does not yield reliable results for $\langle A\rangle$
or $\langle A\rangle_0$ separately, even though these are the values that
would be have to be measured experimentally in order to determine
$A_{\text{imp}}$.

\subsection{Numerical Evaluation of Impurity Thermodynamic Properties}
\label{subsec:thermo_num}

In Section~\ref{sec:results}, we present thermodynamic properties obtained
via the direct numerical evaluation of Eqs.~(\ref{F_imp}) and~(\ref{chi_imp}).
This subsection briefly reviews some of the technical details of these
calculations.

The general strategy for computing thermodynamic properties using the
discretized Hamiltonians $H_N$ is as follows:
One first selects a value for the dimensionless parameter $\bar{\beta}$.
Then for each iteration $N=0$,~1,~\dots, one assigns the result of
Eq.~(\ref{A_imp}) to the temperature defined through Eq.~(\ref{beta_N}) by the
condition $\beta_N = \bar{\beta}$.
This gives the quantity $\langle {\cal A} \rangle_{\text{imp}}$
at a sequence of temperatures $T_N$ satisfying Eq.~(\ref{T_N}).
The $T_N$'s are equally spaced at intervals of $\frac{1}{2}\ln\Lambda$ on
a logarithmic scale.
If desired, this ``grid'' of temperatures can be refined by using several
different values of $\bar{\beta}$ at each iteration.
The choice $\bar{\beta}=\bar{\beta}_0 \Lambda^{-j/2M}$ for $j=0$, \dots~$M$
(we have used $M = 4$) proves convenient because the corresponding
temperature grid $\{T_{N,j}\}$ contains the redundancies $T_{N,0} = T_{N+1,M}$.
The discrepancy between the two independent evaluations of a thermodynamic
property at the same temperature provides a useful measure of the error in
the result.

Since the smallest energy scale of $H_N$ is of order unity, one expects
Eq.~(\ref{A_imp}) to provide increasingly reliable results for $\Lambda > 1$
as $\bar{\beta}$ becomes much smaller than unity.
However, there is another factor which militates against taking
$\bar{\beta}\ll 1$.
Limitations of computer time and memory permit the retention only of those
states having an energy within $E_c$ of the ground state.
In order to minimize the contribution of the missing states to
$\langle {\cal A} \rangle_{\text{imp}}$, one wants $\bar{\beta}E_c$ to be as
large as possible.
In practice, therefore, $\bar{\beta}$ is chosen as a compromise to take
a value somewhat smaller than one.
The results presented in Section~\ref{sec:results} were calculated for
$\Lambda=3$~or~$9$, retaining all eigenstates up to a dimensionless
energy $25$ and using values of $\bar{\beta}$ between~$0.6$
and~$0.6\Lambda^{-1/2}$.
It is shown in Ref.~\onlinecite{KWW} how one can calculate corrections
to compensate for such relatively large values of $\bar{\beta}$.
Our studies indicate that while the corrections are formally of order
$\bar{\beta}/\Lambda \approx 0.1$--$0.2$, they have small prefactors
which reduce the overall shift in $C_{\text{imp}}$ and
$\chi_{\text{imp}}$ to less than one part in~$10^3$.
Since this level of error is smaller than that arising from the discretization
of the conduction band, we have neglected the $\bar{\beta}/\Lambda$
corrections.

As a practical matter, Eqs.~(\ref{S,C_imp}) are not used directly to evaluate
$S_{\text{imp}}$ and $C_{\text{imp}}$.
A more accurate evaluation of the entropy, which avoids numerical
differentiation, exploits the relation
\begin{equation}
   S_{\text{imp}} = k_B \left( \langle\beta{\cal H}/Z\rangle_{\text{imp}}
                        - \beta F_{\text{imp}} \right).
\end{equation}
It is likewise possible to obtain the specific heat without differentiation,
through the equation
\begin{equation}
   C_{\text{imp}}
        = k_B \left[\Bigl\langle(\beta{\cal H}/Z)^2\Bigr\rangle_{\text{imp}}
        - \Bigl(\langle\beta{\cal H}/Z\rangle_{\text{imp}}\Bigr)^2 \right],
\end{equation}
but the results turn out to be rather prone to discretization error.
All plots of the specific heat presented below were instead obtained using
$C_{\text{imp}}=T\partial S_{\text{imp}}/\partial T$ with a simple
two-point approximation to the derivative.

As mentioned above, the thermodynamic quantities presented in this paper
were obtained using $\Lambda = 3$ or $\Lambda = 9$ and an energy cutoff
$E_c \ge 25$.
With these choices, the primary source of error is the discretization.
One of the main effects of working with a value of $\Lambda$ greater than unity
is the introduction into $\langle {\cal A} \rangle_{\text{imp}}$ of
oscillations which are periodic in $\ln T$.
The oscillations have a period $\ln\Lambda$ and a magnitude proportional to
$\exp(-\pi^2/\ln\Lambda)$.
Oliveira and Oliveira have shown \cite{Oliveira} that these oscillations can
be greatly reduced by averaging values of
$\langle {\cal A} \rangle_{\text{imp}}$
computed for different band discretization parameters $z$ (see
Section~\ref{subsec:discrete}).
We have employed four $z$'s ($0.5$, $0.75$, $1$, $1.25$) in obtaining
the results presented below.

Another consequence of the band discretization is a reduction in the effective
coupling between impurity and delocalized degrees of freedom.\cite{Oliveira:81}
Study of a discretized resonant-level model with a power-law scattering rate
\cite{Ingersent:unpub} indicates that the most faithful description of the
continuum problems described by Eqs.~(\ref{H_h:f}) and~(\ref{H_s:f}) is
obtained by premultiplying the parameters $\Gamma_0$, $\rho_0 J_0$, and
$\rho_0 V_0$ entering the discretized calculations by a factor
\begin{eqnarray}
A(\Lambda,r)
&=& \left[ \frac{1\!-\!\Lambda^{-(2+r)}}{2+r} \right]^{1+r} \!
    \left[ \frac{1+r}{1\!-\!\Lambda^{-(1+r)}} \right]^{2+r}
    \!\! \ln\Lambda,                                    \label{A} \\[1ex]
&\approx& 1 + {\cal O}(\ln\Lambda)^2
    \qquad \text{for $\Lambda\rightarrow 1$}.               \nonumber
\end{eqnarray}
For $r=0$, Eq.~(\ref{A}) reduces to the standard result given in
Ref.~\onlinecite{Oliveira:81}.
In the remainder of this paper, we quote the continuum equivalent of
each coupling.
Thus, any numerical data labeled with a particular value of $\Gamma_0$,
$\rho_0 J_0$, or $\rho_0 V_0$ were actually computed by substituting
$A\Gamma_0$, $A\rho_0 J_0$, or $A\rho_0 V_0$ into Eqs.~(\ref{A_couplings})
or Eqs.~(\ref{K_couplings}).
Note that parameters describing the impurity alone, i.e., $\epsilon_d$ and $U$
entering Eqs.~(\ref{A_couplings}), do not have to be corrected.

The measures outlined in the preceding paragraphs greatly reduce, but cannot
completely eliminate, discretization errors in the computed thermodynamic
properties.
We estimate on the basis of limited calculations performed for other values
of $\Lambda$ that for $\Lambda = 3$, the overall error in the impurity
contributions to the susceptibility~$T\chi$ or the entropy~$S$ is less
than~5\%, while that for the specific heat~$C$ is less than~10\%.
It should be emphasized that these are errors in absolute
quantities at finite temperatures.
Zero-temperature properties and exponents describing ratios of properties at
different temperatures or couplings generally have much smaller errors
(below~1\%).
Indeed, fixed-point properties can be computed to better than~1\% using values
of $\Lambda$ as large as $10$, with a considerable reduction in the numerical
effort compared to that required for $\Lambda=3$.

\subsection{Perturbative Evaluation of Impurity Thermodynamic Properties}
\label{subsec:thermo_pert}

In the vicinity of any of the fixed points described in
Section~\ref{sec:fixed}, perturbation theory can be applied to the
appropriate effective Hamiltonian to obtain analytical expressions
for thermodynamic quantities as functions of the couplings $\tilde{\gamma}$
which parametrize the deviation from the fixed point [see Eq.~(\ref{delH})].
Once perturbative expressions have been obtained for the discretized version of
the problem ($\Lambda>1$), they can be extrapolated to the continuum limit
($\Lambda=1$).

A similar perturbative treatment of the standard Kondo and Anderson models is
described in detail in Refs.~\onlinecite{Wilson} and~\onlinecite{KWW},
respectively.
The extension to systems with a power-law scattering rate is
conceptually straightforward but algebraically laborious.
One novel feature is that in certain physical regimes the dominant
temperature-dependences derive from second-order corrections to the
fixed-point properties, whereas in the standard case ($r=0$) is is not
necessary to go beyond first order in perturbation theory.

The remainder of this section summarizes properties of the
five distinct fixed points discussed in Section~\ref{sec:fixed}.
Perturbative corrections to the fixed-point properties are presented for each
of the three stable (or conditionally stable) regimes.
Certain technical details have been relegated to Appendix~\ref{app:thermo}.

\subsubsection{Free-impurity fixed point}

The free-impurity fixed point describes a decoupled orbital which
has four degenerate states.
The impurity contributes to the thermodynamic properties as follows:
\begin{equation}
S_{\text{imp}} = k_B \ln 4, \quad
C_{\text{imp}} = 0, \quad
\frac{k_B T \chi_{\text{imp}}}{(g\mu_B)^2} = \frac{1}{8}.
\end{equation}

\subsubsection{Valence-fluctuation fixed point}

The valence-fluctuation fixed point describes a decoupled orbital which
has three degenerate states, double occupation being forbidden.
The impurity properties include
\begin{equation}
S_{\text{imp}} = k_B \ln 3, \quad
C_{\text{imp}} = 0, \quad
\frac{k_B T \chi_{\text{imp}}}{(g\mu_B)^2} = \frac{1}{6}.
\end{equation}

\subsubsection{Local-moment regime}

One can apply standard perturbative methods \cite{Wilson,KWW} to the effective
Hamiltonian
$H_N = H_{\text{\it WC},N}^{\star}+\sum_{\gamma}\tilde{\gamma}O_{\gamma}$,
where the fixed-point Hamiltonian $H_{\text{\it WC},N}^{\star}$ is given by
Eq.~(\ref{H_weak}) and the perturbations
$O_{\gamma}$ are those defined in Eq.~(\ref{local_delH}).
For the nondegenerate Anderson model or the exactly screened Kondo model one
finds, to lowest order in each of the perturbative couplings, that
\begin{eqnarray}
\lefteqn{-\frac{F_{\text{imp}}}{k_B T} =
     \ln2 - 8\tilde{t}_1\beta_N \Lambda^{-(1+r)N/2}
     \sum_j \alpha_{0j} \alpha_{1j} p_j}        \nonumber \\
&\;\;+& 2 (\tilde{V}^2 + \case{3}{16}\tilde{J}^2)
     \beta_N \Lambda^{-r N} \left\{ \sum_{j,k}
     \frac{\alpha_{0j}^2 \alpha_{0k}^2}{\eta^{\star}_j + \eta^{\star}_k}
     \right.                                    \label{F_LM:sum} \\
&\;\;-& \left. \sum_{j} \alpha_{0j}^4 p_j
     \left( \frac{1}{\eta^{\star}_j} + \beta_N \bar{p}_j \right)
    - 4 \sum_{j \neq k}
     \frac{\alpha_{0j}^2 \alpha_{0k}^2 \eta^{\star}_j p_j}%
     {{\eta^{\star}_j}^2-{\eta^{\star}_k}^2} \right\}           \nonumber
\end{eqnarray}
and
\begin{eqnarray}
\lefteqn{\frac{k_B T \chi_{\text{imp}}}{(g\mu_B)^2} = \frac{1}{4}
     + \frac{\tilde{J}}{2} \, \beta_N \Lambda^{-r N/2}
     \sum_j \alpha_{0j}^2 p_j \bar{p}_j}                        \nonumber \\
&\;\;-& 2 \tilde{t}_1 \beta_N \Lambda^{-(1+r) N/2}
     \sum_j \alpha_{0j} \alpha_{1j} p_j\bar{p}_j(\bar{p}_j-p_j)
                                                                \nonumber \\
&\;\;+& 4 \tilde{U}_0 \beta_N \Lambda^{-(1+2r) N/2} \left(
    \sum_j \alpha_{0j}^2 p_j \bar{p}_j \right)^2       \label{chi_LM:sum} \\
&\;\;-& \frac{\tilde{V}^2}{2} \beta_N \Lambda^{-rN} \left\{
     4 \sum_{j\neq k} \frac{\alpha_{0j}^2 \alpha_{0k}^2 \eta^{\star}_j}%
     {{\eta^{\star}_j}^2\!-\!{\eta^{\star}_k}^2}
     p_j \bar{p}_j (\bar{p}_j\!-\!p_j) \right.                  \nonumber \\
&& \quad \left. +\, \sum_j \alpha_{0j}^4 p_j\bar{p}_j
     \left[ \frac{\bar{p}_j\!-\!p_j}{\eta^{\star}_j}
     - \beta_N(1\!-\!6p_j\bar{p}_j) \right] \right\} .          \nonumber
\end{eqnarray}
Here,
\begin{equation}
p_j \equiv 1 - \bar{p}_j = \frac{e^{- \beta_N \eta^{\star}_j}}
       {1 + e^{- \beta_N \eta^{\star}_j}}                       \label{p_j}
\end{equation}
is the occupation probability of a fermionic state having
energy $\eta^{\star}_j$,
and the indices $j$ and $k$ run over the range $1$ to $(N\!+\!1)/2$,
inclusive.
We have omitted the lowest order contribution of $O_{U_0}$ to
$F_{\text{imp}}$ because it is highly irrelevant ($\propto T^{3+4r}$).
The equations above are written for $N$~odd;
similar expressions hold for $N$~even.

In the continuum limit ($N\rightarrow\infty$, $\Lambda\rightarrow 1$,
and $\beta_N\ll 1$), the sums entering Eqs.~(\ref{F_LM:sum})
and~(\ref{chi_LM:sum}) can be evaluated algebraically
(see Appendix~\ref{app:thermo}).
The resulting equations, valid for all positive $r$ {\em except\/} $r=1$, are
\begin{eqnarray}
\lefteqn{-\frac{F_{\text{imp}}}{k_B T} =
        \ln 2
        - 4\tilde{t}_1 \sqrt{(1\!+\!r)(3\!+\!r)} \, \phi(1\!+\!r)
        \left( \frac{k_B T}{D} \right)^{1+r}}           \nonumber \\
&\;\;\;\;+& 2 (\tilde{V}^2\!+\!\case{3}{16}\tilde{J}^2) (1\!+\!r)^2 \,
        \psi(r) \, \phi(r_1\!+\!r)
        \left( \frac{k_B T}{D} \right)^{r_1+r}          \label{F_LM:cont}
\end{eqnarray}
and
\begin{eqnarray}
\frac{k_B T \chi_{\text{imp}}}{(g \mu_B)^2} &=& \frac{1}{4} +
    \frac{\tilde{J}}{4} \, \bar{\phi}(1\!+\!r) \left( \frac{k_B T}{D} \right)^r
                                                        \nonumber \\
&-& \tilde{t}_1 \sqrt{(1\!+\!r)(3+r)} \, \bar{\phi}(1\!+\!r)
    \left( \frac{k_B T}{D} \right)^{1+r}                \nonumber \\
&+& \tilde{U}_0 \left[ \bar{\phi}(1\!+\!r) \right]^2
    \left( \frac{k_B T}{D} \right)^{1+2r}               \nonumber \\
&+& \frac{\tilde{V}^2}{2} (1\!+\!r)^2 \, \psi(r) \, \bar{\phi}(r_1\!+\!r)
    \left( \frac{k_B T}{D} \right)^{r_1+r} .            \label{chi_LM:cont}
\end{eqnarray}
Both $\phi(x)$ and $\bar{\phi}(x)$, defined in Eqs.~(\ref{phi's}),
vary smoothly with $x$ and are of order unity over the range $0 \le x \le 3$.
By contrast, the function
\begin{equation}
\psi(r) = \left\{ \begin{array}{ll}
    \displaystyle
    \frac{\pi}{2} \tan \frac{r\pi}{2} \qquad &
    \text{for $0\le r<1$}, \\[2ex]
    \displaystyle
    (r-1)^{-1} & \text{for $r\ge 1$},
    \end{array} \right.                                 \label{lambda}
\end{equation}
has a simple pole at $r=1$.

Similar calculations can be performed for the $s=1$ and two-channel Kondo
models.
To summarize the results, the fixed-point impurity properties of the Anderson
model and of the two $s=\frac{1}{2}$ Kondo models are just those expected for
a decoupled spin-$\frac{1}{2}$ impurity:
\begin{equation}
S_{\text{imp}} = k_B \ln 2, \quad
C_{\text{imp}} = 0, \quad
\frac{k_B T \chi_{\text{imp}}}{(g\mu_B)^2} = \frac{1}{4}.
                                                        \label{LM_props}
\end{equation}
The corresponding results for the underscreened model are
\begin{equation}
S_{\text{imp}} = k_B \ln 3, \quad
C_{\text{imp}} = 0, \quad
\frac{k_B T \chi_{\text{imp}}}{(g\mu_B)^2} = \frac{2}{3}.
                                                        \label{LM_u_props}
\end{equation}

In all four models, the leading corrections at low temperatures
take the form
\begin{equation}
\Delta S_{\text{imp}},\,\Delta C_{\text{imp}} \propto T^{r_1+r}, \quad
\Delta(T\chi_{\text{imp}}) \propto T^r          \label{LM_dev}
\end{equation}
for all positive $r\not= 1$.

For the special case $r=1$, the second-order terms in $F_{\text{imp}}$ and
$\chi_{\text{imp}}$ acquire logarithmic corrections, necessitating the
replacement
\begin{equation} \psi(r) \longrightarrow
       \ln \left( \frac{D}{\bar{\beta} k_B T} \right)
                                                \label{lambda:rep}
\end{equation}
in Eqs.~(\ref{F_LM:cont}) and~(\ref{chi_LM:cont}).
Here, $\bar{\beta}$ is the small parameter introduced in Eq.~(\ref{T_N}), the
precise value of which cannot be determined uniquely within the present
formalism.
(See Appendix~\ref{app:thermo} for further discussion.)
As a result, there is some uncertainty in the thermodynamic properties, but it
is clear that $\Delta S_{\text{imp}}$ and $\Delta C_{\text{imp}}$ must contain
contributions proportional to $T^2\log T$ and others varying like $T^2$.
The ratio of the prefactors of these terms will determine whether or not the
logarithmic correction is observable at temperatures of physical interest.
We shall return to this point in Section~\ref{subsec:Ander_res}.

\subsubsection{Symmetric strong-coupling regime}
\label{subsubsec:thermo_SSC}

The methods of the previous section can be applied to compute impurity
properties at the symmetric strong-coupling fixed point, plus the leading
corrections for those cases in which the fixed point is stable, i.e.,
for the particle-hole-symmetric Anderson model and the exactly screened Kondo
model, both with $r<\frac{1}{2}$.
Here, first-order perturbation theory suffices, yielding the expressions
\begin{eqnarray}
-\frac{F_{\text{imp}}}{k_B T} &=& \ln 4
    + \sum_{l} 4 \ln \bar{q}_l - \sum_{j} 4 \ln\bar{p}_j \nonumber \\
&& - \, 8 \tilde{t}_2 \beta_N \Lambda^{-(1-r) N/2}
    \sum_{l} \beta_{1l} \beta_{2l} q_l                  \label{F_SSC:sum}
\end{eqnarray}
and
\begin{eqnarray}
\frac{k_B T \chi_{\text{imp}}}{(g \mu_B)^2} &=&
        \frac{1}{8} + \sum_{l} q_l \bar{q}_l
        - \sum_{j} p_j \bar{p}_j                        \nonumber \\
&& - \, 2 \tilde{t}_2 \beta_N \Lambda^{-(1-r) N/2}
    \sum_{l} \beta_{1l} \beta_{2l}
    q_l \bar{q}_l (\bar{q}_l - q_l)      \nonumber \\
&& + \, 4 \tilde{U}_1 \beta_N \Lambda^{-(1 - 2r) N/2}
    \left[ \sum_{l} \beta_{1l}^2 q_l \bar{q}_l \right]^2 ,
                                                        \label{chi_SSC:sum}
\end{eqnarray}
where $p_j$ and $\bar{p}_j$ are given by Eq.~(\ref{p_j}), and
$q_l$ and $\bar{q}_l$ are the analogous quantities defined for the
eigenenergies of $H_{\text{\it SSC},N}^{\star}$:
\begin{equation}
q_l \equiv 1 - \bar{q}_l = \frac{e^{- \beta_N \omega^{\star}_l}}
       {1 + e^{- \beta_N \omega^{\star}_l}}.                    \label{q_j}
\end{equation}
The index $l$ runs from $1$ to $(N\!-\!1)/2$, inclusive,
while $j$ still runs from $1$ to $(N\!+\!1)/2$.

The sums entering Eqs.~(\ref{F_SSC:sum}) and~(\ref{chi_SSC:sum}) can be
performed algebraically for values of $\Lambda$ close to unity.
Extrapolation of the resulting expressions (see Appendix~\ref{app:thermo})
to the continuum limit gives
\begin{eqnarray}
-\frac{F_{\text{imp}}}{k_B T}
&=& r_1 \ln 4
    - 4 \tilde{t}_2 b_1 b_2 \times \nonumber \\
& & \sqrt{(1\!-\!r)(3\!-\!r)} \, \phi(1\!-\!r) \,
    \left(\frac{k_B T}{D}\right)^{1-r}                  \label{F_SSC:cont}
\end{eqnarray}
and
\begin{eqnarray}
\lefteqn{\frac{k_B T \chi_{\text{imp}}}{(g \mu_B)^2} = \frac{r_1}{8}
    - \tilde{U}_1 \left[ b_1 \bar{\phi}(1-r) \right]^2
    \left( \frac{k_B T}{D} \right)^{1-2r}}              \nonumber \\
&\qquad-& \tilde{t}_2 b_1 b_2 \sqrt{(1\!-\!r)(3\!-\!r)} \, \bar{\phi}(1\!-\!r)
    \left( \frac{k_B T}{D} \right)^{1-r} ,              \label{chi_SSC:cont}
\end{eqnarray}
where
\begin{equation}
b_n^2(r) = \lim_{\Lambda\rightarrow 1} \;
    \frac{2\beta_n^2}{(2n+1-r) \ln\Lambda}.
\end{equation}
It is found numerically that $b_1$ and $b_2$ are of order unity, at least for
all $0 \le r \le \frac{1}{2}$, the range over which the symmetric
strong-coupling fixed point is stable.

Thus, the fixed-point impurity properties of the Anderson and exactly screened
Kondo models are
\begin{equation}
S_{\text{imp}} = 2 r_1 \; k_B \ln 2, \quad
\frac{k_B T \chi_{\text{imp}}}{(g\mu_B)^2} = \frac{r_1}{8};
                                                        \label{SSC_props}
\end{equation}
those of the underscreened Kondo model are
\begin{equation}
S_{\text{imp}} = (1 \!+\! 2 r_1) \; k_B \ln 2, \quad
\frac{k_B T \chi_{\text{imp}}}{(g\mu_B)^2} = \frac{2 \!+\! r_1}{8};
                                                        \label{SSC_u_props}
\end{equation}
and those of the overscreened model are
\begin{equation}
S_{\text{imp}} = (1 \!+\! 4 r_1) \; k_B \ln 2, \quad
\frac{k_B T \chi_{\text{imp}}}{(g\mu_B)^2} = \frac{1 \!+\! r_1}{4}.
                                                        \label{SSC_o_props}
\end{equation}
In all four models, $C_{\text{imp}}=0$ at the fixed point, and the leading
corrections to the fixed-point properties vary as
\begin{equation}
\Delta S_{\text{imp}},\,\Delta C_{\text{imp}} \propto T^{1-r}, \quad
\Delta(T\chi_{\text{imp}}) \propto T^{1-2r}.            \label{SSC_dev}
\end{equation}

The fixed-point properties and temperature exponents above agree with those
obtained for $0<r<1$ by Chen and Jayaprakash\cite{Chen} for the exactly
screened Kondo model and by Bulla {\em at al.}\cite{Bulla} for the Anderson
model.
[At extremely low temperatures ($T/D<10^{-10}$), the latter authors identify a
$T^{1-2r}$ variation in the specific heat.
Such a term can arise only from the operator $O_{U_1}$ defined in
Eqs.~(\ref{symm_delH}).
Under conditions of strict particle-hole symmetry, however, $O_{U_1}$
cannot contribute to the specific heat.\cite{all_r}
We suspect, therefore, that the $T^{1-2r}$ behavior is a numerical artefact.]
In addition, Bulla {\em et al.\/} find that for $0<r<1$ the impurity spectral
function $A(\omega)$ varies like $|\omega|^{-r}$ in the limit
$\omega\rightarrow 0$, in contrast to the Lorentzian form found in the
standard case $r=0$.

At the symmetric strong-coupling fixed point, one sees that each conduction
band makes a contribution to the entropy and to the susceptibility equal to
$r_1$ times that of an isolated level described by Eq.~(\ref{H_d:def}) with
$\epsilon_d = U = 0$.
Furthermore, the impurity density of states,
$\rho_{\text{imp}}(\epsilon) = \pi^{-1}\partial\delta_0/\partial\epsilon$
computed using Eq.~(\ref{delta_0}), has a delta-function peak of weight
$r_1$ at $\epsilon=0$ (see also Ref.~\onlinecite{Chen}).
These observations suggest the phenomenological interpretation that a fraction
$r_1$ of a conduction electron from each band occupies a decoupled level of
zero energy, the remaining fraction $1-r_1$ presumably being absorbed into a
many-body resonance centered on the Fermi energy.

The properties described above appear to be highly anomalous.
It should be emphasized, though, that since the impurity has no internal
degree of freedom at the fixed point (at least in the Anderson and exactly
screened Kondo models), it acts only to exclude conduction electrons from its
immediate vicinity.
The fixed point behaviors are simply those of independent electrons
subjected to a phase shift.
Indeed, the noninteracting \mbox{($U=0$)} limit of the Anderson model
reproduces the low-energy phase shifts of Eq.~(\ref{delta_0}) and hence yields
precisely the thermodynamic properties described in Eqs.~(\ref{SSC_props}).
Furthermore, the noninteracting model is shown in Appendix~\ref{app:reslevel}
to exhibit a spectral function $A(\omega)\propto|\omega|^{|1-r|-1}$, in
agreement with the $U>0$ results of Ref.~\onlinecite{Bulla}.
We conclude that {\em the symmetric strong-coupling fixed point embodies a
natural generalization of standard Fermi-liquid physics to gapless hosts.}

\subsubsection{Asymmetric strong-coupling/frozen-impurity regime}
\label{subsubsec:thermo_ASC}

The thermodynamic properties at the frozen-impurity fixed point of the
Anderson model and at the asymmetric strong-coupling fixed point of the
exactly screened Kondo model are simply
\begin{equation}
S_{\text{imp}} = 0, \quad
C_{\text{imp}} = 0, \quad
T\chi_{\text{imp}} = 0.                                 \label{ASC_props}
\end{equation}
The corresponding properties of the underscreened Kondo model are those
of a free spin-one-half, given by Eqs.~(\ref{LM_props}).
The overscreened Kondo model has a decoupled flavor-one-half degree of freedom,
which contributes
\begin{equation}
S_{\text{imp}} = \ln 2, \quad
C_{\text{imp}} = 0, \quad
T\chi_{\text{imp}} = 0.                                 \label{ASC_o_props}
\end{equation}

The corrections to the fixed-point values can be obtained from the corrections
at the local-moment fixed point by the replacements $\tilde{J}\rightarrow 0$,
$\tilde{V}\rightarrow\tilde{V}_2$, $\tilde{t}_1\rightarrow\tilde{t}_3$,
and $\tilde{U}_0\rightarrow\tilde{U}_2$.
The results are of the form
\begin{equation}
\Delta S_{\text{imp}},\,\Delta C_{\text{imp}} \propto T^{r_1+r}, \quad
\Delta(T\chi_{\text{imp}}) \propto T^{r_1+r}.           \label{ASC_dev}
\end{equation}
Note that these expressions do not extrapolate to $r=0$, in which limit the
leading corrections are {\em linear\/} in $T$.

For $r=1$, the thermodynamic properties above should be supplemented by
terms proportional to $T^2\log T$.
In this same case, Cassanello and Fradkin\cite{Cassanello} have found
logarithmic corrections to the Kondo temperature, the static susceptibility,
and the $T$-matrix.
These authors point out that under such circumstances, multiple
energy scales enter the problem and physical quantities are no longer
controlled by the Kondo scale alone.

The results of the previous two paragraphs imply that for the Anderson and
exactly screened Kondo models, both $C_{\text{imp}}$ and $T\chi_{\text{imp}}$
approach zero as $T^{r_1+r}$.
This is the only instance among all the fixed points and models discussed in
this paper in which a system described by a scattering exponent $r>0$ exhibits
a nontrivial impurity Wilson ratio,
\begin{equation}
R_W = \lim_{T\rightarrow 0} \frac{4\pi^2}{3}
        \frac{k_B^2 T \chi_{\text{imp}}}{(g\mu_B)^2 C_{\text{imp}}}.
                                                        \label{Rw_imp}
\end{equation}

Examination of Eqs.~(\ref{F_LM:cont}) and~(\ref{chi_LM:cont}) shows that for
$r\ge 1$, $R_W$ is a function of $\tilde{V}_2^2/\tilde{t}_3$ as well as of $r$,
and is thus expected to depend on the bare couplings ($\Gamma_0$,
$\epsilon_d$, and $U$ for the Anderson model; $J_0$ and $V_0$ for the Kondo
model).
For $0<r<1$, however, the leading contribution to both the susceptibility and
the specific heat is proportional to $\tilde{V}_2^2$ alone, so the Wilson ratio
takes a universal value,
\begin{equation}
R_W(r) = \frac{\pi^2 \bar{\phi}(2r)}{6r(1\!+\!2r)\phi(2r)}.
                                                        \label{ASC_Rw}
\end{equation}
This expression will be compared with the host Wilson ratio~$R_W^{(0)}(r)$
in the next section.

\section{Numerical Results}
\label{sec:results}

This section presents numerical~RG results obtained using the formalism
described in the earlier parts of this paper.
We concentrate primarily on pure power-law scattering rates of the form of
Eq.~(\ref{Gamma_pure}).
At the end of the section, we discuss the effect of various modifications
to the scattering rate, including the introduction of particle-hole asymmetry
and the restriction of the power-law variation to a pseudogap region around
the Fermi energy.

For simplicity, we shall henceforth set $k_B=g\mu_B=1$.
(We remind the reader that the $g$-factor is assumed to be the same for
localized and conduction electrons.)

\subsection{Anderson model}
\label{subsec:Ander_res}

This subsection treats the nondegenerate Anderson model,
Eq.~(\ref{H_And:def}), restricted to the domain $U>0$.
For pure power-law scattering rates of the form of
Eq.~(\ref{Gamma_pure}) it is necessary to consider only
$\epsilon_d\ge -U/2$ (see Section~\ref{subsec:discrete2}).
Most of the results will be presented for the extreme cases
$U=-2\epsilon_d$ and $U=\infty$ which, respectively, preserve and maximally
break particle-hole symmetry.

We first examine in some detail the properties of the model for a fixed
exponent, $r=0.2$.
In particular, we show how the variation of the thermodynamic properties with
decreasing temperature can be interpreted in terms of crossovers between
various of the fixed-point regimes enumerated in Section~\ref{sec:fixed}.
This interpretation can in some instances be corroborated by computing the
relative populations of the four impurity configurations.
We then explore some of the systematic changes that take place as
$r$~is varied, focusing on the progressive damping of impurity charge
fluctuations and the consequent suppression of the Kondo effect.
Finally, we relate the preceding results to the simple scaling theory of
Ref.~\onlinecite{Buxton}.

\subsubsection{Pure power-law scattering rate with $r=0.2$}
\label{subsubsec:r=0.2}

\begin{figure}[t]
\centerline{
\vbox{\epsfxsize=75mm \epsfbox{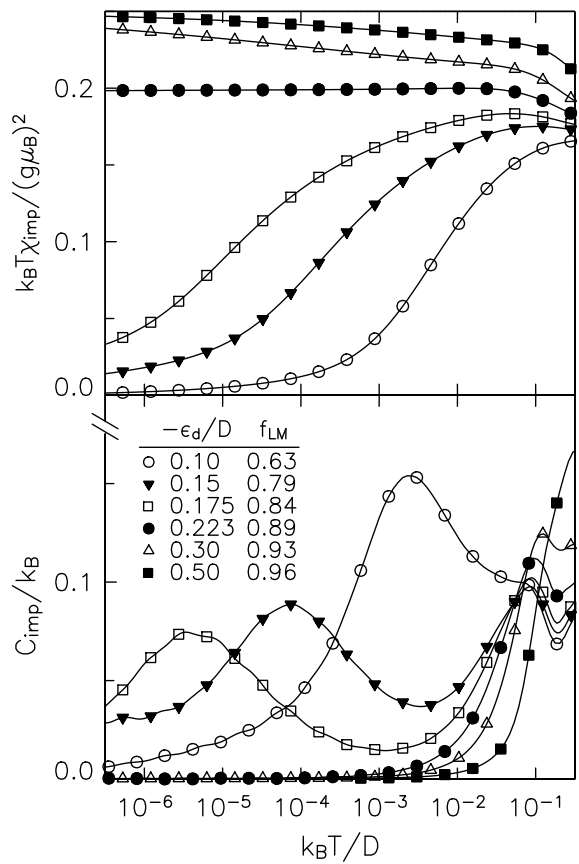}}
}
\vspace{2ex}
\caption{
Impurity susceptibility $T \chi_{\text{imp}}$ and specific heat
$C_{\text{imp}}$ vs temperature~$T$ for the infinite-$U$ Anderson model with
values of $\epsilon_d$ as labeled and a pure power-law scattering rate
specified by $r=0.2$ and $\Gamma_0=0.1D$.
The legend shows the local-moment fraction $f_{\text{LM}}$
[see Eq.~(\protect\ref{f_LM:def})] for each curve.
}
\label{fig:r=0.2_inf}
\end{figure}

\begin{figure}[t]
\centerline{
\vbox{\epsfxsize=75mm \epsfbox{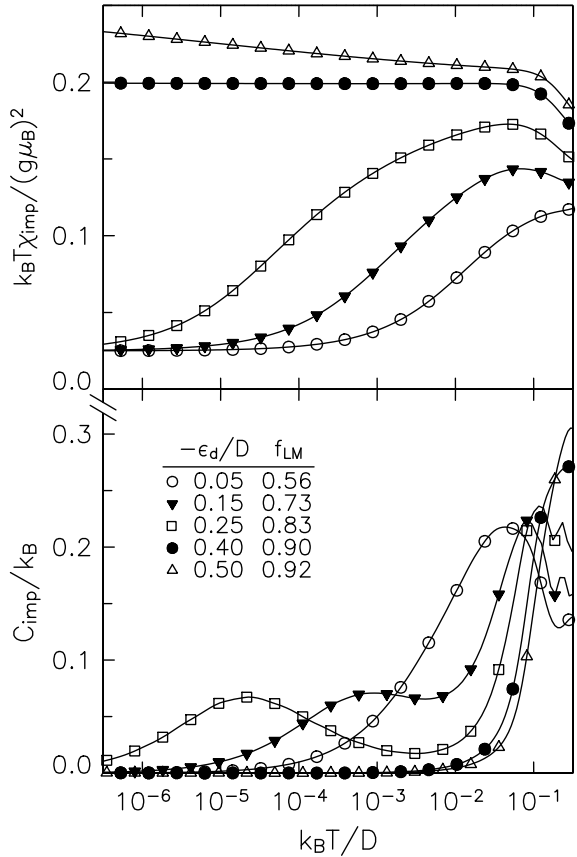}}
}
\vspace{2ex}
\caption{
Impurity susceptibility $T \chi_{\text{imp}}$ and specific heat
$C_{\text{imp}}$ vs temperature~$T$ for the symmetric Anderson model with
values of $\epsilon_d$ as labeled and a pure power-law scattering
rate specified by $r=0.2$ and $\Gamma_0=0.1D$.
}
\label{fig:r=0.2_symm}
\end{figure}

Figures~\ref{fig:r=0.2_inf} and~\ref{fig:r=0.2_symm} provide an idea of the
range of possible behaviors in the temperature-variation of the impurity
magnetic susceptibility and specific heat.
In these plots, $U$~and~$\Gamma_0$ are held fixed, and each curve
represents a different value of~$\epsilon_d$.
(To prevent overcrowding, we generally place a symbol at only one in every six
temperature points along each curve when plotting thermodynamic quantities.
The line connecting the symbols results from a fit through the
complete data set.)

The case $U=\infty$ is shown in Fig.~\ref{fig:r=0.2_inf}.
At very high temperatures, $T\gg D$ (not shown), the properties are close
to those of the valence-fluctuation fixed point (Section~\ref{subsec:valence}):
\mbox{$T\chi_{\text{imp}}\approx 1/6$} and \mbox{$C_{\text{imp}}\approx 0$}.
Once $T$~drops below~$D$, the properties become sensitive to the position of
the impurity level relative to the Fermi level.
If $\epsilon_d$~is positive or weakly negative (e.g., see the curve
for $\epsilon_d/D=-0.1$), $T\chi_{\text{imp}}$ falls monotonically with
decreasing temperature, indicating a crossover from valence fluctuation
to the frozen-impurity regime (Section~\ref{subsec:frozen}).
This crossover is accompanied by a peak in the specific heat representing a
loss of impurity entropy equal to~$\ln 3$.

For more negative values of $\epsilon_d$, $T\chi_{\text{imp}}$ initially rises
as $T$ falls below $D$, but at lower temperatures it drops back towards
zero (see the curves for $\epsilon_d/D=-0.15$ and~$-0.175$
in Fig.~\ref{fig:r=0.2_inf}).
The rise can be associated with a crossover from valence fluctuation to
local-moment behavior (Section~\ref{subsec:local}), even though
$T\chi_{\text{imp}}$ does not climb all the way to~$1/4$, the value
characterizing a free spin $s=\frac{1}{2}$.
The subsequent drop in $T\chi_{\text{imp}}$ signals a second crossover to the
frozen-impurity fixed point as the impurity becomes Kondo-screened.
The specific heat shows two well-defined peaks, corresponding to the two-stage
quenching of the impurity entropy from~$\ln 3$ at the valence-fluctuation
fixed point to~$\ln 2$ in the local-moment regime to zero at strong coupling.
This double-peak structure may be taken as a signature of the Kondo
effect, just as it is in a system with a flat scattering rate.

If the impurity level lies far below the Fermi energy ($\epsilon_d/D=-0.3$
and $-0.5$ in Fig.~\ref{fig:r=0.2_inf}), then as the temperature falls,
$T \chi_{\text{imp}}$ rises monotonically towards the local-moment value
of~$1/4$.
The crossover between from valence fluctuation to local-moment behavior is
marked by a single peak in the specific heat as the impurity loses the
entropy associated with the empty-impurity configuration.
In contrast to the standard case $r=0$, where the system eventually flows to
strong coupling for any choice of bare impurity parameters, these curves
show that for $r=0.2$ and $\epsilon_d$~sufficiently negative, an
unscreened spin survives on the impurity site down to absolute zero.
This is made possible by the local stability of the local-moment fixed point
for all $r>0$ (see Section~\ref{subsec:local}).

The final curve shown in Fig.~\ref{fig:r=0.2_inf} ($\epsilon_d/D=-0.223$)
quickly reaches a plateau at $T\chi_{\text{imp}}\approx 0.2$, and remains
there down to at least $T=10^{-6}D$.
This behavior is not compatible with any of the fixed points described in
Sections~\ref{sec:fixed} and~\ref{sec:thermo}.
Moreover, it is achieved only by careful tuning of~$\epsilon_d$ for
given $r$,~$U$ and~$\Gamma_0$.
We therefore interpret it as evidence for an unstable, intermediate-coupling
fixed point --- the manifestation in the Anderson model of the fixed point
identified by Withoff and Fradkin \cite{Withoff} in the Kondo model.
This identification is discussed further in
Section~\ref{subsubsec:intermediate}.

As mentioned in the introduction to this subsection, it can also be useful to
examine the ground-state impurity configuration.
At temperatures sufficiently high that the system is in the free-impurity
($T>U$) or valence-fluctuation ($T<U$) regime, one expects the Anderson model
to exhibit charge fluctuations between subspaces labeled by different values
of~$n_d$.
However, once $T$~drops below the effective values of both
$\Gamma$~and~$|\epsilon_d|$, real charge transfer is frozen out and
(at least for $\epsilon_d\ge -U/2$), the local level should be well-described
by one of three configurations:
(i) a {\em local moment\/} (only states with $n_d=1$ are significantly
populated);
(ii) an {\em empty impurity\/} ($\langle n_d\rangle\approx 0$);
(iii) a {\em mixed-valence configuration\/} (involving significant occupation
of states with more than one $n_d$~value).

The three low-temperature impurity configurations can be
differentiated using the local-moment fraction,
\begin{equation}
   f_{\text{LM}} = \lim_{T\rightarrow 0}
        \langle n_d-2n_{d\uparrow}n_{d\downarrow}\rangle.  \label{f_LM:def}
\end{equation}
For $U=\infty$, $f_{\text{LM}}$ is identical to the ground-state impurity
occupancy $\langle n_d\rangle$.
For $U=-2\epsilon_d$, however, $\langle n_d\rangle$ always equals one (due to
particle-hole symmetry), whereas $f_{\text{LM}}$ varies from zero to
one-half to one as $\epsilon_d$~is changed from~$+\infty$ to~$0$ to~$-\infty$.

The legend on Fig.~\ref{fig:r=0.2_inf} lists the local-moment fraction for
each of the curves.
The value $f_{\text{LM}}=0.63$ for \mbox{$\epsilon_d/D=-0.1$} places the
impurity within the mixed-valence range.
All the remaining curves have $f_{\text{LM}}>0.75$, signaling the
existence of a well-developed local moment.
It is interesting to compare this information with that provided by the
thermodynamic properties.
Note that the cases \mbox{$\epsilon_d/D=-0.10$} and~$-0.175$ result in flow to
the same fixed point, despite having the different ground-state impurity
configurations.
Conversely, the local-moment configuration present for
\mbox{$\epsilon_d/D=-0.175$}, $-0.223$, and~$-0.30$ is nonetheless associated
with three distinct fixed-point behaviors.
These observations serve to emphasize that~$f_{\text{LM}}$, which probes only
the local impurity configuration in the ground state of the system, is
complementary to the fixed-point analysis, which is based on the low-energy
excitations above the many-body ground state.

The susceptibility plots for a symmetric Anderson impurity in
Fig.~\ref{fig:r=0.2_symm} exhibit many of the features found for $U=\infty$.
In particular, large, negative values of~$\epsilon_d$ drive the system to the
local-moment regime, and there exists a critical value
\mbox{($\epsilon_d/D=-0.40$)} for which $T\chi_{\text{imp}}$ remains
approximately equal to~$0.2$ down to very low temperatures.
However, the susceptibility curves for $\epsilon_d/D>-0.40$ head towards
a low-temperature limit of $T\chi_{\text{imp}}\approx 0.025$ instead of zero.
This is precisely the behavior expected at the symmetric strong-coupling fixed
point (see Sections~\ref{subsec:symm_strong} and~\ref{subsubsec:thermo_SSC}),
which exhibits a susceptibility $T\chi_{\text{imp}}=r_1/8$, where
$r_1=\min(r,1)$.
This departure from the case $U=\infty$ demonstrates the importance of
particle-hole (a)symmetry in determining the low-temperature properties.

The specific heat curves in Fig.~\ref{fig:r=0.2_symm} are qualitatively
similar to those in Fig.~\ref{fig:r=0.2_inf}, a double peak again indicating
the formation and subsequent Kondo-screening of a local moment.
However, differences in quantitative features such as the area under each peak
reflect the fact that the cases $U=\infty$ and $U=-2\epsilon_d$ have different
fixed points in both the high- and low-temperature limits.

\begin{figure}[t]
\centerline{
\vbox{\epsfxsize=75mm \epsfbox{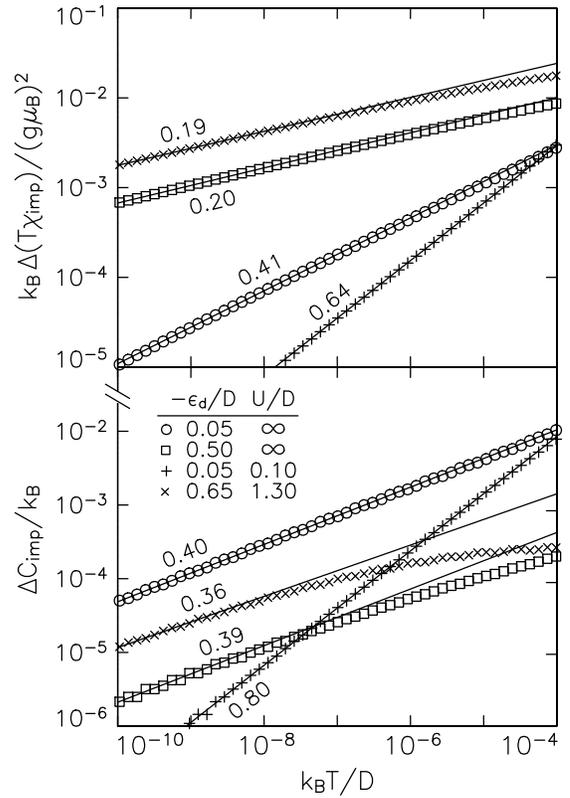}}
}
\vspace{2ex}
\caption{
Deviation of properties from their fixed-point values:
impurity susceptibility $\Delta(T \chi_{\text{imp}})$ and specific heat
$\Delta C_{\text{imp}}$ vs temperature~$T$ for the Anderson model with a pure
power-law scattering rate specified by $r=0.2$ and $\Gamma_0=0.1D$.
Straight lines represent fits to the data points over three decades of
temperature.
Each curve is labeled with its fitted slope (estimated error $\pm 0.01$).
}
\label{fig:r=0.2_app}
\end{figure}

\begin{figure}[!t]
\centerline{
\vbox{\epsfxsize=70mm \epsfbox{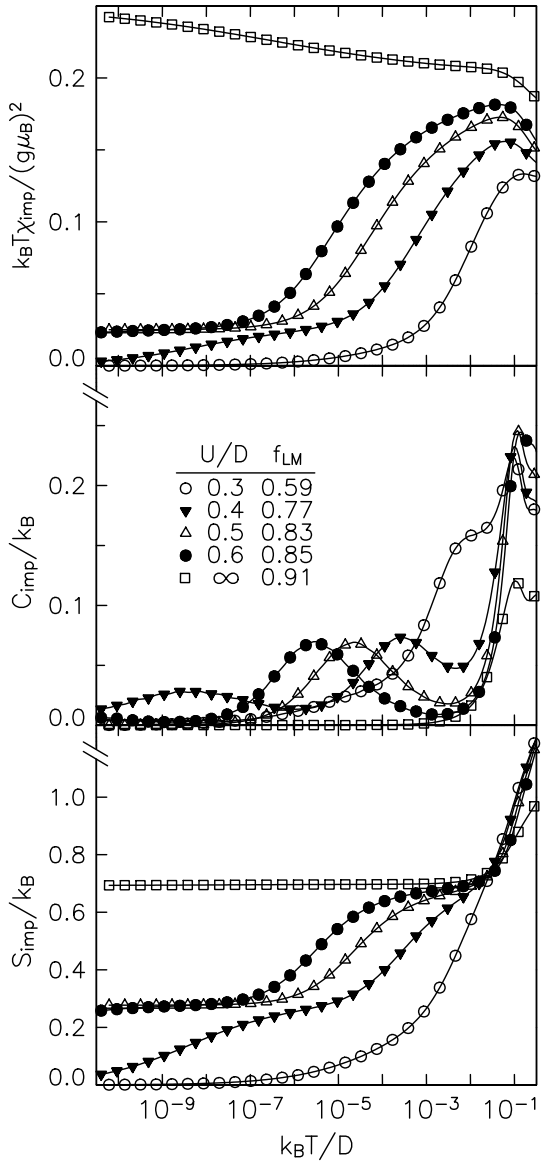}}
}
\vspace{2ex}
\caption{
Impurity susceptibility $T \chi_{\text{imp}}$, specific heat
$C_{\text{imp}}$, and entropy $S_{\text{imp}}$ vs temperature~$T$ for
the Anderson model with $\epsilon_d=-0.25D$, values of $U$ as labeled,
and a pure power-law scattering rate specified by $r=0.2$ and
$\Gamma_0=0.1D$.
}
\label{fig:r=0.2_varU}
\end{figure}

The approach to the low-temperature fixed points is highlighted in
Fig.~\ref{fig:r=0.2_app}, which shows the deviations of $T\chi_{\text{imp}}$
and $C_{\text{imp}}$ from their zero-temperature limits.
Examples are provided of both the infinite-$U$ and symmetric cases, and for
flow both to strong coupling ($\epsilon_d/D=-0.05$) and to the local-moment
fixed point ($\epsilon_d/D=-0.50$, $-0.65$).
(A symbol is placed at every second data point in this figure.)

Each curve in Fig.~\ref{fig:r=0.2_app} is labeled with an exponent obtained
by fitting the low-temperature data to a power law in~$T$.
The exponents for the $U=\infty$ curves are all close to $r$~or~$2r$,
in good quantitative agreement with Eqs.~(\ref{LM_dev}) and~(\ref{ASC_dev}).
For a symmetric impurity there are greater departures from the asymptotic
forms in Eqs.~(\ref{LM_dev}) and~(\ref{SSC_dev}).
The exponent of $0.64\pm 0.01$ for $\Delta(T\chi_{\text{imp}})$ in the
symmetric strong-coupling regime ($\epsilon_d=-0.10D$) probably
reflects the admixture of a substantial residual $T^{1-r}$ contribution into
the leading $T^{1-2r}$ term [see Eq.~(\ref{chi_SSC:cont})].
The deviation from $T^{2r}$~behavior in $\Delta C_{\text{imp}}$ near the
local-moment fixed point ($\epsilon_d=-0.65D$) can also be attributed
to incomplete convergence; fits limited to the lowest decade of temperatures
for which reliable data are available yield an exponent of $0.39\pm 0.02$,
completely consistent with perturbation theory.

Figure~\ref{fig:r=0.2_varU} shows that the low-temperature state can also
be tuned by varying $U$ at fixed $\Gamma_0$ and $\epsilon_d$.
Particularly interesting are the three curves that show signs of entry to the
local-moment regime and subsequent Kondo screening.
The middle curve ($U/D=0.5$) displays the anomalous properties associated with
the symmetric strong-coupling fixed point
($T\chi_{\text{imp}}=r_1/8$, $S_{\text{imp}}=r_1\ln4$).
The curves on either side ($U/D = 0.4$ and $0.6$) eventually enter the
frozen-impurity regime ($T\chi_{\text{imp}}=0$, $S_{\text{imp}}=0$),
but only after lingering close to the symmetric fixed point
over some range of temperatures.
(The effect is especially pronounced for $U/D=0.6$.
This curve has only just begun to move away from the symmetric fixed point
at the lowest temperatures shown.)

\subsubsection{Trends with increasing $r$}
\label{subsubsec:trends}

\begin{figure}[t]
\centerline{
\vbox{\epsfxsize=75mm \epsfbox{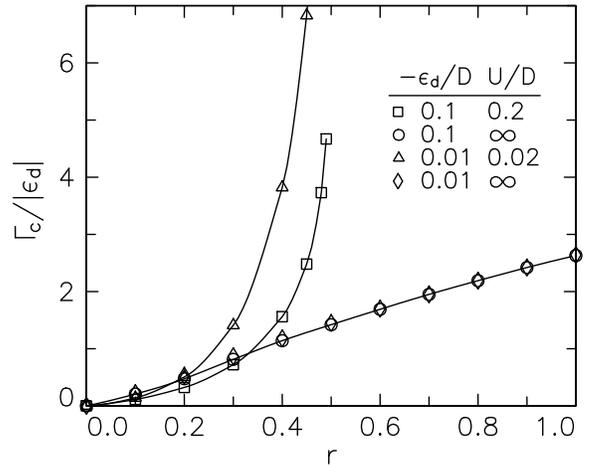}}
}
\vspace{2ex}
\caption{
Position of the intermediate-coupling fixed point, $\Gamma_c$ vs~$r$
for the Anderson model with a pure power-law scattering rate.
Solid lines are provided as a guide to the eye.
}
\label{fig:Gamma_c}
\end{figure}

In the paragraphs that follow, we examine some of the systematic trends that
arise when $r$ is varied.
We begin with the position of the unstable intermediate-coupling fixed point
which separates the stable weak- and strong-coupling basins of attraction.
Discussion of the physical properties of this fixed point will be deferred
until Section~\ref{subsubsec:intermediate}.

As shown in Figs.~\ref{fig:r=0.2_inf}--\ref{fig:r=0.2_varU}, the intermediate
fixed point can be located by adjusting $\epsilon_d$ at fixed $\Gamma_0$ and
$U$, or by tuning~$U$ while holding $\Gamma_0$~and~$\epsilon_d$ constant.
The third possibility is to define a critical scattering rate~$\Gamma_c$, such
that for $\Gamma_0>\Gamma_c$ the system flows to strong coupling, while for
$\Gamma_0<\Gamma_c$ the low-temperature physics is governed by the
local-moment fixed point.
Figure~\ref{fig:Gamma_c} plots $\Gamma_c/|\epsilon_d|$ as a function of~$r$
for two fixed values of~$\epsilon_d$, and for both $U=\infty$ and
$U=-2\epsilon_d$.
The two infinite-$U$ curves coincide almost perfectly and are roughly linear
in~$r$.
For a symmetric impurity, the dependence on $r$~and~$\epsilon_d$ is more
complicated, the most notable feature being the divergence of the critical
scattering rate as $r\rightarrow\frac{1}{2}$, beyond which point the
strong-coupling fixed point is completely inaccessible.
These trends will be discussed further in Section~\ref{subsubsec:KandA}.

\begin{figure}[t]
\centerline{
\vbox{\epsfxsize=75mm \epsfbox{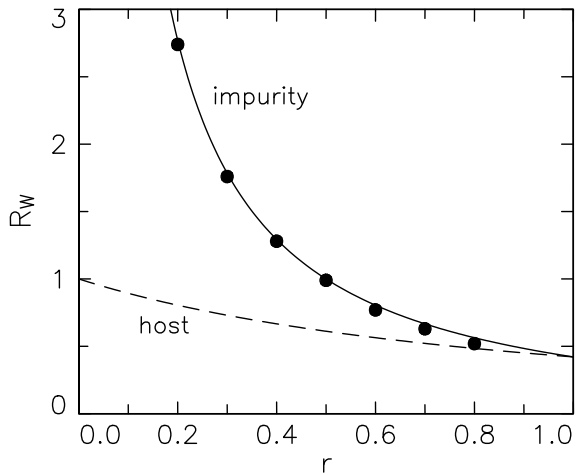}}
}
\vspace{2ex}
\caption{
Wilson ratio $R_W$ vs~$r$ for the infinite-$U$ Anderson model with a power-law
scattering rate.
Individual points, representing the impurity Wilson ratio defined in
Eq.~(\protect\ref{Rw_imp}), were determined numerically using a discretization
parameter $\Lambda=3$.
The solid line shows the perturbative result of Eq.~(\protect\ref{ASC_Rw}) for
the continuum limit, $\Lambda\rightarrow 1$.
The dashed line plots the host Wilson ratio in the absence of impurities,
calculated assuming a power-law density of states
[see Eq.~(\protect\ref{Rw_host})].
}
\label{fig:Rw}
\end{figure}

\begin{figure}[t]
\centerline{
\vbox{\epsfxsize=75mm \epsfbox{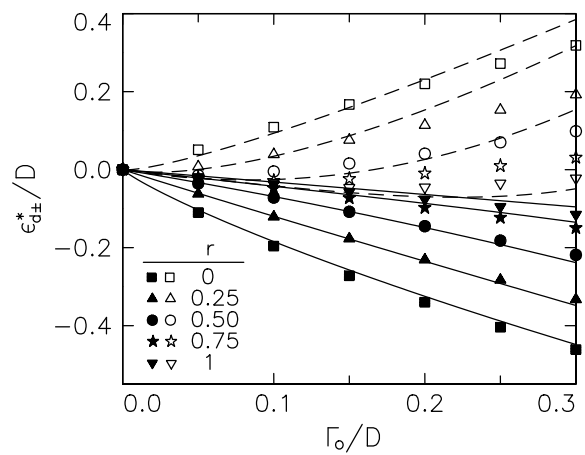}}
}
\vspace{2ex}
\caption{
Boundaries of the mixed-valence regime, $\epsilon_{d+}^{\star}$ (open symbols,
dashed lines) and $\epsilon_{d-}^{\star}$ (filled symbols, solid lines) vs
$\Gamma_0$ for the infinite-$U$ Anderson model with several different pure
power-law scattering rates.
Individual points were determined by the criteria $f_{\text{LM}}=0.75$, $0.25$
while the lines were obtained using poor-man's scaling
(see Section~\protect\ref{subsubsec:scaling}).
Dashed lines from top to bottom correspond to $r=0$, $0.25$, $0.5$ and $0.75$.
For $r\ge1$, the scaling theory predicts
$\epsilon_{d+}^{\star}=\epsilon_{d-}^{\star}$.
}
\label{fig:edstar}
\end{figure}

We now turn to the strong-coupling behavior governed by the frozen-impurity
fixed point.
The perturbation theory of Section~\ref{sec:thermo} indicates that in this
regime the impurity Wilson ratio $R_W$ defined in Eq.~(\ref{Rw_imp}) takes a
universal (although $r$-dependent) value over the range $0<r<1$.
The Wilson ratio can be obtained numerically from the computed values of
$T\chi_{\text{imp}}$ and $C_{\text{imp}}$.
To the accuracy that we can achieve, our results for $0.2\le r \le 0.8$
confirm that there is indeed a single value of $R_W$ for each $r$.

Figure~\ref{fig:Rw} compares the best value of $R_W$ (i.e., the value with
the smallest estimated error) determined numerically using a discretization
parameter $\Lambda=3$ with the continuum perturbative value,
Eq.~(\ref{ASC_Rw}).
The two sets of results agree to within~1.5\% for $r \le 0.5$, 5\%~for
$r=0.6$ and 8\%~for $r = 0.8$.
(As shown in Section~\ref{subsubsec:thermo_ASC}, the subleading corrections to
$T\chi_{\text{imp}}$ and $C_{\text{imp}}$ are smaller than the leading terms
by a factor proportional to $T^{1-r}$.
Thus, as $r$ increases $R_W$ must be calculated at progressively lower
temperatures, producing larger errors.)
Fig.~\ref{fig:Rw} also plots the host Wilson ratio $R_W^{(0)}$, defined in
Eqs.~(\ref{Rw_host}).
If the deviation from unity of the ratio $R_W/R_W^{(0)}$ can be taken as a
measure of the impurity-induced many-body effects, it appears that these
effects weaken or even vanish as $r$~approaches~1.

As noted above, the strong-coupling regime can be divided --- based
on the ground-state impurity configuration --- into empty-impurity,
mixed-valence and local-moment subregions.
Figure~3 of Ref.~\onlinecite{Buxton} shows the variation of the impurity
occupancy with $\epsilon_d$ for $U=\infty$, $\Gamma_0=0.1D$ and for values
of $r$ ranging from $0$ to $2$.
In all cases, $\langle n_d\rangle$ ($\equiv f_{\text{LM}}$ for $U=\infty$)
increases from zero to one as the impurity level moves from far below to far
above the Fermi energy.
The effect of a power-law scattering rate is to narrow the range of
$\epsilon_d$ over which the crossover takes place from an empty impurity,
through mixed valence, to a full local moment.

In order to quantify the narrowing of the mixed-valence regime,
we define $\epsilon_{d+}^{\star}$ and $\epsilon_{d-}^{\star}$ as the values
of the impurity energy which result in a ground-state occupation
$f_{\text{LM}} = 0.25$ and $0.75$, respectively.
These energies, which we take to represent upper and lower bounds
on the mixed-valence regime, are plotted in Fig.~\ref{fig:edstar}.
It is clear that the mixed-valence region of parameter space shrinks
monotonically as the power $r$ increases at fixed $\Gamma_0$.

\begin{figure}[t]
\centerline{
\vbox{\epsfxsize=75mm \epsfbox{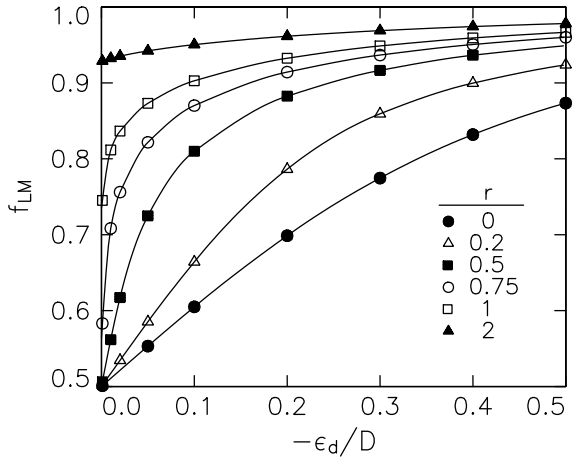}}
}
\vspace{2ex}
\caption{
Local-moment fraction $f_{\text{LM}}$ vs impurity energy~$\epsilon_d$ for the
symmetric Anderson model with pure power-law scattering rates specified by
$\Gamma_0=0.1D$ and several different values of $r$.
The lines are provided as a guide to the eye.
}
\label{fig:flm}
\end{figure}

\begin{figure}[t]
\centerline{
\vbox{\epsfxsize=75mm \epsfbox{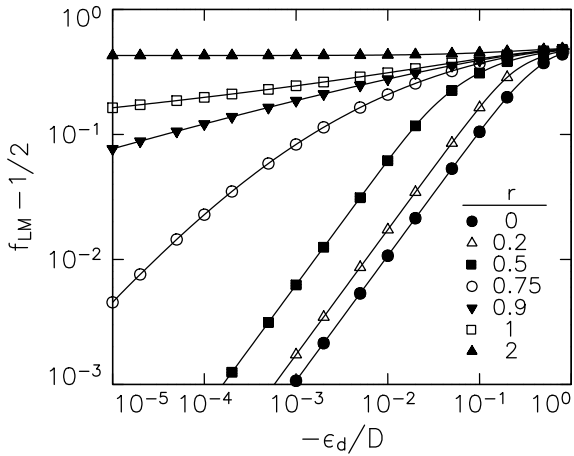}}
}
\vspace{2ex}
\caption{
Same as Fig.~\protect\ref{fig:flm}, plotted on a log-log scale.
}
\label{fig:logflm}
\end{figure}

The local-moment fraction for a symmetric impurity with $\Gamma_0=0.1D$
is plotted as a function of the impurity energy in Fig.~\ref{fig:flm}.
(Only negative values of $\epsilon_d$ are shown.
Due to particle-hole symmetry, $f_{\text{LM}}$ is mapped to $1-f_{\text{LM}}$
when the sign of $\epsilon_d$ is reversed.)
Just as for the infinite-$U$ case, the range of $\epsilon_d$ over which
$f_{\text{LM}}$ takes values between 0.75 and 0.25 decreases dramatically
with increasing $r$.

Figure~\ref{fig:logflm} plots $f_{\text{LM}}-\frac{1}{2}$ against~$-\epsilon_d$
on a log-log scale.
For $r\le 0.5$, $f_{\text{LM}}$ clearly dips downward to approach the
value~$\frac{1}{2}$ linearly as $\epsilon_d\rightarrow 0$.
The curves for $r=0.75$ and~$0.9$ also show signs of the same behavior,
although the linear regime is pushed to much smaller~$|\epsilon_d|$.
By contrast, the curvature of the data for \mbox{$r=1$} and \mbox{$r=2$}
suggests that $f_{\text{LM}}$ approaches a value greater than~$\frac{1}{2}$ as
$\epsilon_d\rightarrow 0^-$, and hence undergoes a discontinuous jump when
the impurity level passes through the Fermi energy.

Figures~\ref{fig:edstar}--\ref{fig:logflm} show that as the exponent~$r$
describing the power-law scattering rate increases, there is a progressive
shrinking of the mixed-valence region of parameter space in favor of the
local-moment and empty-impurity regimes.
This trend is a natural consequence of the depression of the scattering
rate near the Fermi level, which strongly inhibits mixing between the impurity
level and low-energy conduction states.

\begin{figure}[t]
\centerline{
\vbox{\epsfxsize=75mm \epsfbox{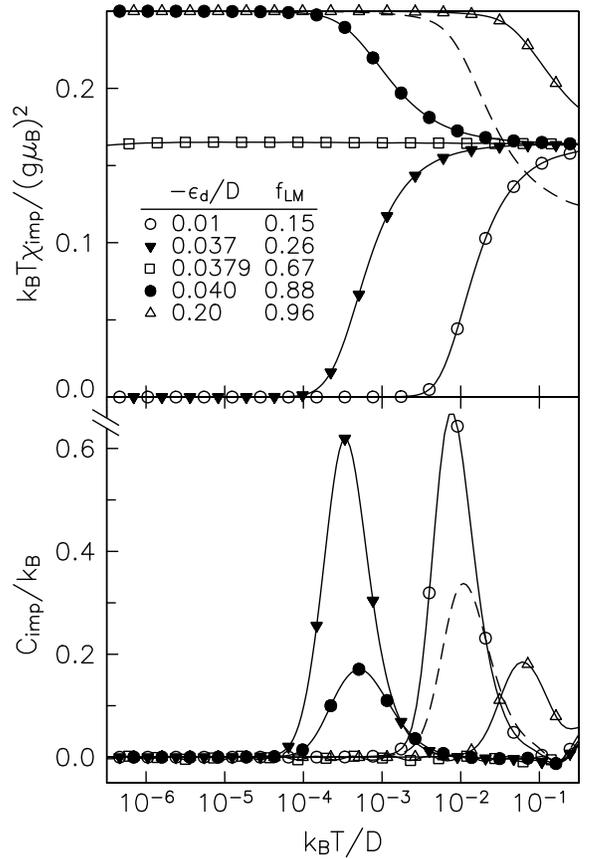}}
}
\vspace{2ex}
\caption{
Impurity susceptibility $T \chi_{\text{imp}}$ and specific heat
$C_{\text{imp}}$ vs temperature~$T$ for the Anderson model with a linear
scattering rate specified by $r=1$ and $\Gamma_0=0.1D$.
Data points connected by solid lines correspond to $U=\infty$ and
values of $\epsilon_d$ shown in the legend.
The dashed line shows the symmetric case $U/2=-\epsilon_d = 0.04D$.
}
\label{fig:r=1}
\end{figure}

The numerics also reveal a second trend with increasing~$r$, namely the
progressive disappearance of the Kondo effect.
For $r\agt 0.5$ it proves almost impossible to find any region of
parameter space within which the thermodynamic properties indicate that
entry to the local-moment regime is followed by a crossover to either of the
two strong-coupling fixed points.
This is illustrated in Fig.~\ref{fig:r=1}, which shows the impurity
susceptibility and specific heat for a linear scattering rate.
As noted above, the case $r=1$ is of particular interest because it may
describe a magnetic impurity in a $d$-wave superconductor
\cite{Borkowski:92,Cassanello} and in flux phases of two-dimensional
electrons.

Consider first the solid curves in Fig.~\ref{fig:r=1} representing the case
$U=\infty$.
For $\epsilon_d/D\ge -0.037$, the impurity level is almost unoccupied in the
ground state ($f_{\text{LM}} \alt 1/4$), and the system crosses
directly from valence fluctuation to the frozen-impurity regime;
there is neither a peak in $T\chi_{\text{imp}}$ nor a double peak in
$C_{\text{imp}}$ to signal the Kondo effect.
For $\epsilon_d/D \le -0.040$, the renormalization is from valence fluctuation
to the stable local-moment fixed point, and the $T=0$ ground state has an
unquenched spin at the impurity site.
Between these behaviors lies an unstable intermediate-coupling fixed point,
obtained by tuning the impurity level to $\epsilon_d/D \approx -0.0379$.
The absolutely flat $T\chi_{\text{imp}}$ curve in this case indicates that the
fixed point is reached directly from the high-temperature regime, rather than
from the local-moment regime as was found for $r=0.2$ (see
Figs.~\ref{fig:r=0.2_inf} and~\ref{fig:r=0.2_symm}).

Figure~\ref{fig:r=1} also provides one representative example of
the properties exhibited by a symmetric impurity with a linear scattering
rate (dashed line).
As far as we have been able to determine, any negative value of
$\epsilon_d$, however small in magnitude, results in flow to the
local-moment fixed-point.
This observation is consistent both with the absence of any finite critical
coupling $\Gamma_c$ for $r>\frac{1}{2}$ (as shown in Fig.~\ref{fig:Gamma_c})
and with the evidence that there is a jump in the local-moment fraction at
$\epsilon_d=0$ for $r\ge 1$ (see the discussion of Fig.~\ref{fig:logflm}).

\begin{figure}[t]
\centerline{
\vbox{\epsfxsize=75mm \epsfbox{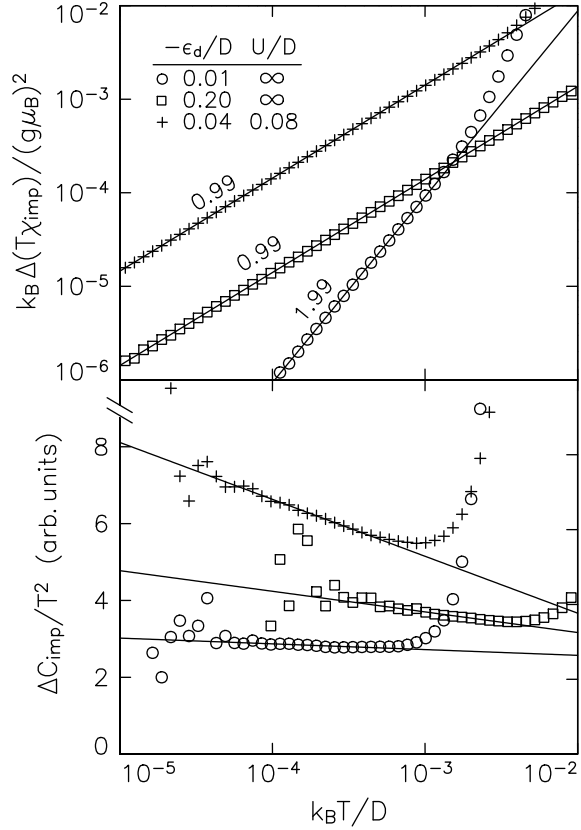}}
}
\vspace{2ex}
\caption{
Deviation of properties from their fixed-point values for the Anderson model
with a linear scattering rate specified by $r=1$ and $\Gamma_0=0.1D$.
Top: Impurity susceptibility $\Delta(T \chi_{\text{imp}})$ vs temperature~$T$.
Straight lines represent fits to the data over the range $k_B T/D< 10^{-3}$.
Each curve is labeled with its fitted slope (estimated error $\pm 0.01$).
Bottom: Specific heat $\Delta C_{\text{imp}}/ T^2$ vs~$T$, with
straight-line fits to the data over selected temperature ranges.
}
\label{fig:r=1_app}
\end{figure}

Figure~\ref{fig:r=1_app} illustrates the approach of the $r=1$ thermodynamic
properties to their zero-temperature values.
As discussed in Section~\ref{subsec:thermo_pert}, a linear scattering rate
admits logarithmic corrections to power-law behaviors at both the
local-moment and frozen-impurity fixed points.
For $T\alt 10^{-3}D$, $\Delta T\chi_{\text{imp}}$ is linear in $T$
on approach to the local-moment fixed point [as predicted by
Eq.~(\ref{chi_LM:cont})] and quadratic (with no significant component of
$T^2\log T$) in the frozen-impurity regime.
The specific heat is expected to behave as $T^2\log T$ at both fixed points.
The deviations from a pure $T^2$ form are only weak for the frozen-impurity
example ($\epsilon_d=-0.01D$) but are considerably stronger for the two flows
to the local-moment fixed point.
In all three cases, the lowest decade of temperature before the data become
too noisy for a reliable fit seems to be consistent with a $T^2\log T$ form,
although higher-accuracy calculations would be needed to rule out completely
any other behavior.

\subsubsection{Comparison with scaling theory}
\label{subsubsec:scaling}

\begin{figure}[t]
\centerline{
\vbox{\epsfxsize=75mm \epsfbox{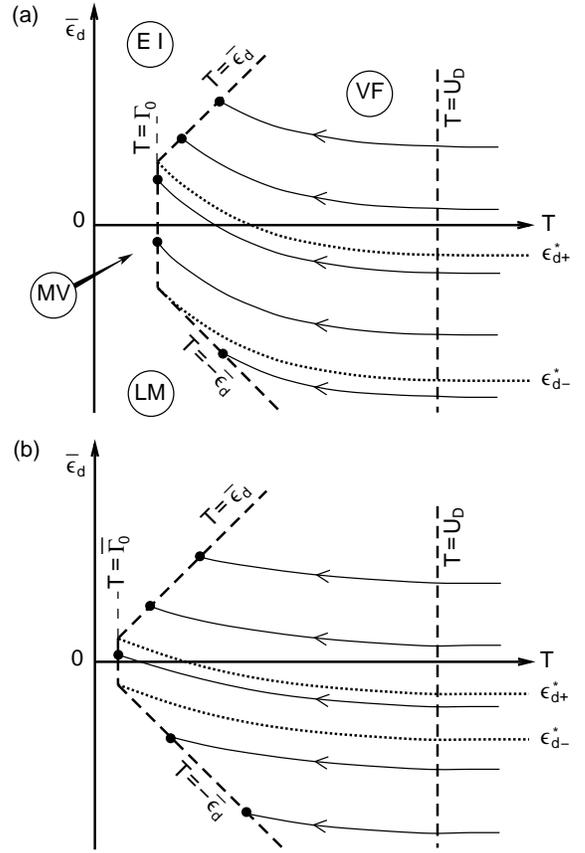}}
}
\vspace{2ex}
\caption{
Scaling of the effective impurity energy~$\bar{\epsilon}_d$ with the
temperature~$T$, shown schematically for (a)~$r=0$; (b)~\mbox{$0<r<1$}.
Renormalization of~$\bar{\epsilon}_d$ from its bare value~$\epsilon_d$ begins
on entry into the valence-fluctuation~(VF) regime, at
$D\approx U_D=\text{min}(U,D)$.
Scaling ends at a crossover to local-moment~(LM), empty-impurity~(EI) or
mixed-valence~(MV) behavior.
A power-law scattering rate flattens the trajectories~$\bar{\epsilon}_d(T)$
and also moves the boundary of the MV regime to the left, reducing the
range of bare impurity energies
($\epsilon_{d-}^{\star}<\epsilon_d<\epsilon_{d+}^{\star}$) which yield
MV behavior.
}
\label{fig:scaling}
\end{figure}

The numerical results presented above can be compared with the predictions of
an approximate analytical treatment of the problem based on the poor-man's
scaling technique first developed for the Kondo problem.\cite{Anderson:scaling}
Jefferson\cite{Jefferson} and Haldane\cite{Haldane} applied this method to
the nondegenerate Anderson model with a flat scattering rate ($r=0$).
The main effect of many-body interactions was found to be a
temperature-dependent shift in the effective energy of a nonsymmetric impurity
level from its bare value~$\epsilon_d$ to
\begin{equation}
   \bar{\epsilon}_d(T) = \epsilon_d  + \frac{\Gamma}{\pi}
        \ln\left(\frac{U_D}{T}\right),  \quad r = 0.
                                                        \label{epsilon(T):r=0}
\end{equation}
Here $U_D = \min(U,D)$ is the energy scale below which many-body effects
come into play.
Neither $U$~nor~$\Gamma$ is significantly renormalized.
(Note that particle-hole symmetry prevents renormalization of $\epsilon_d$
for a symmetric impurity.
In this case, the upper bound~$U_D$ on the range of~$T$ within which
renormalization occurs essentially coincides with the lower
bound~$|\epsilon_d|$.)

In this scaling picture, real charge fluctuations on the impurity site are
expected to become frozen out around a temperature
$T_F = \max(|\bar{\epsilon}_d|,\Gamma).$
(As usual, we assume that \mbox{$U\ge 2|\epsilon_d|$}.)
If $T_F = -\bar{\epsilon}_d \gg \Gamma$, then only the singly
occupied configurations will be significantly populated, and the impurity
will possess a local moment, while an empty impurity will result if
$T_F = +\bar{\epsilon}_d \gg \Gamma$.
Finally, if $T_F = \Gamma \gg |\bar{\epsilon}_d|$, then the
ground state will have mixed valence.
Figure~\ref{fig:scaling}(a) provides a schematic representation of the
renormalization of $\epsilon_d$ in the case $r=0$ and illustrates
the crossover from valence fluctuation into the three low-temperature regimes.

The poor-man's scaling treatment was recently extended to a power-law
scattering rate.\cite{Buxton}
The depression of the scattering rate at low energies can be represented as
a renormalization of the parameter~$\Gamma_0$ entering Eq.~(\ref{Gamma_pure}):
\begin{equation}
\bar{\Gamma}_0(T) = \Gamma_0 \cdot (T/D)^r, \quad T < D.
                                                        \label{Gamma(T)}
\end{equation}
This in turn feeds back to produce a smaller renormalization of~$\epsilon_d$
than occurs in the case $r=0$:
\begin{equation}
\bar{\epsilon}_d(T) = \epsilon_d  + \frac{\Gamma_0}{\pi r}
                \left[ \left( \frac{U_D}{D} \right)^r -
                \left( \frac{T}{D} \right)^r \right ], \quad r > 0.
                                                        \label{epsilon(T)}
\end{equation}
The crossover temperature characterizing the freezing-out of real
charge fluctuations on the impurity site must be redefined to be
\begin{equation}
   T_F = \max(|\bar{\epsilon}_d(T_F)|, \bar{\Gamma}_0(T_F)).
                                                        \label{T_F}
\end{equation}

The scaling of~$\bar{\epsilon}_d(T)$ for $r>0$ is shown schematically
in Fig.~\ref{fig:scaling}(b).
Note that the trajectories are flatter than in~(a), reflecting the
reduction in the many-body shift of~$\epsilon_d$.
Moreover, the energy-dependence of the scattering rate pushes the
vertical line $T=\bar{\Gamma}_0(T)$ significantly to the left,
thereby shrinking the range of $\epsilon_d$~values which result in a
mixed-valence ground state.
(For $r>1$ the vertical line is driven to $T=0$, and the
mixed-valence regime disappears altogether.)

{\em Bounds on the mixed-valence region.\/}
Within the poor-man's scaling approach, upper and lower bounds on the
mixed-valence regime, $\epsilon_{d+}^{\star}$~and~$\epsilon_{d-}^{\star}$,
respectively, can be defined as the bare values of~$\epsilon_d$ for which the
scaling trajectories pass through the upper and lower intersections between
dashed lines in Fig.~\ref{fig:scaling}, i.e., as roots of the implicit
equation
\begin{equation}
   \Gamma_0(|\bar{\epsilon}_d(T_F)|) =
      \pm a_{\pm} \bar{\epsilon}_d(T_F),        \label{epsilon_pm_implicit}
\end{equation}
where $a_{+}$~and~$a_{-}$ are positive constants of order unity.
The solutions of Eq.~(\ref{epsilon_pm_implicit}) are
\begin{equation}
    \frac{\epsilon_{d\pm}^{\star}}{D} =
        \left( \frac{a_{\pm}}{\pi r} \pm 1 \right)
        \left( \frac{\Gamma_0}{a_{\pm}D} \right)^{1/(1-r)}
        -\frac{\Gamma_0}{\pi r D} \left( \frac{U_D}{D} \right)^r.
                                                \label{epsilon_d*}
\end{equation}

Figure~\ref{fig:edstar} superimposes these bounds for the mixed-valence region
in the case $U=\infty$ (plotted as lines) on those defined in
Section~\ref{subsubsec:r=0.2} based on the computed value of the local-moment
fraction $f_{\text{LM}}$ (individual symbols).
The choice $a_{-}=0.9$ brings the alternative definitions
of~$\epsilon_{d-}^{\star}$ into good quantitative agreement.
The dashed lines representing~$\epsilon_{d+}^{\star}$ in Fig.~\ref{fig:edstar},
computed for $a_{+}=0.6$, fit the numerical data reasonably well for small
values of~$\Gamma_0$ but deviate considerably for stronger impurity
scattering.
This discrepancy is not especially surprising, given the approximations
inherent to poor-man's scaling and the degree of arbitrariness present in
both definitions of~$\epsilon_{d+}^{\star}$.

For the symmetric case shown in Figs.~\ref{fig:flm} and~\ref{fig:logflm},
the criterion $f_{\text{LM}}=0.75$ for the border of the local-moment regime
can be compared with the scaling definition
$\bar{\Gamma}_0(|\epsilon_{d-}^{\star}|) = -a_{-}\epsilon_{d-}^{\star}$.
(Recall that the level energy does not renormalize at particle-hole symmetry,
so $\bar{\epsilon}_d = \epsilon_d$.)
A value of $a_{-}\approx 0.4$ seems to yield reasonable agreement with the
numerics, at least for $r\le 0.5$.

{\em Exchange on entry to the local-moment regime.\/}
The scaling theory also sheds light on the disappearance of the
Kondo effect with increasing~$r$.
On entry to the local-moment regime, the Anderson model can be mapped onto
the Kondo problem by projecting into the subspace in which $n_d=1$.
Applying Eqs.~(\ref{SW}), one obtains \cite{Buxton} a dimensionless exchange
having a power-law energy-dependence,
\begin{equation}
\sqrt{\rho(\epsilon)\rho(\epsilon')} J(\epsilon,\epsilon') =
        \rho_0 J_0 \, |\epsilon \, \epsilon'/D^2|^{r/2},
                                                \label{J_pure}
\end{equation}
where
\begin{equation}
\rho_0 J_0 = \frac{2\bar{\Gamma}_0(T_F)}{\pi}
             \left(\frac{1}{T_F}+\frac{1}{U-T_F}\right).        \label{J_SW}
\end{equation}
In this instance, Eq.~(\ref{T_F}) reduces to $T_F=-\bar{\epsilon}_d(T_F)$.

For given impurity parameters ($\epsilon_d$, $U$, and $\Gamma_0$),
the exchange decreases with increasing $r$, as shown in Fig.~2 of
Ref.~\onlinecite{Buxton}.
Both the depression of $\Gamma(T_F)$ and the weaker renormalization of
$\epsilon_d$ (which increases $T_F$) contribute to this effect.
A conservative bound on the multiplicative reduction factor for $\rho_0 J_0$,
obtained by neglecting the renormalization of $\epsilon_d$ altogether, is
$|\epsilon_d/D|^r$.
This sharp reduction of $\rho_0 J_0$ with increasing $r$ militates strongly
against any Kondo effect, because (as first shown by Withoff and Fradkin
\cite{Withoff}), an impurity spin becomes screened only if $J_0 > J_c$, where
$\rho_0 J_c \approx r$.

In summary, the poor-man's scaling analysis captures the essential features of
the Anderson problem with a power-law scattering rate, and provides a
convenient theoretical framework for understanding the numerical results.
A number of quantitative features, however, are not accounted for correctly
within the scaling approach.

\subsection{Screened Kondo model}
\label{subsec:screened}

The conventional $s=\frac{1}{2}$ (``exactly screened'') Kondo model with a
power-law exchange coupling [Eq.~(\ref{J_pure})] has been discussed
extensively in
Refs.~\onlinecite{Withoff,Borkowski:92,Cassanello,Chen,Ingersent:scaling,%
Ingersent:NRG} and reviewed briefly in the Introduction above.
(Some of the papers cited \cite{Withoff,Borkowski:92,Cassanello} formally
treat the degenerate Anderson model using large-$N$ methods; however, the focus
throughout is the Kondo physics of the local-moment regime.)
We recall that the novel feature of the model is the existence of a finite
exchange coupling $J_c(r,V_0)$ which separates a region of parameter space
within which the impurity spin becomes asymptotically free ($J_0<J_c$) from
another in which the impurity moment is quenched.
The latter case is governed by two distinct fixed points:
under conditions of strict particle-hole symmetry,\cite{Chen} the
low-temperature susceptibility is a universal function of $T/T_K$,
where $T_K\propto |J_0-J_c|^{1/r}$; otherwise, the properties are
determined by two independent energy scales.\cite{Ingersent:NRG}

Rather than attempting to provide a comprehensive treatment of the exactly
screened Kondo model, which would necessarily duplicate much previously
published work, we focus in this section on three topics that have received
little attention.
First, we address the properties of the Kondo model at the critical coupling.
We show that over a range of $r$ there in fact exist two distinct
intermediate-coupling fixed points --- one accessible only under conditions
of exact particle-hole symmetry, the other reached when this symmetry is
broken.
Second, we present results for a local response function which is a
candidate order parameter for the critical behavior at the
intermediate-coupling fixed point(s).
Finally, we examine the relationship between the Kondo and Anderson
models in systems with a power-law scattering rate.
Although each RG~fixed point of the Kondo model is equivalent to a fixed point
of the Anderson model, we argue that two models are independent to a greater
extent than is the case for $r=0$.

\subsubsection{Properties of the intermediate-coupling fixed point}
\label{subsubsec:intermediate}

\begin{figure}[t]
\centerline{
\vbox{\epsfxsize=75mm \epsfbox{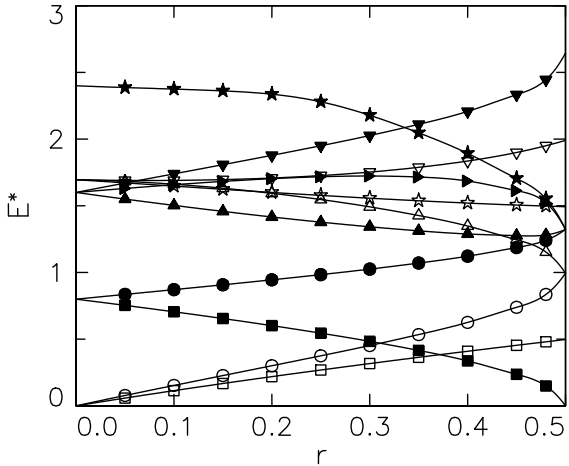}}
}
\vspace{2ex}
\caption{
Low-lying eigenvalues $E^{\star}$ at the particle-hole-symmetric
intermediate-coupling fixed point of the exactly screened Kondo model,
plotted vs~the exponent $r$ describing the power-law exchange.
Data computed with $\Lambda=3$ are shown both for $N$~even (open symbols) and
for $N$~odd (filled symbols).
Solid lines are provided as a guide to the eye.
The curves are extrapolated to the weak-coupling values at $r=0$ and to
symmetric strong coupling at $r=0.5$.
}
\label{fig:Jc_levels}
\end{figure}

\begin{figure}[t]
\centerline{
\vbox{\epsfxsize=75mm \epsfbox{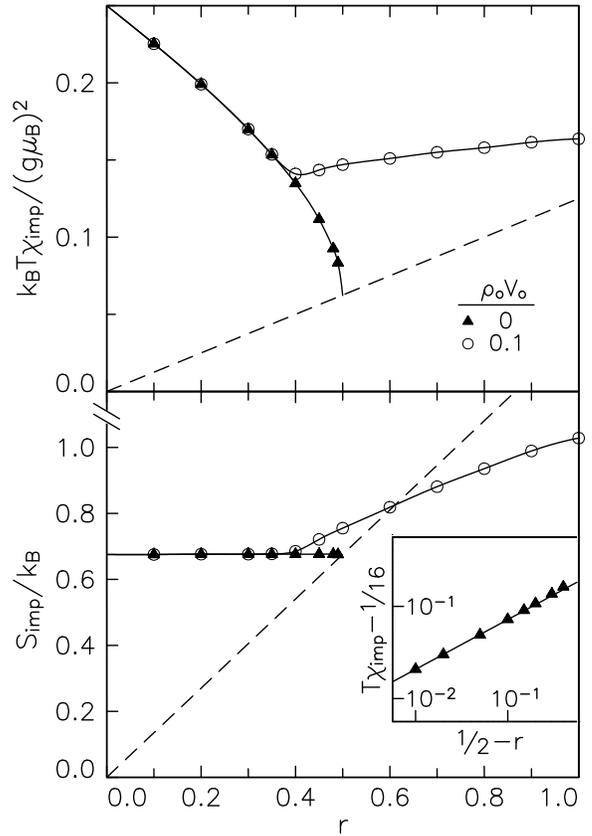}}
}
\vspace{2ex}
\caption{
Impurity susceptibility $T \chi_{\text{imp}}$ and entropy $S_{\text{imp}}$
at the intermediate-coupling fixed points of the exactly screened Kondo model,
plotted vs~$r$.
Solid lines are provided as a guide to the eye.
The curves are extrapolated to weak-coupling at $r=0$, and for $V_0=0$ are
extrapolated to the symmetric strong-coupling values [given by
Eq.~(\protect\ref{SSC_props}), dashed lines] at $r=\frac{1}{2}$.
Inset: Approach of $T \chi_{\text{imp}}$ to its value at $r=\frac{1}{2}$,
plotted on a log-log scale.
}
\label{fig:Jc_props}
\end{figure}

Unlike the weak- and strong-coupling fixed points discussed in
Section~\ref{sec:fixed}, the intermediate-coupling fixed points of the
Anderson and Kondo Hamiltonians are not amenable to conventional perturbative
methods, and our knowledge of their properties comes entirely from numerical
studies.
This subsection presents numerical~RG results for the Kondo model obtained by
tuning the exchange parameter $J_0$ at fixed $r$ and $V_0$ to lie very close
to the border between the strong- and weak-coupling basins of attraction.
(These results will be compared with those for the Anderson model in
Section~\ref{subsubsec:KandA}.)

Since we focus on zero-temperature (fixed-point) properties, it is not
necessary to average over different discretizations of the conduction band
(as described in Section~\ref{subsec:thermo_num}).
The results discussed below were obtained using a single discretization,
corresponding to $z=1$ in Eqs.~(\ref{+bins}) and~(\ref{-bins}).

We begin by considering the particle-hole-symmetric problem corresponding to
$V_0=0$, for which the critical coupling $J_c(r)$ is finite only over the
range $0<r<r_{\text{max}}$, where
$r_{\text{max}}=\frac{1}{2}$.\cite{Chen,Ingersent:scaling,Ingersent:NRG}
When the bare exchange $J_0$ is tuned precisely to $J_c(r)$, the problem
flows to the intermediate-coupling fixed point mentioned previously.
Figure~\ref{fig:Jc_levels} plots the lowest-lying eigenenergies at this
fixed point over the range $0.05\le r\le 0.48$, while Fig.~\ref{fig:Jc_props}
shows (filled symbols) the impurity contribution to the total susceptibility
and entropy for $0.1\le r\le 0.49$.
As $r$ is varied from zero to $\frac{1}{2}$ each curve interpolates smoothly
between the corresponding weak-coupling and strong-coupling values.
Over the entire range of $r$, the computed entropy at the intermediate fixed
point remains within 1\% of $\ln 2$.
The inset to Fig.~\ref{fig:Jc_props} shows that the approach of
$T\chi_{\text{imp}}$ to its value at $r=\frac{1}{2}$ is described by a
power law: $T\chi_{\text{imp}}-\frac{1}{16}\propto(\frac{1}{2}-r)^{\nu}$
with $\nu=0.54\pm 0.05$.
All the quantities plotted in Figs.~\ref{fig:Jc_levels} and~\ref{fig:Jc_props}
vary linearly at small $r$, and are consistent with the divergence of the
critical coupling $J_c$ as $r\rightarrow r_{\text{max}}\equiv\frac{1}{2}$.

Away from particle-hole symmetry, a finite critical coupling $J'(r,V_0)$
can be identified for any $r>0$.
For $r<\frac{1}{2}$, $J'$ deviates smoothly from its particle-hole-symmetric
value $J_c(r)$ as $V_0$ is increased from zero.
The initial slope $dJ'/dV_0$ can be of either sign, but for strong potential
scattering (which disfavors the presence of a single conduction electron at
the impurity site), $J'$ invariably rises sharply (see Fig.~1 of
Ref.~\onlinecite{Ingersent:NRG}).

The nature of the fixed point reached for $J_0 = J'$ is found to be
fundamentally different for small and large values of $r$.
Particle-hole symmetry proves to be marginally irrelevant for
$0<r<r^{\star}\approx 0.4$, over which range the intermediate fixed point
is identical to that obtained for $V_0 = 0$.
For $r>r^{\star}$, by contrast, all positive bare values of $V_0$ result
in flow to a new, particle-hole-asymmetric fixed point located at
$V_0=V_c$, $J_0 = J_c' \equiv J'(r,V_c)$.
Thus, {\em over the limited range $r^{\star} < r < r_{\text{max}}$, there exist
two distinct intermediate-coupling fixed points,} located at $J_c$ and $J_c'$.
Potential scattering is marginally relevant at the former, whereas the
coupling $V_0-V_c$ is marginally irrelevant at the latter.
(Yet another fixed point is reached for all $V_0 < 0$, but since it is
trivially related to that for $V_0>0$ by particle-hole exchange, we treat
this pair as being physically equivalent.)

Support for the statements contained in the previous paragraph comes from the
impurity susceptibility and entropy shown in Fig.~\ref{fig:Jc_props}.
Properties computed for a fixed value of the potential scattering,
$\rho_0 V_0=0.1$, are essentially indistinguishable from their
$V_0=0$ counterparts for all $r\alt 0.4$.
Beyond this point, $T\chi_{\text{imp}}$ and $S_{\text{imp}}$ at the
particle-hole-asymmetric fixed point rise monotonically with increasing $r$.
The properties for $\rho_0 V_0=0.5$ (not shown in Fig.~\ref{fig:Jc_props})
are identical to those for \mbox{$\rho_0 V_0$ = 0.1}, at least to within our
estimated accuracy.

\begin{figure}[t]
\centerline{
\vbox{\epsfxsize=75mm \epsfbox{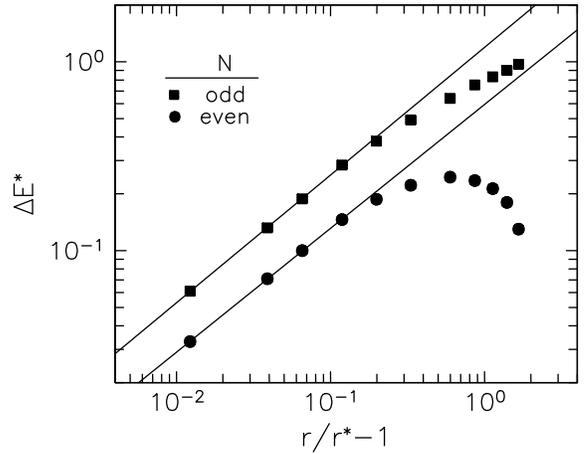}}
}
\vspace{2ex}
\caption{
Energy splittings $\Delta E^*$ of the lowest pair of charge-conjugate
eigenstates at the particle-hole-asymmetric intermediate-coupling fixed point
of the exactly screened Kondo model, plotted vs~the reduced exponent
$r/r^{\star}-1$, for $\Lambda = 3$ and $r^{\star}=0.3754$.
Straight lines show fits to the leftmost four data points for odd- and
even-numbered iterations of the numerical~RG method.
}
\label{fig:Jc_split}
\end{figure}

The many-body spectrum provides additional evidence for the existence of dual
fixed points over the range $r^{\star} < r < r_{\text{max}}$.
Certain pairs of charge-conjugate states which are necessarily degenerate
for $V_0=0$ are split at the asymmetric fixed point.
Figure~\ref{fig:Jc_split} plots the splitting $\Delta E^{\star}$ of the lowest
pair of affected states, both for odd- and even-numbered iterations~$N$,
over the range \mbox{$0.38 \le r \le 1$}.
The vanishing of~$\Delta E^{\star}$ defines the critical value~$r^{\star}$.
For $r\alt 0.45$, we find that
\mbox{$\Delta E^{\star} \propto (r-r^{\star})^{\nu^{\star}}$}, with
\mbox{$r^{\star}=0.375\pm 0.002$} and \mbox{$\nu^{\star}=0.67\pm 0.15$}.
It seems probable that the magnitude of the splitting is
directly proportional to the critical potential scattering, at least
for small~$V_c$, but we have no proof of this conjecture.

\begin{figure}[!t]
\centerline{
\vbox{\epsfxsize=75mm \epsfbox{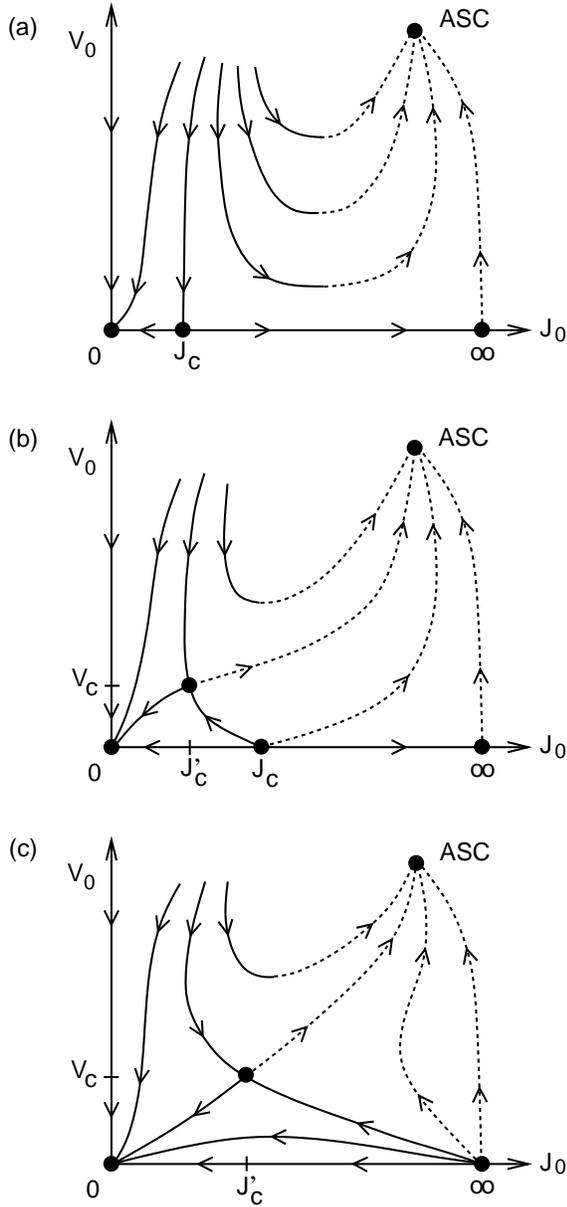}}
}
\vspace{2ex}
\caption{
Schematic renormalization-group flow diagrams for the exactly screened Kondo
model, showing the $J_0$--$V_0$~plane for $V_1=0$ and for fixed $r$:
\mbox{(a) $0 < r < r^{\star}$;}
\mbox{(b) $r^{\star} < r < r_{\text{max}}$;}
\mbox{(c) $r > r_{\text{max}}$.}
Thin lines with arrows show the renormalization of effective couplings with
decreasing temperature.
Filled circles indicate RG~fixed points.
Broken lines represent flows out of the plane towards the asymmetric
strong-coupling (ASC) fixed point, located at $|V_1|=\infty$.
}
\label{fig:flows}
\end{figure}

Figures~\ref{fig:flows}(a), (b)~and~(c) summarize the effect of particle-hole
asymmetry on the exactly screened Kondo model for a fixed value of $r$, where
\mbox{$0<r<r^{\star}$}, \mbox{$r^{\star}<r<r_{\text{max}}$}, and
\mbox{$r>r_{\text{max}}$}, respectively.
These figures sketch the RG~flow of the effective couplings $J_0$ and $V_0$ on
the plane $V_1=0$ for fixed $r$.
[Here $V_1$ measures the strength of potential scattering experienced by
electrons in the Wilson shell $f_1$; see Eq.~(\ref{symm_delH}).
Note that asymmetric strong coupling corresponds to $|V_1|=\infty$.
Out-of-plane flows towards this fixed point are represented by broken lines.]
As $r$ increases from zero, the intermediate-coupling fixed point moves
steadily to the right along the horizontal axis.
At $r=r^{\star}$, the particle-hole-asymmetric fixed point separates from
that for $V_0=0$.
The two fixed points grow further apart as $r$ rises towards
$r_{\text{max}}=\frac{1}{2}$, at which point the symmetric fixed point
merges into the strong-coupling limit.
Beyond $r_{\text{max}}$, the asymmetric fixed point remains at finite
couplings.
We believe that it continues to move upward and to the right with increasing
$r$.

\subsubsection{Local impurity susceptibility}

In a recent paper,\cite{Chen} Chen and Jayaprakash studied a Kondo
impurity with a pure power-law exchange coupling under conditions of strict
particle-hole symmetry.
An interesting feature of this work was a comparison between the impurity
contribution to the total magnetic susceptibility [the quantity
$\chi_{\text{imp}}$ defined in Eq.~(\ref{chi_imp}) above]
and a local susceptibility $\chi_{\text{loc}}$ which directly probes
the magnetic properties of the impurity.
It was argued that the behavior of $\chi_{\text{loc}}$ provides evidence for
the existence of a finite critical coupling $J_c$ for $r>\frac{1}{2}$ ---
even though the thermodynamics provide no sign of such a critical coupling ---
and that the low-temperature physics is fundamentally different for $J_0<J_c$
and $J_0>J_c$.
In this subsection we re-examine this issue, and conclude that there is no
such critical point for $r>\frac{1}{2}$, and that the local properties and
total properties are perfectly consistent with one another.

The local susceptibility $\chi_{\text{loc}}$, denoted
$\langle\langle s_z;s_z\rangle\rangle$ in Ref.~\onlinecite{Chen}, is a
zero-frequency response function defined by the relation
\begin{equation}
\frac{\chi_{\text{loc}}}{g\mu_B} = \left.
-\frac{\partial\langle s_z\rangle}{\partial h} \right|_{h = 0}
 = \lim_{h\rightarrow 0} -\frac{\langle s_z\rangle}{h},
                                                        \label{chi_loc}
\end{equation}
where $s_z$ is the $z$-component of the impurity spin; $h$ is a magnetic
field which acts only on the impurity and which enters the Kondo Hamiltonian
through an additional term $\Delta{\cal H}_K = g\mu_B h s_z$.

\begin{figure}[t]
\centerline{
\vbox{\epsfxsize=75mm \epsfbox{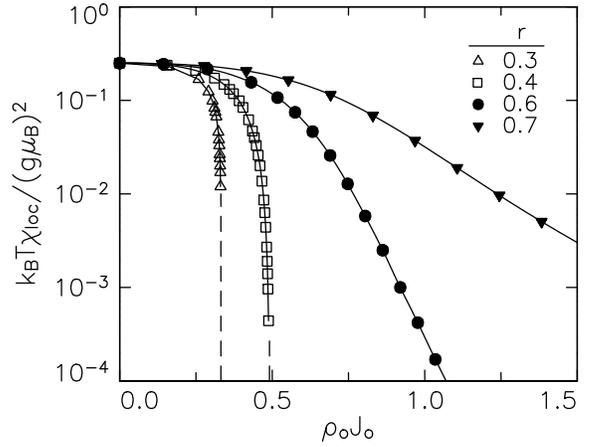}}
}
\vspace{2ex}
\caption{
Low-temperature limit of the local impurity susceptibility
$T \chi_{\text{loc}}$ vs~the dimensionless Kondo coupling~$\rho_0 J_0$
in the exactly screened Kondo model, calculated for different
values of the exponent~$r$ describing the power-law exchange.
The data points were computed using a discretization parameter $\Lambda=9$.
Solid lines are provided as a guide to the eye, and vertical dashed lines
indicate the critical couplings for the cases $r=0.3$ and $r=0.4$.
}
\label{fig:chiloc}
\end{figure}

\begin{figure}[t]
\centerline{
\vbox{\epsfxsize=75mm \epsfbox{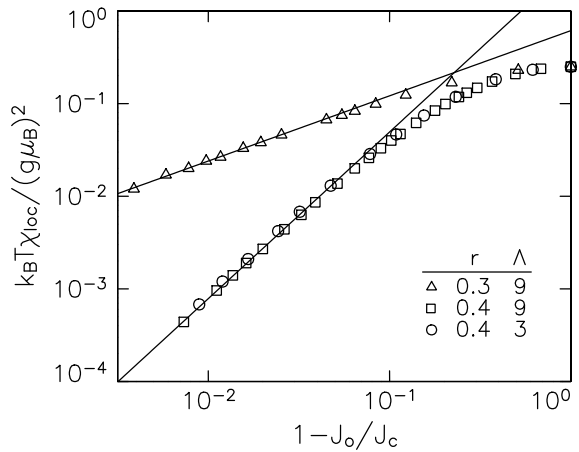}}
}
\vspace{2ex}
\caption{
Same as Fig.~\protect\ref{fig:chiloc}, restricted to $r<\frac{1}{2}$
and plotted on a log-log scale.
Data are shown for two different discretization parameters,
$\Lambda=3$ and $9$.
Straight lines represent fits to those $\Lambda=9$ data points having an
exchange coupling $J_0$ within 3\% of the critical value $J_c$.
}
\label{fig:logchiloc}
\end{figure}

In the standard case $r=0$, $\chi_{\text{loc}}$ closely tracks
$\chi_{\text{imp}}$ as a function of temperature.\cite{Chen:92}
Chen and Jayaprakash's results for $r>0$ can be summarized as follows:

(i) For $r<\frac{1}{2}$ and $J_0<J_c(r)$,
$\lim_{T\rightarrow 0}T\chi_{\text{imp}}=1/4$ while $T\chi_{\text{loc}}$
approaches a smaller, but still nonzero, value.

(ii) For $r<\frac{1}{2}$ and $J_0>J_c(r)$, $T\chi_{\text{imp}}$ heads
to the value $r/8$, whereas $\lim_{T\rightarrow 0}T\chi_{\text{loc}}=0$.

(iii) For $r>\frac{1}{2}$ there exists a finite critical value,
\mbox{$\rho_0 J_c={\cal O}(r)$}, such that the behavior for $J_0<J_c$
is the same as in~(i).

(iv) For $r>\frac{1}{2}$ and $J_0>J_c$, $T\chi_{\text{imp}}$ approaches $1/4$
but $\lim_{T\rightarrow 0}T\chi_{\text{loc}}=0$.

The authors of Ref.~\onlinecite{Chen} interpreted result (iv) as indicating
that the impurity spin is locally quenched even though the total magnetic
susceptibility shows no Kondo effect.
This implies the existence of a third low-temperature regime, in addition to
those governed by the local-moment fixed point [reached in cases (i) and (iii)]
and the symmetric strong-coupling fixed point [the ground state for (ii)].

Since the total thermodynamic properties for $r>\frac{1}{2}$ seem to indicate
flow to the local-moment fixed point for all $J_0$, we have systematically
examined the zero-temperature limit of $T\chi_{\text{loc}}$.
Within the numerical~RG framework the right-hand expression
in Eq.~(\ref{chi_loc}) can be evaluated for a small but finite value of $h$.
(This is the same method as employed by Chen and Jayaprakash.
It should be noted that for $h\not= 0$, the total spin $S$ is no
longer a good quantum number.)
Equation~(\ref{chi_loc}) implies that a finite limiting value of
$T\chi_{\text{loc}}$ reveals itself in a $T^{-1}$ variation of
$\langle s_z\rangle$.
The field $h$ must be chosen sufficiently small that any such $T^{-1}$
regime can be detected before $\langle s_z\rangle$ saturates, as it must
ultimately do since $|s_z|\le\frac{1}{2}$.
Our calculations were performed for $10^{-13} \le g\mu_B h/D \le 10^{-7}$.
As a check, it was established that the results were insensitive to the
precise field used.
In order to reduce the computer time required for these runs, we used
$\Lambda=9$ with $E_c=40$.
As mentioned in Section~\ref{subsec:thermo_num}, it is possible to obtain
accurate zero-temperature results even with such a large discretization
parameter.
We have verified that reducing $\Lambda$ may produce small shifts in the
critical coupling $J_c$, but does not change our essential conclusions
(presented below).

Figure~\ref{fig:chiloc} plots the low-temperature limit of $T\chi_{\text{loc}}$
versus the exchange coupling $\rho_0 J_0$ for four values of~$r$.
Consider first the data for $r=0.3$ and $0.4$, which indicate that
$\lim_{T\rightarrow 0}T\chi_{\text{loc}}$ is finite for
all $J_0<J_c$, and vanishes for $J_0>J_c$.
Here, the critical coupling $J_c$ coincides with that deduced from the
thermodynamic properties or the low-temperature many-body eigenspectrum:
\mbox{$\rho_0 J_c(r=0.3) = 0.343$} and \mbox{$\rho_0 J_c(0.4) = 0.491$}
(both to three significant figures).
As shown in Fig.~\ref{fig:logchiloc}, the form of the curves for $J_0$ just
below $J_c$ is well-described by a power law,
\mbox{$\lim_{T\rightarrow 0}T\chi_{\text{loc}}\propto(J_c-J)^{\nu}$}, with
\mbox{$\nu(r=0.3)=0.70\pm 0.05$} and \mbox{$\nu(0.4)=1.80\pm 0.08$}.
(Figure~\ref{fig:logchiloc} also presents data for $r=0.4$ computed using a
discretization parameter $\Lambda=3$.
The resulting exponent, \mbox{$\nu=1.81\pm 0.05$}, is in close agreement with
that obtained using $\Lambda=9$.)
On the strong-coupling side of the critical point (not shown) it is found that
$\lim_{T\rightarrow 0} \chi_{\text{local}}\propto (J_0-J_c)^{-\lambda}$,
where $\lambda(r=0.3)=4.3\pm 0.1$ and $\lambda(0.4)=3.9\pm 0.1$.

By contrast, the curves in Fig.~\ref{fig:chiloc} representing
\mbox{$r=0.6$} and~$0.7$ offer no hint of critical behavior within a
range of $\rho_0 J_0$ extending well beyond~$r$.
For large~$J_0$, $\lim_{T\rightarrow 0}T\chi_{\text{loc}}$ falls off in
roughly exponential fashion.
It appears probable that for all $J_0$, $T\chi_{\text{loc}}$ heads to a
nonzero value, describing incomplete quenching of the impurity spin.

These observations are consistent with our previous
conclusion,\cite{Ingersent:NRG} based on the finite-size spectrum and the
computed thermodynamic properties, that the low-temperature behavior of the
particle-hole-symmetric Kondo model for $r>\frac{1}{2}$ is described quite
straightforwardly by the weak-coupling (local-moment) limit.
For $r<\frac{1}{2}$, the approach of $J_0$ to its critical value from below
is signaled by the vanishing of $T\chi_{\text{loc}}$, while the divergence of
$\chi_{\text{loc}}$ marks the approach from above.
The behavior of $T\chi_{\text{loc}}$ makes it a candidate order parameter for
describing the $J_c$ critical point.
Further investigation of this possibility is under way.\cite{KI&QS:unpub}

\subsubsection{Relationship between the Kondo and Anderson models}
\label{subsubsec:KandA}

The mapping between the Anderson and Kondo models via the Schrieffer-Wolff
transformation [Eqs.~(\ref{SW})] can be formally justified only for
$-\epsilon_d, U\gg \Gamma$ and $0\le \rho_0 J_0, |\rho_0 V_0| \ll 1$.
In these limits, the physical properties of the Kondo model must be
equivalent to those presented in Section~\ref{subsec:Ander_res} provided that
one equates the half-bandwidth $D$(Kondo) with $T_F$(Anderson), the
temperature at which charge fluctuations freeze out.
For the standard case $r=0$, the properties of the Kondo model are universal
for all $J_0>0$ and can be fully explored from within the Anderson model.
The relation between the two models becomes nontrivial, however, in the
presence of a power-law scattering rate, which introduces a new energy
scale $J_c$ into the problem.

\begin{figure}[t]
\centerline{
\vbox{\epsfxsize=75mm \epsfbox{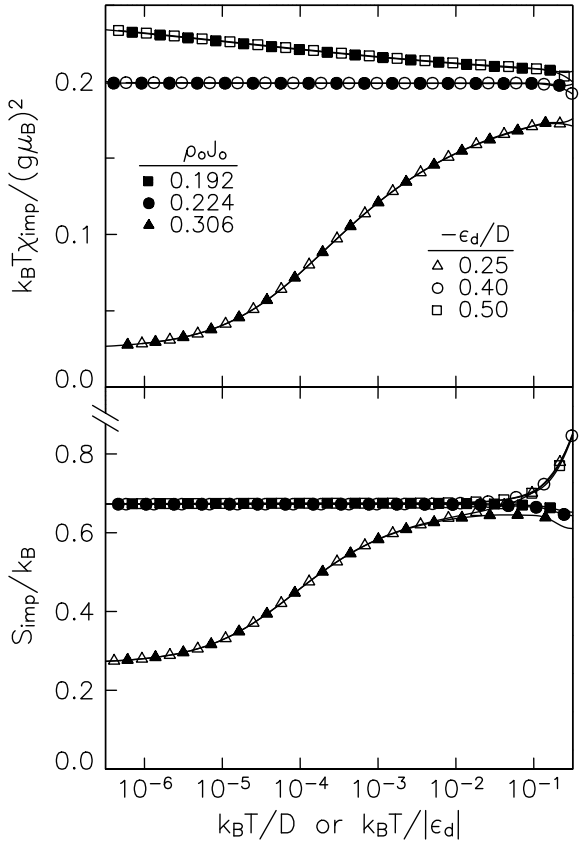}}
}
\vspace{2ex}
\caption{
Impurity susceptibility $T \chi_{\text{imp}}$ and entropy $S_{\text{imp}}/k_B$
for $r=0.2$, plotted vs~either $T/D$ (screened Kondo model with $V_0=0$, solid
symbols) or $T/|\epsilon_d|$ (symmetric Anderson model with $\Gamma_0 = 0.1D$,
open symbols).
In four of the six cases shown, the entropy rapidly converges with
decreasing temperature to the value $k_B\ln 2$.
}
\label{fig:r=0.2_KandA}
\end{figure}

For small, positive values of $r$, Eqs.~(\ref{SW}) can yield values of $J_0$
greater than $J_c$, as well as values less than $J_c$.
One therefore expects all behaviors of the Kondo model to be reproduced by
the Anderson Hamiltonian.
This is confirmed by Fig.~\ref{fig:r=0.2_KandA}, which superimposes
thermodynamic properties computed for the particle-hole-symmetric Kondo and
Anderson models with $r=0.2$.
The figure shows flows to weak coupling and to symmetric strong coupling,
as well as the critical behavior associated with the intermediate-coupling
fixed point.
Each Anderson curve --- computed for one of the parameter sets from
Fig.~\ref{fig:r=0.2_symm}, and plotted versus $T/|\epsilon_d|$ because
$T_F\approx|\epsilon_d|$ in these examples --- is reproduced almost
exactly by the Kondo model with a suitable choice of $J_0$.

For $r\agt 0.5$, by contrast, it is impossible to attain values of the
exchange $J_0>J_c$ under the Schrieffer-Wolff transformation, and the
strong-coupling behavior of the Kondo model (e.g., the
particle-hole-asymmetric thermodynamic properties in Figs.~2 and~3 of
Ref.~\onlinecite{Ingersent:NRG}) cannot be reproduced.
This does not mean that strong coupling is unattainable within the Anderson
model; only that the route to strong coupling via the local-moment regime is
blocked.
A direct crossover from valence fluctuation to asymmetric strong coupling (or,
equivalently, to the frozen-impurity regime) takes place for all $r>0$ whenever
the impurity level is placed above, or only slightly below, the Fermi energy.

We now turn to a comparison between the intermediate-coupling fixed points of
the Kondo and Anderson models.
A number of pieces of evidence point to the complete equivalence of these
fixed points:
(1)~The thermodynamic properties coincide to within the accuracy of
our calculations.
Figure~\ref{fig:r=0.2_KandA} illustrates this explicitly for
$r=0.2$, while Figs.~\ref{fig:r=1} and~\ref{fig:Jc_props} show that
for $r=1$, $T\chi_{\text{imp}}\approx 0.164$ in both models.
Similar agreement is found for other values of $r$, as well.
(2)~An extensive comparison of the energies and quantum numbers of the
many-body eigenstates indicates that the low-energy spectra at the Kondo
and Anderson fixed points are identical.
In each model, there is a range of exponents, $r^{\star}<r<r_{\text{max}}$,
over which there are two distinct intermediate-coupling fixed points
(see Section~\ref{subsubsec:intermediate}).
Within this range, the spectra at the particle-hole symmetric fixed points of
the two models match, as do the levels at the nonsymmetric fixed points.
(3)~It is shown in the next paragraphs that for small $r$, the critical
coupling $\Gamma_c$ of the Anderson model corresponds to an effective exchange
coupling very close to the value of~$J_c$ measured directly in the Kondo model.
(We argue below that the positions of the two fixed points are not expected
to coincide for larger values of~$r$.)

\begin{figure}[t]
\centerline{
\vbox{\epsfxsize=75mm \epsfbox{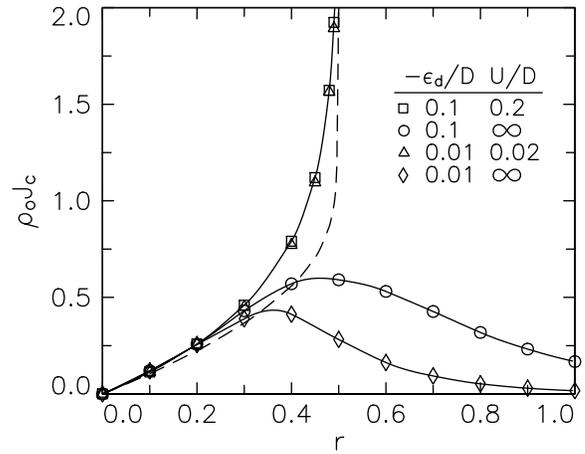}}
}
\vspace{2ex}
\caption{
Position of the intermediate-coupling fixed point $\rho_0 J_c$ (defined in
the text) vs~$r$ for the Anderson model with a pure power-law scattering rate.
Solid lines are provided as a guide to the eye.
The dashed line shows the corresponding quantity computed for the
particle-hole-symmetric Kondo model (data from
Ref.~\protect\onlinecite{Ingersent:NRG}).
}
\label{fig:Jc_A}
\end{figure}

The position of the two intermediate-coupling fixed points can be compared
using the scaling theory discussed in Section~\ref{subsubsec:scaling}.
Equations~(\ref{Gamma(T)})--(\ref{T_F}) and~(\ref{J_SW})
may be combined to convert the critical value of the bare scattering rate,
$\Gamma_c$ plotted in Fig.~\ref{fig:Gamma_c}, into the dimensionless exchange
$\rho_0 J_c$ on entry to the local-moment regime.
Figure~\ref{fig:Jc_A} shows that this transformation collapses the two
data sets for $U=-2\epsilon_d$ onto a single curve.
A dashed line shows the critical exchange computed directly within the
particle-hole-symmetric Kondo model.\cite{Ingersent:NRG,rescaling}
The critical couplings of the two models are in close agreement for small $r$.
The significant differences that develop for larger $r$, in a range where
$\Gamma_c$ becomes of order $|\epsilon_d|$, may be attributed to a
breakdown of the Schrieffer-Wolff transformation as charge fluctuations become
important.
(In Ref.~\onlinecite{Bulla} a similar comparison is made between the
effective values of $\Gamma_c$ in the Anderson and Kondo models.)

The infinite-$U$ curves plotted in Fig.~\ref{fig:Jc_A} lie close to each other,
and to the data for a symmetric impurity, only for $r\alt 0.25$.
For larger $r$, $J_c(\Gamma_c)$ turns sharply downward, a feature not seen in
the particle-hole-asymmetric Kondo model (see Fig.~1 of
Ref.~\onlinecite{Ingersent:NRG}).
It is clear from Fig.~\ref{fig:Gamma_c} that the position of the
critical point is much better described by a relation of the form
$\Gamma_c\propto r|\epsilon_d|$ than by $\rho_0 J_c \propto r$.
Again, charge fluctuations on the impurity site account for these differences.
The Kondo intermediate-coupling fixed point is always approached from within
the local-moment regime.
For large $r$, however, the $\Gamma_c$ fixed point of the Anderson model is
instead reached directly from the high-temperature regime
(see Fig.~\ref{fig:r=1}, for instance).
Indeed, comparison of $\epsilon_d$ with $\epsilon_{d-}^{\star}$ given by
Eq.~(\ref{epsilon_d*}) shows that the value of $\Gamma_c$
places the system outside the local-moment regime for all $r\agt 0.25$
in the case \mbox{$\epsilon_d=-0.1D$}, and for all $r\agt 0.15$ in the case
\mbox{$\epsilon_d=-0.01D$}.
We conclude, therefore, that in all situations where the mapping of the
Anderson model onto the Kondo model is justified, the position of the
intermediate fixed point determined in the two problems is in good agreement.

In summary, the RG~fixed points of the Kondo model appear to form a true
subset of those of the Anderson model.
For $r\agt 0.5$, though, certain paths between these fixed points that
can be followed in the pure-spin problem cannot be realized once charge
fluctuations are allowed.
In this sense, the Kondo model with power-law scattering has an existence
independent of the Anderson model.

\subsection{Underscreened Kondo model}
\label{subsec:underscreened}

In this section we present results for the Kondo Hamiltonian describing the
interaction of a spin-one impurity with a single band of electrons.
We focus on the differences between this model and the conventional
(exactly screened) problem addressed in the previous section.

The weak-coupling limit of the $s=1$ model shares many features with
the conventional case.
In particular, the fixed-point at $J_0=0$ is marginally unstable for
$r=0$ but (according to the analysis in Section~\ref{subsec:local}) it is
stable for all $r>0$.
For small positive values of $r$, one can apply poor-man's scaling to
demonstrate \cite{Ingersent:unpub} the existence of an intermediate-coupling
fixed point at $\rho_0 J_c \approx r$.

At antiferromagnetic strong coupling, the $s=1$ model behaves very differently
from the $s=\frac{1}{2}$ problem.
The larger impurity spin is ``underscreened'' and retains a net spin
$\tau=\frac{1}{2}$.
For $r=0$ this limit is known to be marginally stable,\cite{Nozieres} but
(as shown in Section~\ref{subsec:symm_strong}) for any $r>0$ the residual
impurity spin destabilizes the symmetric strong-coupling fixed point.

Given the instability of the fixed points at $J_0=J_c$ and $J_0=\infty$, one
might expect any $J_0>J_c$ to produce flow to asymmetric strong coupling.
At particle-hole symmetry this option is ruled out, however, suggesting
the existence of a stable fixed point at some exchange coupling that we shall
denote $J^{\star}$, where $J_c<J^{\star}<\infty$.
Our numerical~RG calculations support this conjecture, at least for small $r$.

\begin{figure}[t]
\centerline{
\vbox{\epsfxsize=75mm \epsfbox{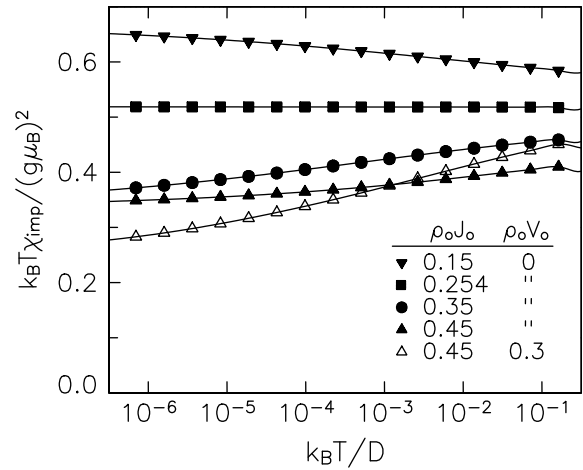}}
}
\vspace{2ex}
\caption{
Impurity susceptibility $T \chi_{\text{imp}}$ vs temperature~$T$ for the
$s=1$ Kondo model with pure power-law scattering specified by $r=0.2$.
}
\label{fig:r=0.2_u}
\end{figure}

\begin{figure}[t]
\centerline{
\vbox{\epsfxsize=75mm \epsfbox{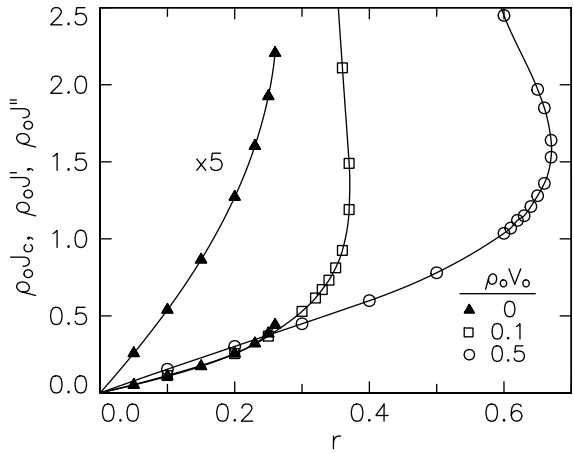}}
}
\vspace{2ex}
\caption{
Critical couplings vs~$r$ for the $s=1$ Kondo model with pure power-law
scattering.
Solid lines are provided as a guide to the eye.
The particle-hole-symmetric fixed-point coupling~$J_c$ (filled symbols) is
plotted both to scale and magnified~$\times 5$.
For nonzero potential scattering (open symbols) there can be zero, one, or
two critical couplings ($J' < J''$), depending on the value of $r$.
}
\label{fig:Jc_u}
\end{figure}

\begin{figure}[t]
\centerline{
\vbox{\epsfxsize=75mm \epsfbox{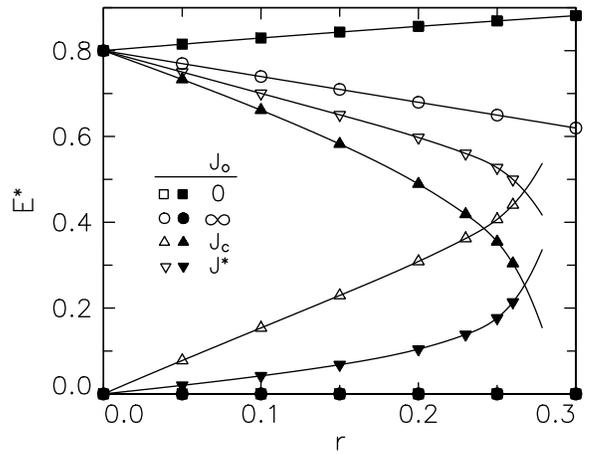}}
}
\vspace{2ex}
\caption{
Lowest eigenvalues $E^{\star}$ of the discretized $s=1$ Kondo model
($\Lambda=3$) plotted vs~$r$.
Data are shown for the four fixed points of the particle-hole-symmetric
problem, both for $N$~even (open symbols) and for $N$~odd (filled symbols).
Solid lines are provided as a guide to the eye.
The $J_c$ and $J^{\star}$ curves are extrapolated at $r=0$ to the weak-coupling
and strong-coupling values, respectively.
}
\label{fig:Jc_u_levels}
\end{figure}

\begin{figure}[t]
\centerline{
\vbox{\epsfxsize=75mm \epsfbox{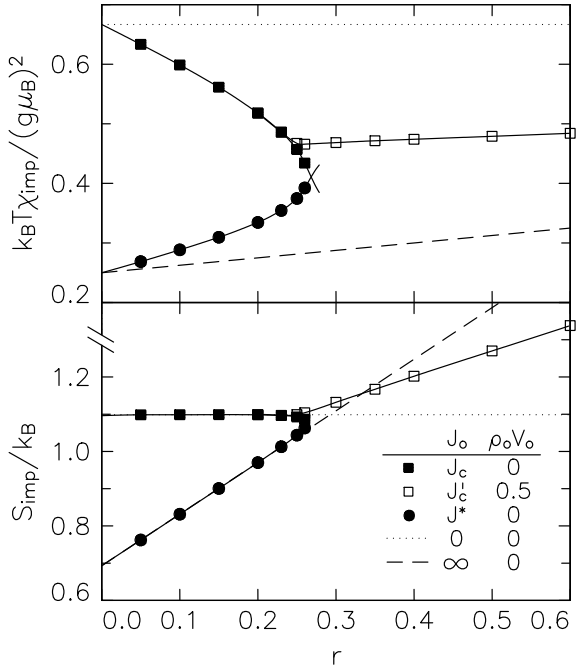}}
}
\vspace{2ex}
\caption{
Impurity susceptibility $T \chi_{\text{imp}}$ and entropy $S_{\text{imp}}$
at the two intermediate-coupling fixed points of the $s=1$ Kondo model,
plotted vs~$r$.
Solid lines are provided as a guide to the eye.
The $J_c$ and $J^{\star}$ curves are extrapolated at $r=0$ to the values for
weak coupling [given by Eq.~(\protect\ref{LM_u_props})] and strong coupling
[Eq.~(\protect\ref{SSC_u_props})], respectively.
}
\label{fig:Jc_u_props}
\end{figure}

\begin{figure}[!t]
\centerline{
\vbox{\epsfxsize=75mm \epsfbox{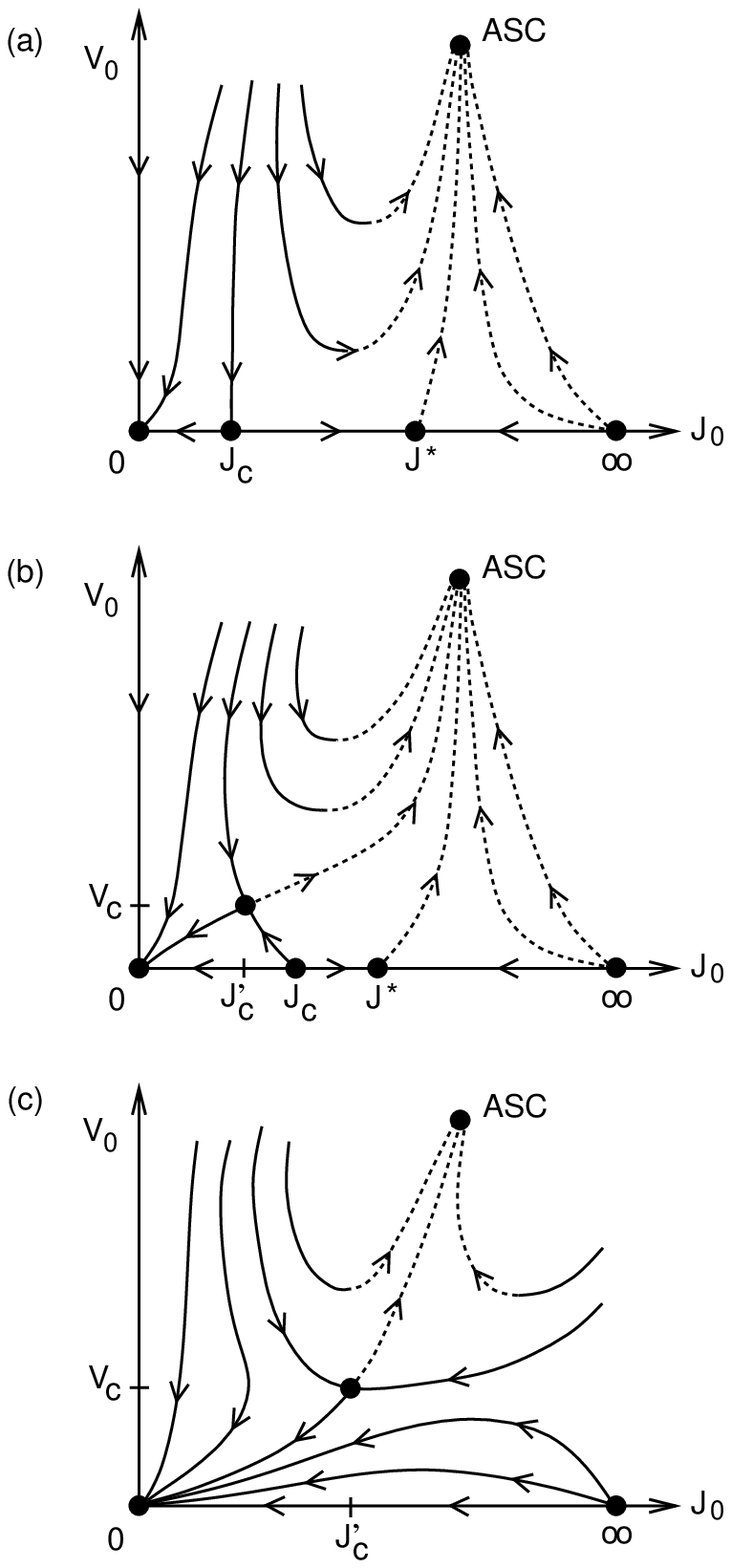}}
}
\vspace{2ex}
\caption{
Schematic renormalization-group flow diagrams for the $s=1$ Kondo model,
showing the $J_0$--$V_0$ plane for $V_1=0$ and fixed~$r$:
\mbox{(a) $0 < r < r^{\star}$;}
\mbox{(b) $r^{\star} < r < r_{\text{max}}$;}
\mbox{(c) $r > r_{\text{max}}$.}
See Fig.~\protect\ref{fig:flows} for an explanation of the symbols.
}
\label{fig:u_flows}
\end{figure}

\begin{figure}[t]
\centerline{
\vbox{\epsfxsize=75mm \epsfbox{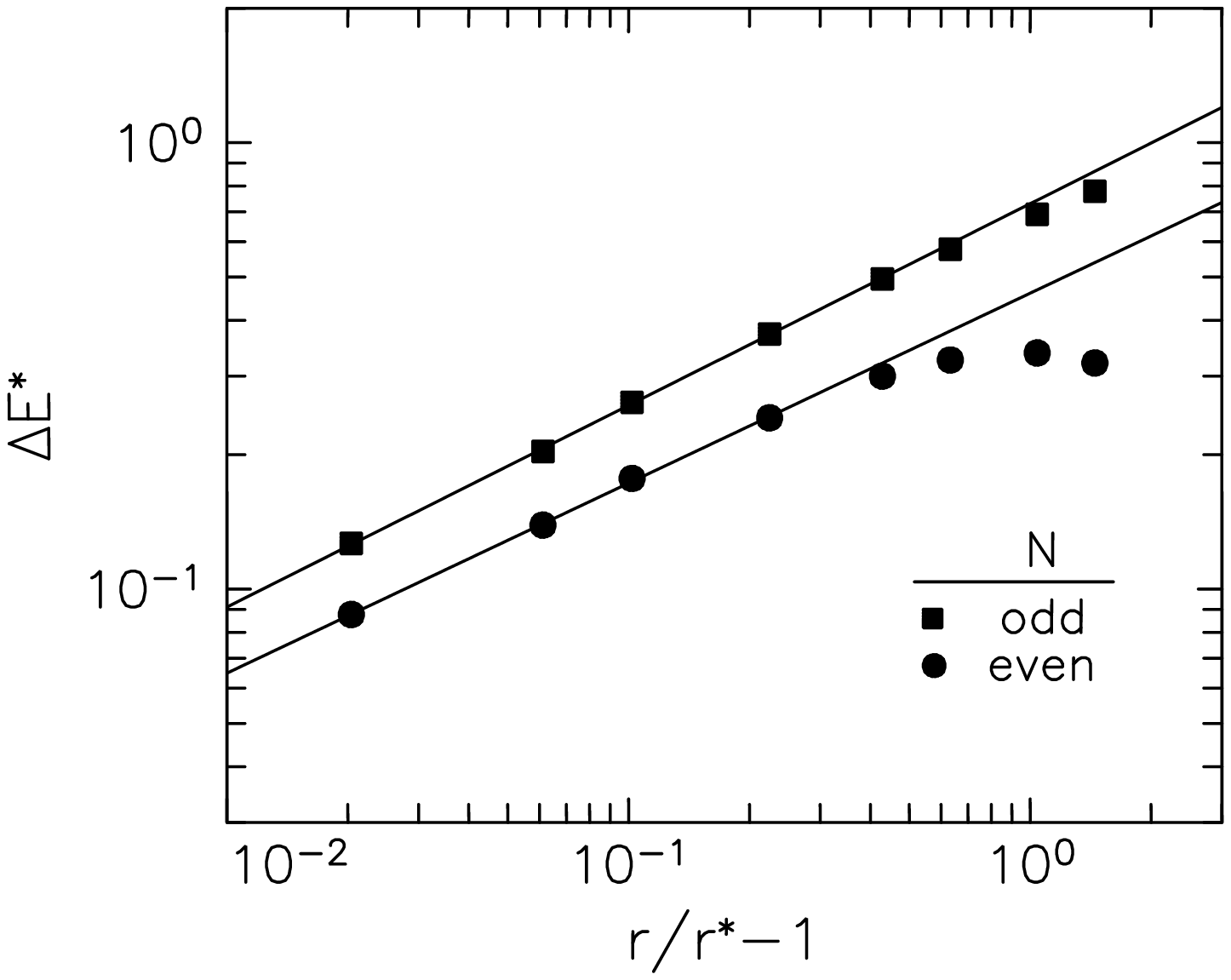}}
}
\vspace{2ex}
\caption{
Energy splittings $\Delta E^*$ of the lowest pair of charge-conjugate
eigenstates at the particle-hole-asymmetric intermediate-coupling fixed point
of the $s=1$ Kondo model, plotted vs~the reduced exponent
$r/r^{\star}-1$, for $\Lambda = 3$ and $r^{\star}=0.245$.
Straight lines show fits to the leftmost four data points for odd- and
even-numbered iterations of the numerical~RG method.
}
\label{fig:Jc_u_split}
\end{figure}

We begin by considering the particle-hole symmetric case, $V_0=0$.
Figure~\ref{fig:r=0.2_u} shows the impurity susceptibility computed for
$r=0.2$.
For all values of the exchange coupling $J_0$, $T\chi_{\text{imp}}$ varies
rather slowly with temperature and heads to a nonzero $T=0$ limit.
However, closer inspection reveals three qualitatively distinct cases.
Tuning $\rho_0 J_0$ to $0.254$ yields a flat curve reminiscent of
the (unstable) $J_c$ fixed point of the exactly screened model
(see Fig.~\ref{fig:r=0.2_KandA}), whereas any smaller exchange produces a
monotonic rise in $T\chi_{\text{imp}}$ towards the weak-coupling value of
$2/3$.
For $\rho_0 J_0 > 0.254$, $T\chi_{\text{imp}}$ falls towards a low-temperature
limit of approximately~$0.33$, significantly higher than the strong-coupling
value, $(2+r_1)/8=0.275$.
We interpret this as one piece of evidence for the $J^{\star}$ fixed point
alluded to in the previous paragraph.

Behavior qualitatively similar to that shown in Fig.~\ref{fig:r=0.2_u}
is found for all \mbox{$0<r < r_{\text{max}}$}, where
\mbox{$0.26 < r_{\text{max}} < 0.27$}.
For all $r>r_{\text{max}}$, by contrast, the thermodynamics contain no
signature of any intermediate-coupling fixed point.
Instead, the system flows to weak coupling ($T\chi_{\text{imp}}=2/3$) for all
values of the bare exchange $J_0$.

Some of the systematic trends with increasing $r$ are shown in
Figs.~\ref{fig:Jc_u}--\ref{fig:Jc_u_props}.
Figure~\ref{fig:Jc_u} plots (filled symbols) the position $\rho_0 J_c$ of the
unstable intermediate-coupling fixed point as a function of $r$.
For small $r$, $\rho_0 J_c\approx r$, as expected from poor-man's scaling,
but the curve turns upward and then abruptly terminates at $r=r_{\text{max}}$.

The critical coupling $J^{\star}$ corresponding to the second
intermediate-coupling fixed point cannot be determined directly
because the low-temperature behavior does not undergo any qualitative change
as~$J_0$ passes through~$J^{\star}$.
However, one can estimate the value of~$J^{\star}$ by examining the low-energy
many-body spectrum.
The eigenvalue of the first excited state at each of the four fixed points of
the symmetric underscreened problem ($J_0=0$, $J_c$, $J^{\star}$ and~$\infty$)
is plotted versus~$r$ in Fig.~\ref{fig:Jc_u_levels}.
For small~$r$, the deviation of the $J_c$~levels from the weak-coupling
energies is linear in~$r$, consistent with the relation $\rho_0 J_c\approx r$.
It appears that $J^{\star}$~is infinite at $r=0$ (at which point the
extrapolated levels for the stable intermediate fixed point coincide with the
strong-coupling values), and falls progressively as $r$~increases.
The energies at the two intermediate fixed points are projected to cross at a
value of~$r$ between~0.26 and~0.27 --- precisely the range in which these fixed
points disappear.

Figure~\ref{fig:Jc_u_props} shows the impurity susceptibility and entropy
at the two intermediate-coupling fixed points.
For each property, the $J_c$ and $J^{\star}$ curves deviate with increasing
$r$ from the weak- and strong-coupling limits, respectively.
Just as for the eigenenergies, the two sets of curves appear to
cross in the range $0.26<r<0.27$.
Both $T\chi_{\text{imp}}$ and $S_{\text{imp}}$ vary linearly
for small $r$.
Empirically, the susceptibilities for $r\alt 0.15$ are well-described
by the formulae
\begin{equation}
T\chi_{\text{imp}}(J_c) = \frac{2}{3} - 0.7 r, \quad
T\chi_{\text{imp}}(J^{\star}) = \frac{1}{4} + 0.4 r,
\end{equation}
while the entropies for $r\alt 0.2$ have an excellent fit to
the weak- and strong-coupling expressions:
\begin{equation}
S_{\text{imp}}(J_c) = \ln 3, \quad
S_{\text{imp}}(J^{\star}) = \ln 2 + r \ln 4.
\end{equation}
We have been unable to reproduce these properties by constructing suitable
phase shifts for noninteracting electrons.
We speculate that these intermediate-coupling fixed points are truly
non-Fermi-liquid in character.

Let us summarize the situation at particle-hole symmetry.
For $0<r<r_{\text{max}}$ there are two intermediate-coupling fixed points ---
one stable ($J^{\star}$), the other unstable ($J_c$).
These fixed points merge at $r=r_{\text{max}}$, above which value they both
disappear, leaving weak coupling as the only stable fixed point.
(We remind the reader that in the exactly screened Kondo model, the
disappearance of the $J_c$ fixed point can be tied directly to the value
$r_{\text{max}}=\frac{1}{2}$ at which the symmetric strong-coupling fixed point
becomes unstable.
We have no such argument to fix the precise value of $r_{\text{max}}$ in the
underscreened model.)

The introduction of potential scattering considerably modifies the picture
presented above.
As shown in Section~\ref{subsec:local}, this perturbation is irrelevant in the
weak-coupling regime.
However, our numerics indicate that it destabilizes the $J^{\star}$ fixed point
--- found in the symmetric problem for all $0<r<r_{\text{max}}$ --- towards
asymmetric strong coupling.
The effect on the $J_c$ fixed point is more subtle, as will be explained in
the paragraphs that follow.
There are several parallels with (but also clear differences from) the behavior
of an exactly screened Kondo impurity described in
Section~\ref{subsubsec:intermediate}.

We first consider values of $r$ less than $r_{\text{max}}$.
For each $r$ in this range and for any $V_0\not= 0$ one can find a critical
coupling $J'(r,V_0)$ such that any $J_0 < J'$ yields weak-coupling behavior
while any $J_0 > J'$ drives the system to asymmetric strong coupling.
(The latter case is exemplified by the curve for $r=0.2$, $\rho_0 J_0=0.45$,
and $\rho_0 V_0 = 0.3$ in Fig.~\ref{fig:r=0.2_u}, which shows
$T\chi_{\text{imp}}$ heading towards its asymmetric strong-coupling value
of $1/4$.)

The behavior for $J_0$ precisely equal to $J'(r,V_0)$ depends on the value of
$r$.
For \mbox{$0<r<r^{\star}\approx 0.245$}, particle-hole asymmetry is irrelevant
on the separatrix, and the system approaches the fixed point located at
\mbox{$V_0 = 0$}, \mbox{$J_0 = J_c\equiv J'(r,0)$}.
The RG~flows for this case are sketched in Fig.~\ref{fig:u_flows}(a).
For $r^{\star}<r<r_{\text{max}}$, by contrast, the flow along the line
\mbox{$J_0 = J'(r,V_0)$} is towards a new intermediate-coupling fixed point at
\mbox{$V_0 = V_c$}, \mbox{$J_0 = J_c'(r) \equiv J'(r,V_c)$}, as shown in
Fig.~\ref{fig:u_flows}(b).
(As was the case for a screened impurity spin, there is actually a pair
of $J_c'$~fixed points at \mbox{$V_0 = \pm V_c$}.
Throughout this section, these two fixed points --- which are related by
particle-hole interchange --- are treated as one, and all properties
discussed will be assumed to depend only on the absolute value of $V_0$.)

For $r>r_{\text{max}}$, there is a range of potential scatterings $|V_0|<V_c$
over which the low-temperature physics is governed by the weak-coupling fixed
point, whatever the bare exchange coupling $J_0$, i.e., no critical exchange
$J'(r,V_0)$ can be found.
For $|V_0| > V_c$, by contrast, there are {\em two\/} critical couplings.
The system flows to weak coupling both for $J_0 < J'(r,V_0)$ and for
$J_0 > J''(r,V_0)$, while asymmetric strong coupling is reached for couplings
which fall between these critical values.
If $J_0$ is tuned precisely to $J'$ or to $J''$, the system flows to
an intermediate-coupling fixed point which we take to be located
at $V_0 = V_c$, $J_0 = J_c'(r)\equiv J'(r,V_c)\equiv J''(r,V_c)$.
The RG~flows that we deduce for this range of $r$ are sketched in
Fig.~\ref{fig:u_flows}(c).

It was shown in Section~\ref{subsec:symm_strong} that there are two equally
relevant operators in the vicinity of symmetric strong coupling
\mbox{$(J_0=\infty,V_0=0)$}.
Figure~\ref{fig:u_flows} illustrates the competition between $O_{J_1}$, which
drives the system towards weak coupling, and $O_{V_1}$, which causes flow
towards asymmetric strong coupling.
For \mbox{$r<r_{\text{max}}$}, the $J_c$ and $J^{\star}$ fixed points
block flow along the axis $V_0 = 0$, allowing potential scattering to
dominate the low-temperature behavior.
For \mbox{$r>r_{\text{max}}$}, by contrast, the sole surviving
intermediate-coupling fixed point at \mbox{$(J'_c, V_c)$} stifles the growth
of particle-hole asymmetry and instead steers the system to weak coupling.

We now present some of the numerical evidence in support of the picture
laid out above.
Figure~\ref{fig:Jc_u} plots $J'$ for fixed $\rho_0 V_0$ and for $0<r\le 0.7$.
For $r > r_{\text{max}}$, the second critical coupling $J''$ is also plotted.
One sees that $J''$ is typically very large (greater than the bandwidth),
implying that the upper bound on the asymmetric strong-coupling regime is
unlikely to be accessible in practice.
We identify the rightmost point on each critical curve (the meeting of the
$J'$ and $J''$ curves) with the $J_c'$ fixed point.
Note that this particular plot yields $r$ and $\rho_0 J_c'$ as functions of
$\rho_0 V_c$.
However, by inverting the procedure to make $r$ the independent variable, one
can deduce that both $J_c'$ and $V_c$ are increasing functions of $r$, at
least over the parameter range shown.
One can deduce, for instance, that \mbox{$\rho_0 V_c(r=0.38)\approx 0.1$}
and \mbox{$\rho_0 V_c(r=0.68)\approx 0.5$}.
These rather large values of $V_c$ suggest that for $r\agt 0.5$, flow
to asymmetric strong coupling (i.e., Kondo-screening of the impurity) can be
achieved only under conditions of strong particle-hole asymmetry.

Figure~\ref{fig:Jc_u_split} shows the energy splitting $\Delta E^{\star}$ of
the lowest pair of charge-conjugate states at the $J_c'$ fixed point for
$0.25 \le r \le 0.6$.
For $r\alt 0.3$, this splitting is well-fit by
\mbox{$\Delta E^{\star} \propto (r-r^{\star})^{\nu^{\star}}$}, with
\mbox{$r^{\star}=0.245\pm 0.002$} and \mbox{$\nu^{\star}=0.44\pm 0.15$}.
For $r<r^{\star}$ the fixed-point spectrum is observed always to be
particle-hole-symmetric.

The impurity susceptibility and entropy at the various intermediate-coupling
fixed points are compared in Fig.~\ref{fig:Jc_u_props}.
Over the rather narrow range $r^{\star}<r<r_{\text{max}}$ in which the
$J_c$ and $J_c'$ fixed points coexist, their properties are seen to diverge
steadily.
For larger $r$, both $T\chi_{\text{imp}}(J_c')$ and $S_{\text{imp}}(J_c')$
rise monotonically.

Finally, we note that all properties of the $J_c'$ fixed point shown in
Figs.~\ref{fig:Jc_u},~\ref{fig:Jc_u_props}, and~\ref{fig:Jc_u_split} appear
to vary smoothly as $r$ passes through $r_{\text{max}}$, even though the
fixed point is defined in a different manner for $r<r_{\text{max}}$ and
$r>r_{\text{max}}$.
This serves as an indication that the fixed-point couplings $(J_c', V_c)$
evolve continuously across the border between the regimes shown in
Figs.~\ref{fig:u_flows}(b) and~(c).

\subsection{Overscreened Kondo Model}
\label{subsec:overscreened}

In this section we present results for the Kondo Hamiltonian describing the
interaction of a spin-one-half impurity with two degenerate bands or channels
of electrons [Eqs.~(\ref{H_c2:def}) and~(\ref{H_s2:def})].
The weak-coupling properties of the two-channel problem are very similar to
those presented in the previous two subsections --- including the existence
for small $r$ of an unstable fixed point at $\rho_0 J_c\approx r$.
We focus on the intermediate-to-strong-coupling regime, where the three
models differ markedly.

At strong coupling, the impurity is ``overscreened'' by the two conduction
bands and retains a net spin of one-half.
For $r=0$, this limit is marginally unstable,\cite{Nozieres} giving rise to
a stable, intermediate-coupling fixed point at $\rho_0 J^{\star}={\cal O}(1)$.
For $r>0$, the symmetric strong-coupling fixed point is outright unstable due
to the residual impurity degree of freedom, so one might again expect flow to
some $J^{\star}>J_c$, just as in the underscreened model.

The existence of two intermediate-coupling fixed points was predicted
previously using an extension of Withoff and Fradkin's poor-man's scaling
analysis \cite{Withoff} to the \mbox{$N_c$-channel} Kondo
problem.\cite{Ingersent:scaling}
The dimensionless Kondo coupling $\rho_0 \bar{J}$ was found to rescale
from its bare value $\rho_0 J_0$ according to the equation
\begin{equation}
\frac{d(\rho_0 \bar{J})}{d\ln T} = r\rho_0 \bar{J}
      - (\rho_0 \bar{J})^2 + c (\rho_0 \bar{J})^3.    \label{scaling3}
\end{equation}
The coefficient $c$ is a complicated function of $r$~and~$N_c$, which
reduces to $c=N_c/2$ in the limits $r\ll 1$ and $N_c\gg 1$.
Then Eq.~(\ref{scaling3}) has fixed points satisfying
$d(\rho_0 \bar{J})/d\ln T=0$ at $\rho_0 \bar{J} = 0$, $\infty$, and
\mbox{$(1\pm\sqrt{1-2N_c r})$}.
For small $r$, the intermediate-coupling fixed points are located at
\mbox{$\rho_0 J_c\approx r$} (unstable) and
\mbox{$\rho_0 J^{\star}\approx 2/N_c-r$} (stable).
However, these two fixed points merge at \mbox{$r=1/2N_c$},
\mbox{$\rho_0 \bar{J}=1/N_c$}.
For \mbox{$r>1/2N_c$} there is no intermediate-coupling fixed point and the
RG~trajectories flow directly from strong coupling to weak coupling.

Although the two-channel case does not strictly satisfy the condition
$N_c\gg 1$, the predictions of scaling theory are well borne out by
numerical~RG calculations.
Moreover, the intermediate fixed points turn out to survive the inclusion
of potential scattering (which was not taken into account in
Ref.~\onlinecite{Ingersent:scaling}).

The addition of a second conduction band greatly increases the size of the
basis of the discretized version of the Kondo model, and hence the
computer time required for a solution of the problem.
We have found it impractical to compute thermodynamic properties using a
discretization parameter $\Lambda=3$ while keeping all states up to an
energy cutoff $E_c \ge 25$, as was done in the single-channel problems.
This prevents reliable determination of the temperature-dependence of the
thermodynamic properties.
However, experience indicates that fixed-point properties can be computed
accurately for values of $\Lambda$ as large as $10$.
Below we present many-body eigenstates computed for $\Lambda=3$, but
thermodynamic properties obtained using $\Lambda=9$ and $E_c = 25$.

\begin{figure}[t]
\centerline{
\vbox{\epsfxsize=75mm \epsfbox{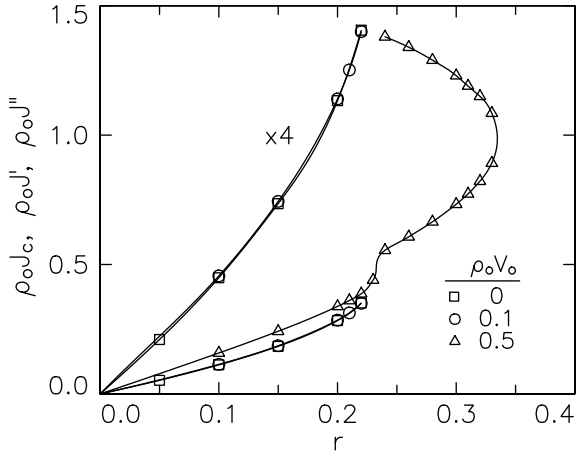}}
}
\vspace{2ex}
\caption{
Critical couplings for the two-channel Kondo model with pure power-law
scattering.
Solid lines are provided as a guide to the eye.
The data for $\rho_0 V_0 = 0$ and $0.1$ are plotted both to scale and
magnified~$\times 4$.
For nonzero potential scattering there can be zero, one, or two critical
couplings ($J' < J''$), depending on the values of $r$ and $V_0$.
}
\label{fig:Jc_o}
\end{figure}

\begin{figure}[t]
\centerline{
\vbox{\epsfxsize=75mm \epsfbox{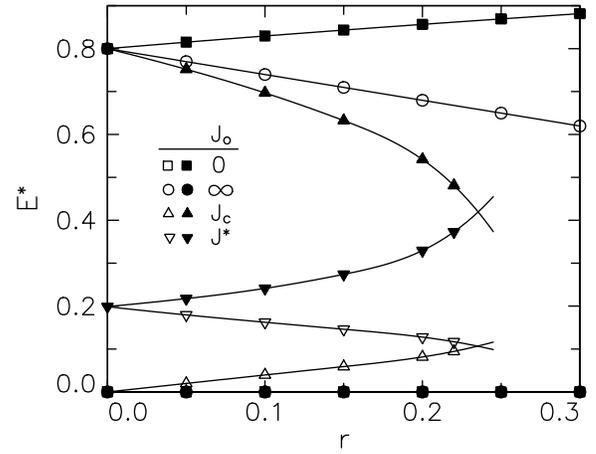}}
}
\vspace{2ex}
\caption{
Lowest eigenvalues $E^{\star}$ of the discretized two-channel Kondo model
($\Lambda=3$) plotted vs~the exponent~$r$ describing the power-law exchange.
Data are shown for the four fixed points of the particle-hole-symmetric
problem, both for $N$~even (open symbols) and for $N$~odd (filled symbols).
Solid lines are provided as a guide to the eye.
The $J_c$ curves are extrapolated at $r=0$ to the weak-coupling values.
}
\label{fig:Jc_o_levels}
\end{figure}

\begin{figure}[t]
\centerline{
\vbox{\epsfxsize=75mm \epsfbox{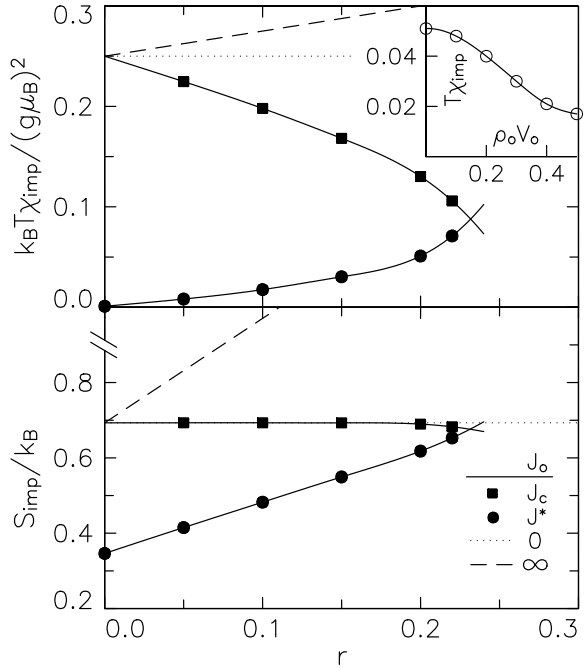}}
}
\vspace{2ex}
\caption{
Impurity susceptibility $T \chi_{\text{imp}}$ and entropy $S_{\text{imp}}$
at the two intermediate-coupling fixed points of the two-channel Kondo model,
plotted vs~$r$.
Solid lines are provided as a guide to the eye.
The $J_c$ curves are extrapolated at $r=0$ to the values for weak coupling,
given by Eq.~(\protect\ref{LM_props}).
The strong-coupling properties [Eq.~(\protect\ref{SSC_o_props})] are
also plotted.
Inset: Impurity susceptibility (in the same units as the main figure)
at the $J^{\star}$ fixed point vs~potential scattering $\rho_0 V_0$, for
$r=0.2$ and $\rho_0 J_0 = 0.6$.
}
\label{fig:Jc_o_props}
\end{figure}

\begin{figure}[!t]
\centerline{
\vbox{\epsfxsize=75mm \epsfbox{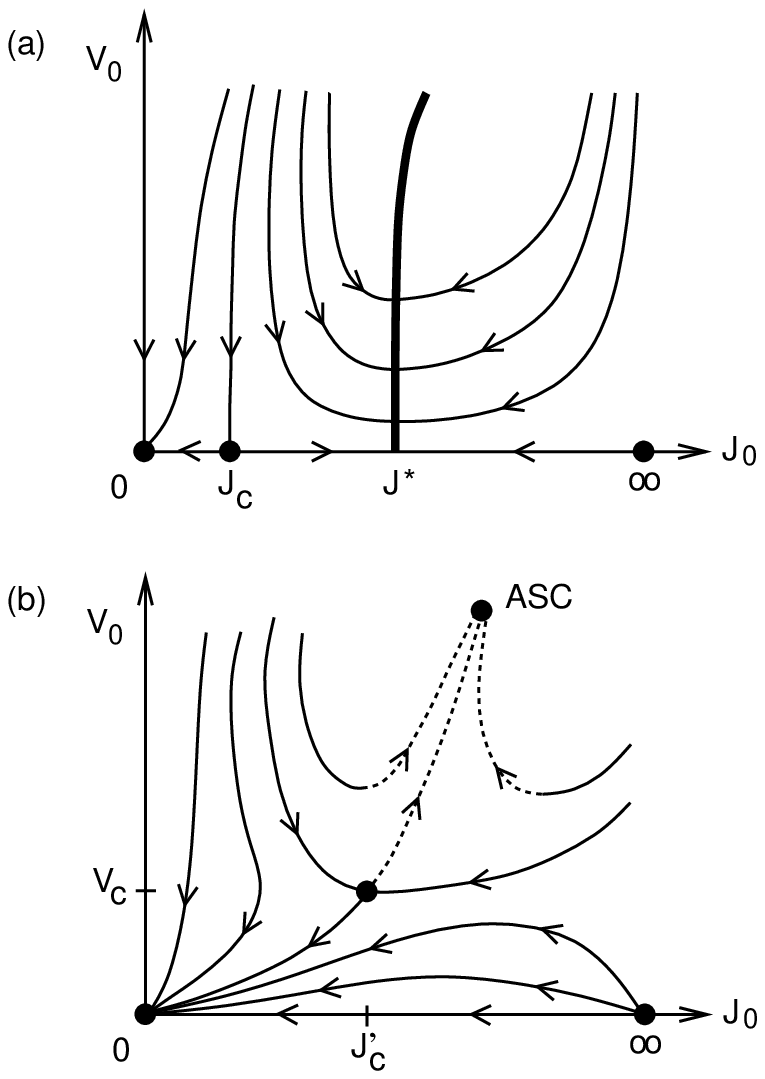}}
}
\vspace{2ex}
\caption{
Schematic renormalization-group flow diagrams for the two-channel Kondo model,
showing the $J_0$--$V_0$ plane for $V_1=0$ and fixed~$r$:
\mbox{(a) $0 < r < r_{\text{max}}$;}
\mbox{(b) $r > r_{\text{max}}$.}
See Fig.~\protect\ref{fig:flows} for an explanation of the symbols.
}
\label{fig:o_flows}
\end{figure}

Figure~\ref{fig:Jc_o} plots the position $\rho_0 J_c$ of the unstable
intermediate-coupling fixed point as a function of $r$.
For small $r$, $\rho_0 J_c\approx r$, as expected from poor-man's scaling.
At particle-hole symmetry, the curve turns upward and then abruptly
terminates at $r=r_{\text{max}}$ where $0.23 < r_{\text{max}} < 0.24$.
In this respect, the underscreened and overscreened problems are
very similar.

Just as for the underscreened problem, one can obtain indirect evidence for
the value of $J^{\star}$ by examining the low-lying many-body spectrum.
The energy of the first excited state at each of the four fixed points of
the symmetric overscreened problem ($J_0=0$, $J_c$, $J^{\star}$ and~$\infty$)
is plotted versus~$r$ in Fig.~\ref{fig:Jc_o_levels}.
The levels at the $J_c$~fixed point progressively diverge from the
weak-coupling energies as $r$~is increased from zero, consistent with the
relation \mbox{$\rho_0 J_c\approx r$} reported in the previous paragraph.
The smooth evolution of the levels at the stable intermediate fixed point
suggests that $J^{\star}$ decreases continuously with increasing~$r$ from
its value for a constant scattering rate, $\rho_0 J^{\star}(r=0)\approx 1$.
The energies at the two intermediate fixed points can be extrapolated to
cross in the range $0.23 < r < 0.24$ where both fixed points disappear.

Figure~\ref{fig:Jc_o_props} shows the impurity contributions to the
susceptibility and the entropy at the two intermediate-coupling fixed points.
For each property, the $J_c$ and $J^{\star}$ curves deviate with increasing
$r$ from their weak-coupling and $r=0$ non-Fermi-liquid values, respectively,
and the two curves can be extrapolated to cross in the range $0.23<r<0.24$.
Both $T\chi_{\text{imp}}$ and $S_{\text{imp}}$ vary linearly with $r$
for $r\alt 0.15$.
Empirically, the susceptibilities are well-described by the formulae
\begin{equation}
T\chi_{\text{imp}}(J_c) = \frac{1}{4} - \frac{r}{2}, \quad
T\chi_{\text{imp}}(J^{\star}) = \frac{r}{6},
\end{equation}
while the entropies fit
\begin{equation}
S_{\text{imp}}(J_c) = \ln 2, \quad
S_{\text{imp}}(J^{\star}) = \frac{1}{2} \ln 2 + r \ln 4.
\end{equation}

Two factors greatly impede the study of the effects of potential scattering in
the overscreened Kondo model:
(1)~Away from particle-hole symmetry, the total axial charge
(see Section~\ref{subsec:iterate}) is no longer a good quantum number.
This change roughly doubles the size of the basis at each iteration
of the numerical~RG procedure.
Even working with a discretization parameter as large as $\Lambda=9$, we have
found it feasible to retain only those many-body eigenstates with
scaled eigenvalues $E^{\star} < E_c\approx 15$ (compared to $E_c = 25$
for the particle-hole-symmetric problem).
(2)~The instability of the two-channel Kondo model with respect to channel
asymmetry \cite{Nozieres} is found to rise markedly with increasing
$r$~and~$|V_0|$.
Over much of the parameter space, unavoidable numerical asymmetry at the level
of the machine precision grows to of order unity before the many-body
energy levels get close to the zero-temperature fixed point of the
channel-symmetric problem.
In light of these obstacles, we focus our remarks on the qualitative
features of the RG~flow diagrams for $r\alt 0.3$ and $|\rho_0 V_0| \alt 0.5$.

For all $0 < r < r_{\text{max}}$, there appears to be a critical
coupling $J'(r,V_0)$ for any potential-scattering strength~$V_0$.
Figure~\ref{fig:Jc_o} plots $J'(r)$ for two fixed values of $V_0$.
For $J_0<J'$, the system flows to weak coupling, while for $J_0 = J'$ it
reaches the $J_c$ fixed point of the particle-hole-symmetric problem.
For $J_0 > J'$, the flow is to a generalization of the $J^{\star}$ fixed point
found for $V_0 = 0$.
Specifically, the energy levels are obtained from those of the
$J^{\star}$ fixed point by splitting each pair of charge-conjugate states.
At fixed $J_0$, this splitting grows with increasing $V_0$;
with $V_0$ held fixed and $J_0$ starting at $J'$, the splitting initially grows
from zero as $J_0$ increases, then passes through a maximum, and eventually
falls back towards zero as $J_0\rightarrow\infty$.
From this behavior, we deduce that the RG~flows have the form shown
in Fig.~\ref{fig:o_flows}(a).
To within the accuracy that we can achieve (around 2\%), the impurity
entropy is the same everywhere along the line of fixed points, but
$T\chi_{\text{imp}}$ falls as one moves away from the symmetric fixed point
(see the inset to Fig.~\ref{fig:Jc_o_props}).

The RG~flows for $r>r_{\text{max}}$ [Fig.~\ref{fig:o_flows}(b)] are
qualitatively similar to those of the underscreened $s=1$ Kondo model.
There exists a single intermediate-coupling fixed point at $J_0 = J_c'(r)$,
$V_0 = V_c(r)$.
For all $|V_0|<V_c$ the system flows to weak coupling, whatever the
value of the bare exchange coupling $J_0$.
For $|V_0| > V_c$, by contrast, the flow is to weak coupling for
$J_0 < J'(r,V_0)$ and for $J_0 > J''(r,V_0)$; otherwise the system flows to
asymmetric strong coupling.
Figure~\ref{fig:Jc_o} plots $J'(r)$ and $J''(r)$ for $\rho_0 V_0 = 0.5$.
The absence of any critical coupling for \mbox{$\rho_0 V_0=0.1$} beyond
\mbox{$r=r_{\text{max}}$} indicates that \mbox{$\rho_0 V_c(r)>0.1$} for all
\mbox{$r_{\text{max}}<r<0.4$}.
This observation and the discontinuity in the slope of the $J'(r)$ curve for
$\rho_0 V_0 = 0.5$ both suggest that the properties of the $J_c'$ fixed point
do not vary smoothly as $r$~passes through~$r_{\text{max}}$.

We note that, unlike the exactly screened and underscreened models,
the two-channel Kondo model does not seem to exhibit any range of
exponents $r^{\star}<r<r_{\text{max}}$ within which particle-hole-symmetric
and asymmetric versions of the $J_c$ fixed point coexist.
Due to the numerical difficulties mentioned above, this possibility cannot
be completely ruled out, but any range of dual fixed points is certainly
very narrow ($r_{\text{max}} - r^{\star} < 0.02$).

\subsection{Departures from a pure power-law scattering rate}

Now we consider various changes to the form of the power-law scattering rate
defined in Eq.~(\ref{Gamma_pure}).
We focus on three features which are likely to be present in real
materials:
(i)~removal of the symmetry $\Gamma(-\epsilon)=\Gamma(\epsilon)$;
(ii)~restriction of the power-law variation in $\Gamma(\epsilon)$ to
a region of halfwidth $\Delta\ll D$;
and (iii)~the existence of a small but nonvanishing scattering rate at
the Fermi energy, $\Gamma(0) \not= 0$.
The effects of these modifications can be predicted qualitatively using
poor-man's scaling, and can be investigated in detail via numerical~RG
calculations.
In the latter approach, each change in the form of $\Gamma(\epsilon)$
simply alters the values of $F$ defined in Eq.~(\ref{F}) and the
tight-binding coefficients $e_n$ and $t_n$ entering the discretized
Hamiltonian, Eq.~(\ref{H_c:disc}); otherwise the numerical treatment remains
the same as for a pure power-law scattering rate.
It turns out that the first modification above is relatively inconsequential,
whereas the second can significantly increase the likelihood of observing the
Kondo effect for values of $r\ge \frac{1}{2}$, and the third can produce even
more fundamental departures from the results obtained using
Eq.~(\ref{Gamma_pure}).

\subsubsection{Particle-hole-asymmetric scattering rate}

It was pointed out in Section~\ref{sec:band} that if the scattering rate
satisfies $\Gamma(-\epsilon)=\Gamma(\epsilon)$ for all $\epsilon$,
then the parameters $e_n$ entering Eq.~(\ref{H_c:disc}) are identically zero.
This symmetry is unlikely to be exactly preserved in any real system.
We have studied the effect of various symmetry-breaking perturbations,
the simplest being the modification of Eq.~(\ref{Gamma_pure}) so that the
conduction band extends over energies
\mbox{$-(1+\mu)D \le \epsilon \le (1-\mu)D$}.
Setting $\mu\not= 0$ changes not only the hopping coefficients $t_n$ entering
the discretized conduction band, but also leads to nonzero diagonal
coefficients $e_n$.
This invalidates the asymptotic expressions for the conduction-band
eigenvalues and eigenvectors presented in Section~\ref{sec:band}, and
vitiates the fixed-point stability analysis of Section~\ref{sec:fixed}.
However, numerical~RG results indicate that the physical behaviors of the
Anderson model remain essentially the same.
The band asymmetry prevents the appearance of the symmetric strong-coupling
fixed point, but the other regimes discussed in Section~\ref{sec:fixed}
exist and moreover exhibit the same power-laws in their thermodynamic
properties.
The effect of band asymmetry on the intermediate-coupling fixed points of the
three variants of the Kondo model is essentially equivalent to that of
potential scattering on the same problem with a symmetric band.

\subsubsection{Restricted power-law scattering rate}

In real gapless systems, the power-law variation of the scattering rate or of
the Kondo exchange is unlikely to extend over the entire band in the manner
assumed in Eq.~(\ref{Gamma_pure}).
A more realistic form for the scattering rate rolls over to a roughly constant
value outside a pseudogap of halfwidth~$\Delta$.
This can be approximated by writing
\begin{equation}
\Gamma(\epsilon) = \left\{
    \begin{array}{ll}
       \Gamma_0 |\epsilon/\Delta|^r, \quad & |\epsilon|\le\Delta; \\[1ex]
       \Gamma_0 & \Delta < |\epsilon| \le D; \\[1ex]
       0 & |\epsilon| > D.
    \end{array}
    \right.                                     \label{Gamma_rest}
\end{equation}

The effects of restricting the power-law scattering regime can be
predicted using poor-man's scaling.
Consider a localized level described by the nondegenerate Anderson model.
At temperatures $T\gg\Delta$, the impurity is insensitive to the presence
of the pseudogap, and one expects the standard physics exhibited in a
metallic host.
As the temperature is lowered, the effective position of the impurity level
scales upward according to Eq.~(\ref{epsilon(T):r=0}), while
$\Gamma_0$~remains essentially constant.
Two qualitatively different situations can arise.
In the first, the impurity remains in the valence-fluctuation regime (see
Fig.~\ref{fig:scaling}) all the way down to temperatures $T\ll \Delta$.
In this case, once the temperature falls much below $\Delta$, the system
will behave very much like an Anderson impurity with a pure power-law
scattering rate, the role of the half-bandwidth~$D$ being taken
by $\Delta$ and with $\epsilon_d$ replaced by $\bar{\epsilon}_d(\Delta)$
[given by Eq.~(\ref{epsilon(T):r=0})].
The qualitative effects of the pseudogap should therefore be those
reported in Sections~\ref{subsec:Ander_res} and~\ref{subsec:screened},
although the magnitude of these effects will decrease as the pseudogap narrows.
For example, on any subsequent entry to the local-moment regime, the effective
Kondo exchange~$J$ will be reduced relative to the case $r=0$ by a factor of
at least $|\bar{\epsilon}_d(\Delta)/\Delta|^r$ (compared to a reduction of
$|\epsilon_d/D|^r$ for $\Delta=D$).

Should there exist a solution to Eq.~(\ref{T_F}) such that $T_F\agt\Delta$,
then real charge fluctuations on the impurity site will be frozen out before
the power-law density of states makes its presence felt.
Perhaps the most interesting situation arises when the system enters the
local-moment regime, in which case one can Schrieffer-Wolff-transform to
the Kondo description of the problem.
Over the temperature range $T_F \agt T \agt \Delta$, the conduction band will
begin to screen out the impurity moment; in the scaling picture, the
effective value of the Kondo coupling will renormalize upwards according
to Eq.~(\ref{scaling3}) with $r=0$.
At temperatures $T\ll\Delta$, the impurity maps onto a model with pure
power-law exchange having an effective half-bandwidth $\Delta$ and
an exchange coupling $\rho_0 \bar{J}_0 \approx [\ln(\Delta/T_K^0)]^{-1}$.
Here $T_K^0 = D\exp[-1/(\rho_0 J_0)]$, the Kondo temperature for a system
having a constant density of states $\rho_0$, is assumed to be smaller
than~$\Delta$.
(If $T_K^0>\Delta$, then the impurity will already have entered the
strong-coupling regime before the pseudogap comes into play, in which event
perturbative scaling can provide no insight into the behavior for
$T\le \Delta$.)

The scenario of the previous paragraph implies a significant enlargement of
the region of parameter space within which a fully developed Kondo effect can
take place.
Consider, for instance, the case $U=\infty$, with some fixed value~$t$ of
the hybridization matrix element.
Working to lowest order, let us neglect the many-body renormalization of the
impurity energy (which becomes increasingly weak as $r$ increases;
see Section~\ref{subsubsec:scaling}).
Then the exchange $\rho_0 J_0$ entering the effective low-temperature Kondo
problem is approximately $\rho_0 J^0=t^2/(D|\epsilon_d|)$
for a constant scattering rate ($r=0$), $\rho_0 J^0 |\epsilon_d/D|^r$
for pure power-law scattering ($\Delta=D$), and
$\rho_0 J^0 \left[ 1-\rho_0 J^0 \,\ln(|\epsilon_d|/\Delta) \right]^{-1}$
for restricted power-law scattering.
The last value is {\em enhanced\/} over that obtained with a constant
scattering rate, and for $\Delta$~sufficiently small, $J_0$ will exceed the
threshold $J_c(r)$.
This is true even for $r\agt 0.5$, a range in which no Kondo effect can be
observed in cases of pure power-law scattering.

The predictions made in the preceding paragraphs can be tested against
numerical~RG results.
Restriction of the power-law scattering to a region $|\epsilon|<\Delta$
alters the hopping coefficients $t_n$ entering Eq.~(\ref{H_c:disc}).
The values for small~$n$ [such that the characteristic temperature $T_n$
given by Eq.~(\ref{T_N}) greatly exceeds $\Delta$] become essentially
identical to the corresponding values for the case $r=0$, while the $t_n$'s
for large~$n$ (such that $T_n\ll\Delta$) are still given by Eq.~(\ref{t_lim}).
Since it is the large-$n$ coefficients that determine the low-temperature
behavior of the system, the analysis of the stable fixed points of the
Anderson model in Section~\ref{sec:fixed} remains applicable.

\begin{figure}[t]
\centerline{
\vbox{\epsfxsize=75mm \epsfbox{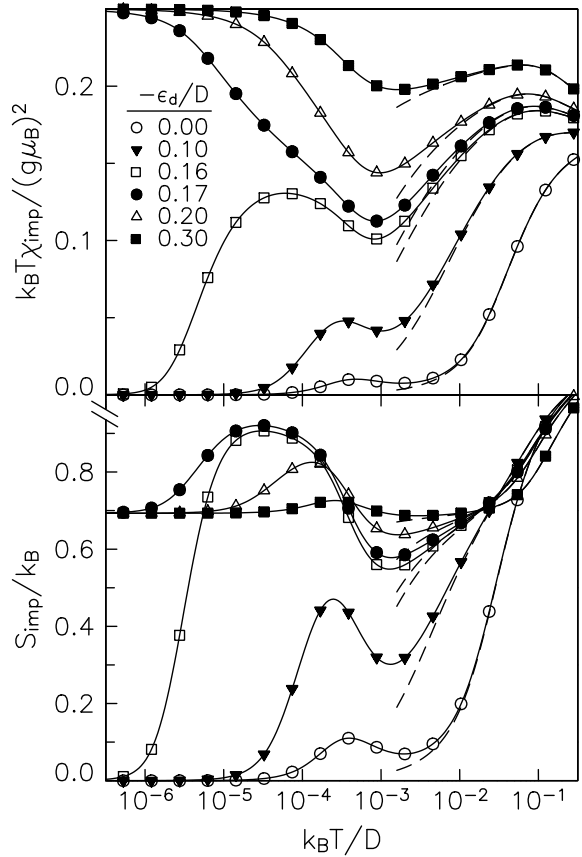}}
}
\vspace{2ex}
\caption{
Impurity susceptibility $T \chi_{\text{imp}}$ and entropy $S_{\text{imp}}$
vs temperature~$T$ for the infinite-$U$ Anderson model with a
restricted linear scattering rate described by Eq.~(\protect\ref{Gamma_rest})
with $r=1$, $\Delta=10^{-3}D$ and $\Gamma_0=0.050D$ (symbols, solid lines).
Data are also shown for the same impurity parameters ($t$ and $\epsilon_d$)
but a constant scattering rate (dashed lines).
}
\label{fig:r=1_rest}
\end{figure}

Figure~\ref{fig:r=1_rest} presents numerical~RG data for the Anderson model
with $r=1$, $\Delta/D=10^{-3}$, and $\Gamma_0/D=0.050$.
[The value of~$\Gamma_0$ is chosen so that if the energy-dependence
of~$\Gamma(\epsilon)$ were to derive solely from the density of states, then
the (energy-independent) hybridization~$t$ would be identical to that used in
Fig.~\ref{fig:r=1}.]
The numerical results bear out quite well the predictions of poor-man's
scaling.
Over the range $T\agt\Delta$, the impurity susceptibility and entropy are very
close to those obtained using the same $t$~and~$\epsilon_d$
but a constant scattering rate (dashed lines).
For $\epsilon_d\le -0.16D$, an initial increase in $T\chi_{\text{imp}}$
signaling entry to the local-moment regime is followed by a downturn as the
conduction band begins to screen the impurity moment;
whereas for $\epsilon_d \ge -0.1D$, $T\chi_{\text{imp}}$ falls monotonically
in the manner characteristic of the mixed-valence and empty-impurity limits.

The pseudogap in the scattering rate begins to make its presence felt about
a decade in temperature above $\Delta$.
For all the values of $\epsilon_d$ shown in Fig.~\ref{fig:r=1_rest}, the
initial effect is to produce an upturn in both $T\chi_{\text{imp}}$ and
$S_{\text{imp}}$, signaling a weakening in the effective coupling between
the impurity level and the conduction band.
For $\epsilon_d\ge -0.16D$, this weakening is reversed as the temperature
decreases further, and the system eventually flows to asymmetric strong
coupling.
In the case $\epsilon_d=-0.16D$, where the impurity clearly enters the
local-moment regime at a temperature well above $\Delta$, the downturn
in each property indicates the existence of a Kondo effect where none
was found in the pure power-law case (see Fig.~\ref{fig:r=1}).
For $\epsilon_d\le -0.17D$, the rise in $T\chi_{\text{imp}}$ towards the
free local-moment value indicates that when the system encounters the power-law
scattering regime, the effective exchange coupling is smaller than the
critical value needed to bring about Kondo-screening of the impurity.
Somewhere in the range $-0.17 < \epsilon_d/D < -0.16$ there presumably exists
a critical impurity energy that places the system precisely at the
intermediate-coupling fixed point, where
$T\chi_{\text{imp}}\approx 1/6$ (see Fig.~\ref{fig:Jc_props}).

We have performed similar calculations for parameters chosen so that real
charge fluctuations remain possible when the temperature becomes comparable
with the width of the pseudogap (i.e., for a larger $\Delta$ or a smaller
$|\epsilon_d|$ and $\Gamma_0$ than used in Fig.~\ref{fig:r=1_rest}).
As surmised in the scaling discussion above, the physical properties depart
less dramatically from those for a pure power-law scattering rate than in
the case shown in Fig.~\ref{fig:r=1_rest}.

We have also investigated the effect of restricted power-law exchange
within the Kondo models.
Results for the exactly screened case appear in
Ref.~\onlinecite{Ingersent:NRG}.
In this and the other variants of the model, the low-temperature behavior can
be accounted for by assuming a pure power-law exchange with the bare
coupling $J_0$ replaced by a (larger) effective value that compensates for
the elimination of conduction band states at energy scales $|\epsilon|>\Delta$.

\subsubsection{Finite Fermi-energy scattering rate}

In the context of $d$-wave superconductivity, it has been predicted
\cite{Borkowski:92} that the pair-breaking effect of any finite concentration
of magnetic impurities will feed back to produce a small but nonzero
quasiparticle density of states at zero energy.
This may be taken into account by introducing a lower cutoff $\Delta'$
on the power-law scattering rate (in addition to the upper cutoff~$\Delta$
considered above), such that
$\Gamma(\epsilon)\approx\Gamma_0(\Delta'/\Delta)^r$ for $|\epsilon|<\Delta'$.

Based on a heuristic scaling approach, similar to that employed in the previous
subsection, it is apparent that the system can exhibit a wide variety of
behaviors, depending on the relative magnitudes of the energy scales in the
problem.
The lower cutoff on the power-law scattering rate cannot significantly affect
the physics until the temperature drops to become of order $\Delta'$.
At this point, the clearest departure from the case $\Delta'=0$ arises if
the impurity is already in, or subsequently enters, its local-moment regime.
In such cases the existence of a nonzero Fermi-surface scattering rate ensures
the eventual appearance of a standard, metallic Kondo effect.
However, for $r\agt 1$ at least, most plausible values of the various
parameters in the model lead to screening of the impurity only
at temperatures that are so low as to be physically irrelevant.

\section{Summary}
\label{sec:summary}

In this work, we have studied four models in which the interaction
between a magnetic impurity level and an otherwise-uncorrelated host
fermion system is described by an energy-dependent scattering rate that
vanishes in power-law fashion at the Fermi level.
Our principal results are summarized in the following paragraphs.

In all four models, the effect of the pseudogap in the low-energy scattering
rate is to create a low-temperature regime which has no counterpart in metals
(where the scattering rate is flat in the vicinity of the Fermi level).
This stable weak-coupling limit, first identified\cite{Withoff} in the exactly
screened Kondo model, is characterized by a free impurity local moment which
retains its Curie susceptibility down to zero temperature.

Depending on the value of $r$, the weak-coupling region of parameter space is
bounded in each model by either one or two unstable intermediate-coupling
fixed points.
A particle-hole-symmetric fixed point of this type exists for all
\mbox{$0<r<r_{\text{max}}$} and is the only such fixed point for
\mbox{$0<r< r^{\star}$}, over which range particle-hole asymmetry is
marginally irrelevant.
For \mbox{$r>r^{\star}$} a second fixed point with different properties is
reached whenever particle-hole asymmetry is present, and for
\mbox{$r>r_{\text{max}}$} it is the only fixed point of this type.
The values of $r_{\text{max}}$~and~$r^{\star}$ are model-dependent,
but in all the four cases that we have studied
$0<r^{\star}\le r_{\text{max}}\le\frac{1}{2}$.

Another common feature of these models is the existence of two distinct
strong-coupling regimes.
The first, symmetric strong coupling, is the natural generalization to $r>0$
of the low-temperature limit of the metallic Kondo problem.
In the Anderson and screened Kondo models, the impurity degree of freedom is
completely quenched, as evidenced by the vanishing of the quantity
$T\chi_{\text{loc}}$, and the sole effect of the magnetic level is to impose
a phase shift on electrons at the Fermi energy.
This phase shift results in anomalous thermodynamic properties, namely those
that would arise as if a fraction $\text{min}(r,1)$ of an electron were to
decouple from each conduction band.
For exponents $r<1$, the phase shift should also result in a zero-temperature
peak in the electrical resistivity.
The magnitude of this peak should decrease with increasing $r$, and for
$r>1$ it should disappear altogether.

Except under artificial conditions of strict particle-hole symmetry, the
symmetric strong-coupling limit is unstable.
Generally, the system is driven to an asymmetric strong-coupling fixed
point at which a Fermi-level phase shift of~$\pm\pi$ implies a vanishing
impurity contribution to thermodynamic and transport properties.
Underscreened and overscreened impurities, however, admit two other
possibilities:
For $r\le 1/4$, one can obtain flow to (or, for the $s=1$ model, near)
a non-Fermi-liquid fixed point, corresponding to an effective
coupling~$J^{\star}$ which decreases with increasing~$r$.
For $r\agt 1/4$, the generic behavior is flow to weak coupling, in which case
the absence of interesting many-body effects invites comparison with the
{\em ferromagnetic\/} Kondo problem in metals.

In addition to characterizing the location, stability, and thermodynamics
of the various fixed points mentioned above, we have also studied the
possible crossovers between regimes dominated by these fixed points.
We have quantified observations made previously based on a perturbative
scaling analysis of the Anderson model that while local-moment formation is
assisted by the presence of a pseudogap in the low-energy scattering rate,
there is a strong suppression of the effective exchange coupling between any
such moment and the delocalized electrons.
As a result, it becomes progressively harder as $r$ increases from zero ---
and essentially impossible for a pure power-law scattering rate with
$r\agt 0.5$ --- to find any choice of model parameters which leads to
Kondo-screening of the impurity moment.
In more realistic situations where the power-law scattering is confined
to a narrow range of energies around the Fermi level, the suppression is less
complete.
Even in such cases, though, there is a significant region of parameter
space in which an unscreened local-moment survives down to zero temperature.

\acknowledgments

We gratefully acknowledge useful discussions with
L.~Borkowski,
C.~Cassanello,
E.~Fradkin,
M.~Hettler,
P.~Hirschfeld,
L.~Oliveira,
and
Q.~Si.
This research was supported in part by NSF Grant No.~DMR93--16587.

\appendix

\section{Resonant Level Model with a Power-Law Mixing Rate}
\label{app:reslevel}

In this appendix, we analyze a noninteracting resonant level model for the
mixing between an impurity level and a spinless conduction band.
This model, described by the Hamiltonian
\begin{equation}
{\cal H} = \epsilon_{\bf k} c^{\dag}_{\bf k} c_{\bf k}
        + \epsilon_d d^{\dag} d + \sum_{\bf k} \frac{t_{\bf k}}{\sqrt{N}}
        (c^{\dag}_{\bf k} d + \text{H.c}),
                                                        \label{H_res:def}
\end{equation}
represents the limit $U=0$ of the Anderson model
[Eq.~(\ref{H_And:def})], in which up and down spins decouple from one another,
and the spin index can therefore be dropped.
The mixing rate,
\begin{equation}
\Gamma(\epsilon) =
    \pi \sum_{\bf k} \frac{|t_{\bf k}|^2}{N} \delta(\epsilon-\epsilon_{\bf k}),
\end{equation}
is assumed to have the pure power-law form given in Eq.~(\ref{Gamma_pure}).
Since the Hamiltonian is quadratic, various properties of the model can be
calculated analytically.

\subsubsection{Impurity self-energy}

The impurity self-energy in the model described by Eq.~(\ref{H_res:def})
is
\begin{equation}
\Sigma(\omega) = \sum_{\bf k}
    \frac{|t_{\bf k}|^2/N}{\omega-\epsilon_k+i0^+} .
\end{equation}
The real and imaginary parts of this quantity are
\begin{eqnarray}
\text{Re} \Sigma(\omega) &=&
    \text{P} \sum_{\bf k} \frac{|t_{\bf k}|^2/N}{\omega-\epsilon_{\bf k}}
    \equiv \frac{1}{\pi} \, \text{P} \int_{-\infty}^{\infty} \!\! d\epsilon \,
    \frac{\Gamma(\epsilon)}{\omega - \epsilon}, \label{Re_Sigma:def} \\
\text{Im} \Sigma(\omega) &=& -\Gamma(\omega).
\end{eqnarray}
Substituting Eq.~(\ref{Gamma_pure}) into Eq.~(\ref{Re_Sigma:def}),
one obtains
\begin{equation}
\text{Re}\Sigma(\omega) = \left\{
    \begin{array}{ll}
    \displaystyle
    \frac{\Gamma_0}{\pi} \ln \left| \frac{D+\omega}{D-\omega} \right|,
    & r = 0, \\[2ex]
    \displaystyle
    -\Gamma(\omega) \tan\frac{r\pi}{2} \, \text{sgn}(\omega),
    & 0 < r < 1, \\[2ex]
    \displaystyle
    -\frac{\Gamma(\omega)}{\pi}
    \ln\left|\frac{D^2-\omega^2}{\omega^2}\right| \text{sgn}(\omega),
    & r = 1, \\[2ex]
    \displaystyle
    -\frac{2\Gamma_0}{(r\!-\!1)\pi}\,\frac{\omega}{D},
    & r > 1.
    \end{array} \right.                         \label{Re_Sigma}
\end{equation}
The expressions for $r=0$ and $r=1$ are exact, but those for other values
of $r$ are approximations which are valid only for $|\omega|< \omega_c(r)$.
Here, $\omega_c(r)$ is a cutoff energy scale which approaches zero as
$r$ approaches $1$ from above or below.
In all cases, $\text{Re}\Sigma(\omega)$ passes through zero at $\omega=0$.
However, $\text{Re}\Sigma(\omega)$ has the same sign as $\omega$ for
$r=0$, whereas the signs are opposite for all $r>0$.

\subsubsection{Impurity spectral function}

The impurity spectral function is
\begin{equation}
A(\omega) = \frac{-\text{Im}\Sigma(\omega) / \pi}%
    {[\omega-\epsilon_d-\text{Re}\Sigma(\omega)]^2+%
    [\text{Im}\Sigma(\omega)]^2}.
\end{equation}

If $\epsilon_d\not= 0$, then for all $r\ge 0$ the spectral function
is featureless in the vicinity of the impurity energy,
\begin{equation}
A(\epsilon_d) = \frac{\Gamma(\epsilon_d) / \pi}%
    {[\text{Re}\Sigma(\epsilon_d)]^2+[\Gamma(\epsilon_d)]^2}
    \approx \text{constant},
\end{equation}
and takes its low-frequency behavior from $\Gamma(\omega)$,
\begin{equation}
A(\omega) \approx \frac{\Gamma(\omega)}{\pi\epsilon_d^2}, \quad
    |\omega| \ll |\epsilon_d|.
\end{equation}

For $\epsilon_d = 0$, by contrast, the spectral function exhibits nontrivial
structure near $\omega=0$:
\begin{equation}
A(\omega) \approx \left\{
    \begin{array}{ll}
    \displaystyle
    \frac{1}{\pi\Gamma_0} \, \frac{\Gamma_0^2}{\Gamma_0^2 + \omega^2},
    & r = 0, \\[3ex]
    \displaystyle
    \frac{1}{\pi \Gamma(\omega)} \cos^2\frac{r\pi}{2}
    & 0 < r < 1, \\[3ex]
    \displaystyle
    \frac{1}{\pi \Gamma(\omega)} \left[ 1 + \left(\frac{D}{\Gamma_0}
    \!-\! \frac{2}{\pi}\ln \left|\frac{\omega}{D} \right|\right)^2
    \right]^{-1}\!\!\!\!,
    & r = 1, \\[3ex]
    \displaystyle
    \frac{\Gamma(\omega)}{\pi (g \omega)^2},
    & r > 1.
    \end{array} \right.                         \label{spectral}
\end{equation}
Here,
\begin{equation}
g = 1 + \frac{2\Gamma_0}{(r\!-\!1)\pi D}.
\end{equation}
For a flat mixing rate ($r=0$), the spectral function consists of the
standard Lorentzian resonance centered on $\omega=0$.
For all $r>0$ this feature is replaced by a power-law cusp,
$A(\omega)\sim |\omega/D|^{|1-r|-1}$, such that the spectral function
diverges for $0<r<2$ but instead vanishes for $r>2$.

\subsubsection{Conduction-band phase shifts}

The mixing term in the Hamiltonian effectively adds one extra state to the
band, centered on an energy $\bar{\epsilon}_d$ which is a root of the
equation $\bar{\epsilon}_d-\epsilon_d-\text{Re}\Sigma(\bar{\epsilon}_d) = 0$.
As a result, each of the original band states is shifted in energy from
$\epsilon$ to $\epsilon-\delta_0(\epsilon)/\pi$.
Here $\delta_0$, the $s$-wave phase shift, satisfies
\begin{equation}
\delta_0(\epsilon) = \text{atan}\left(
\frac{\text{Im}\Sigma(\epsilon)}{\epsilon-\epsilon_d-\text{Re}\Sigma(\epsilon)}
\right).
                                                        \label{delta_0:def}
\end{equation}
Since one expects the band states to be pushed away from the inserted level,
the sign of $\delta_0(\epsilon)$ should be opposite to that of
$\epsilon-\epsilon_d-\text{Re}\Sigma(\epsilon)$.
Coupling to the impurity brings about a change in the density of states,
\begin{equation}
\rho_{\text{imp}}(\epsilon) = \delta(\epsilon-\epsilon_d) +
                \pi^{-1}\partial\delta_0/\partial\epsilon.  \label{rho_imp}
\end{equation}

For the case of a pure power-law scattering rate with $\epsilon_d \not=0$,
$\delta_0(\epsilon)\approx\Gamma(\epsilon)/\epsilon_d$ at low frequencies.
The case $\epsilon_d=0$ is again more interesting:
\begin{equation}
\frac{\delta_0(\epsilon)}{\text{sgn}(-\epsilon)} \approx \left\{
    \begin{array}{ll}
    \displaystyle
    ( 1\!-\!r) \frac{\pi}{2}
    - \frac{|\epsilon|}{\Gamma(\epsilon)} \cos^2\frac{r\pi}{2}, \;
    & 0 \le r < 1, \\[3ex]
    \displaystyle
    \frac{\pi}{2\ln|D/\epsilon|}
    & r = 1, \\[3ex]
    \displaystyle
    \frac{\Gamma(\epsilon)}{g|\epsilon|},
    & r > 1.
    \end{array} \right.                         \label{delta_0:res}
\end{equation}
Equations~(\ref{delta_0:res}) indicate that at $\epsilon=0$, $\delta_0$ jumps
through $(1\!-\!r_1)\pi$ while $\rho_{\text{imp}}$ has a delta-function peak
of weight~$r_1$.
In the case $r=0$, the standard interpretation of the smooth variation of the
density of states is that the impurity level becomes completely absorbed into
the band.
This absorption appears to be incomplete for all $r>0$, with the delta function
in~$\rho_{\text{imp}}$ representing the fraction of the impurity degree of
freedom that remains localized.
For $r\ge 1$, not only does the delta function contain the entire weight of
the original impurity, but a counter-intuitive situation arises in which the
impurity level repels delocalized states close to its renormalized position
(here, $\bar{\epsilon}_d = 0$) {\em less\/} strongly than it repels states
that lie further away in energy.

\section{Details of the Expansion of \mbox{$F_{\text{imp}}$}
and \mbox{$\chi_{\text{imp}}$}}
\label{app:thermo}

This appendix fills in some of the steps in the derivation of the algebraic
expressions for impurity thermodynamic properties presented in
Section~\ref{sec:thermo}.
In particular, we focus on the methods for performing sums over
single-particle eigenstates arising in the perturbative treatment of the
discretized effective Hamiltonians introduced in Section~\ref{sec:fixed}
and on extrapolation of the resulting expressions to the continuum limit.

Consider expansion of the properties in the vicinity of the weak-coupling
fixed point.
(Analogous arguments apply at strong coupling.)
The summands encountered in these calculations can generally be separated into
the product of two parts: the first increasing with the index $j$ which labels
the single-particle eigenvalues, but doing so no faster than $\alpha_{0j}^4$;
the second decreasing for large $j$ at least as fast as
$\exp(-\beta_N\eta^{\star}_j)$.
Provided that $k_B T \ll D$ and $\beta_N\ll 1$, this decomposition ensures that
(a) the summand takes its largest value for $1\ll j\ll N$, in which range the
asymptotic forms given in Eqs.~(\ref{eta*}), (\ref{A_nj}) and~(\ref{alphas})
are essentially exact; and (b) the summand is sufficiently small for
$j={\cal O}(1)$ and for $j={\cal O}(N/2)$ that the range of $j$ can safely be
extended to run from $-\infty$ to $+\infty$.
If, in addition, $\Lambda$ is sufficiently close to unity, the sum over $j$
can be well-approximated \cite{KWW} by an integral over the variable
$u = \beta_N t^{\star} \Lambda^{j-\nu_N}$.
This procedure, which amounts to the replacements
\begin{mathletters}
\begin{eqnarray}
  \sum_{j=1}^{(N+1)/2} &\rightarrow& \frac{1}{\ln \Lambda}
            \int_0^{\infty} \frac{du}{u},               \\
  \eta^{\star}_j &\rightarrow& u/\beta_N,               \\
  \alpha_{nj} &\rightarrow& \alpha_n
     \left( \frac{u}{\beta_N t^{\star}} \right)^{(2n+1+r)/2},
\end{eqnarray}                                          \label{sum_to_int}%
\end{mathletters}%
converts a sum over $j$ to a $\Lambda$-independent integral multiplied by a
simple $\Lambda$-dependent prefactor.

The continuum limit is reached by simultaneously taking $\Lambda\rightarrow 1$
and $N\rightarrow\infty$ in such a manner that $\beta_N$ [defined by
Eq.~(\ref{beta_N})] approaches some value $\bar{\beta}\ll 1$.
For all values $r\not= 1$, the precise value of $\bar{\beta}$ drops out of the
final expression for each thermodynamic property, so this prescription
produces an unambiguous result.
It will be shown below, however, that for linear scattering rates the leading
corrections at the weak-coupling and asymmetric strong-coupling fixed points
depend explicitly on $\ln\bar{\beta}$.
Since $\bar{\beta}$ has no physical meaning for $\Lambda\rightarrow 1$,
the continuum limit of the discretized thermodynamic calculation contains a
degree of ambiguity in this special case.

\subsection{Local-moment regime}
\label{subapp:thermo_LM}

The starting point for computing $F_{\text{imp}}$ in the local-moment regime
is Eq.~(\ref{F_LM:sum}), which contains four separate summations over
single-particle eigenstates.
Consider first the unconstrained double sum over $j$ and $k$.
One can show that this term represents the second-order shift in the
ground-state energy of the system due to the perturbations $O_V$ and $O_J$
[Eqs.~(\ref{local_delH})].
This shift is a temperature-independent quantity which should not contribute
to the impurity specific heat.
Moreover, since the RG~transformation [Eq.~(\ref{H_N})] subtracts
off the ground-state energy at each iteration, such a term will not be
detected numerically and can safely be neglected.

Each of the remaining summations entering $F_{\text{imp}}$ contains at least
one factor of $p_j$, which permits application of the
transformation~(\ref{sum_to_int}).
The last term in Eq.~(\ref{F_LM:sum}), which contains a summation over
indices $j$ and $k\not= j$ requires special attention.
We find it convenient to define
\begin{eqnarray}
\Sigma_{\text{F}}
&=& - \left( \frac{\ln \Lambda}{\alpha_0^2} \right)^2 (t^{\star})^{2+2r}
   \beta_N^{1+r+r_1} \times     \nonumber \\
& & \quad \Lambda^{(r_1-r)N/2} \sum_{j \neq k}
        \frac{\alpha_{0j}^2 \alpha_{0k}^2 \eta^{\star}_j p_j}%
               {{\eta^{\star}_j}^2-{\eta^{\star}_k}^2}  \nonumber \\
&\stackrel{\text{(\ref{sum_to_int})}}{\longrightarrow}&
    \frac{\ln \Lambda}{\alpha_0^2} (t^{\star})^{1+r}
    \beta_N^{r_1-1} \Lambda^{(r_1-r)N/2} \times \nonumber \\
&& \quad \int_0^{\infty} \!\!\!\!\! du \; \frac{u^{1+r}}{e^u + 1}
   \sum_{k=1}^{(N+1)/2}
   \frac{(\beta_N \alpha_{0k})^2}{(\beta_N \eta^{\star}_k)^2-u^2} .
                                                        \label{Sigma_F}
\end{eqnarray}
For $0<r<1$, $\Sigma_{\text{F}}$ will be dominated by contributions from
$u$ and $k$ such that $\beta_N \eta^{\star}_k \approx u = {\cal O}(1)$.
In this case, the sum over $k$ can be converted to an integral over
$v = \beta_N t^{\star} \Lambda^{k-\nu_N}$, yielding
\begin{equation}
\Sigma_F =
    \int_0^{\infty} \!\!\!\! du \int_0^{\infty} \!\!\!\! dv \;
        \frac{v^r u^{1+r}}{v^2\!-\!u^2} \frac{1}{e^u\!+\!1}.
\end{equation}
Making the change of variables $v\rightarrow uy$, one obtains
\begin{equation}
\Sigma_{\text{F}} = \int_0^{\infty} \!\!\!\! du \; \frac{u^{2r}}{e^u+1} \;
    \int_0^{\infty} \!\!\!\! dy \; \frac{y^r}{y^2-1}
    \quad \text{for $r<1$.}                             \label{Sigma_F:r<1}
\end{equation}
The $u$ integral is related to the Riemann zeta function,\cite{Riemann}
while the $y$ integral was evaluated in Ref.~\onlinecite{Withoff}.
As a result, one can write
\begin{equation}
\Sigma_{\text{F}} = \phi(r_1\!+\!r)\,\psi(r),        \label{Sigma_F:res}
\end{equation}
where $\phi(x)$ and $\psi(r)$ are defined in
Eqs.~(\ref{phi's}) and~(\ref{lambda}), respectively.

For $r\ge 1$, the $k$-sum in Eq.~(\ref{Sigma_F}) is dominated by the
largest values of $k$, and is not well-approximated by an integral.
If one neglects $u^2$ in the denominator, the sum can be performed directly
to give Eq.~(\ref{Sigma_F:res}) once again, but with
\begin{equation}
\psi(r) \approx
   \left\{
      \begin{array}{ll}
         \displaystyle
         \frac{\ln\Lambda}{1-\Lambda^{-(r-1)}} (t^{\star})^{r-1},
         \quad & r > 1, \\[3ex]
         \displaystyle
         \frac{1}{2}(N+1) \ln\Lambda , & r = 1.
      \end{array}
   \right.                                      \label{lambda:r>=1}
\end{equation}

These manipulations, when combined with Eq.~(\ref{beta_N}), transform
Eq.~(\ref{F_LM:sum}) to
\begin{eqnarray}
-\frac{F_{\text{imp}}}{k_B T}
&=& \ln 2 - 8 \tilde{t}_1 \, \frac{\alpha_0 \alpha_1}{\ln \Lambda}
    \frac{\phi(1\!+\!r)}{(t^{\star})^{2+r}}
    \left( \frac{k_B T}{\alpha D} \right)^{1+r}         \nonumber \\
&+& 4 (\tilde{V}^2 + \case{3}{16}\tilde{J}^2)
    \left( \frac{\alpha_0^2}{\ln\Lambda} \right)^2 \!\!
    \frac{1}{(t^{\star})^{2+2r}} \times                 \nonumber \\
& & \quad \left[ \ln\Lambda \, r \phi(2r)
    \left( \frac{k_B T}{\alpha D} \right)^{2r} \right.  \nonumber \\
& & \qquad \left. + \, 2\psi(r) \phi(r_1\!+\!r)
    \left( \frac{k_B T}{\alpha D} \right)^{r_1+r} \right].
                                                        \label{F_LM:int}
\end{eqnarray}

Similar methods can be applied to Eq.~(\ref{chi_LM:sum}) for
$\chi_{\text{imp}}$.
Again, evaluation of the double sum requires the most care.
We define
\begin{eqnarray}
\Sigma_{\chi} &=&
   \left(\frac{\ln \Lambda}{\alpha_0^2} \right)^2 (t^{\star})^{2+2r}
   \beta_N^{1+r+r_1} \times                             \nonumber \\
&& \quad \Lambda^{(r_1-r)N/2} \sum_{j\neq k}
        \frac{\alpha_{0j}^2 \alpha_{0k}^2 \eta^{\star}_j}%
        {{\eta^{\star}_k}^2\!-\!{\eta^{\star}_j}^2}
             p_j \bar{p}_j (\bar{p}_j\!-\!p_j)          \nonumber \\
&\stackrel{\text{(\ref{sum_to_int})}}{\longrightarrow}&
   \frac{\ln \Lambda}{\alpha_0^2} (t^{\star})^{1+r}
   \beta_N^{r_1-1} \Lambda^{(r_1-r)N/2} \times  \nonumber \\
&& \int_0^{\infty}\!\!\!\!du \; \frac{u^r e^u (e^u\!-\!1)}{(e^u\!+\!1)^3}
   \sum_{k=1}^{(N+1)/2} \!\!\!
   \frac{(\beta_N \alpha_{0k})^2}{(\beta_N \eta^{\star}_k)^2\!-\!u^2}.
                                                        \label{Sigma_chi}
\end{eqnarray}
For $r < 1$, the $k$-sum in Eq.~(\ref{Sigma_chi}) can be converted to an
integral, yielding
\begin{equation}
\Sigma_{\chi} = \int_0^{\infty} \!\!\! dv \int_0^{\infty} \!\!\! du \;
    \frac{v^r u^{1+r}}{v^2\!-\!u^2}
    \frac{e^u(e^u\!-\!1)}{(e^u\!+\!1)^3} .              \label{Sigma_chi:r<1}
\end{equation}
Letting $v\rightarrow uy$, one obtains
\begin{equation}
\Sigma_{\chi} = \frac{\bar{\phi}(1+r)}{1+r} \psi(r),
                                                        \label{Sigma_chi:res}
\end{equation}
where $\bar{\phi}(x)$ and $\psi(r)$ are defined in
Eqs.~(\ref{phi's}) and~(\ref{lambda}), respectively.
For $r\ge 1$, direct summation neglecting $u^2$ in the denominator gives
Eq.~(\ref{Sigma_chi:res}) with $\psi(r)$ instead defined by
Eq.~(\ref{lambda:r>=1}).

The remaining sums in Eq.~(\ref{chi_LM:sum}) are straightforward to perform.
The resulting expression for the impurity susceptibility is
\begin{eqnarray}
\lefteqn{\frac{k_B T \chi_{\text{imp}}}{(g \mu_B)^2} = \frac{1}{4}
    + \frac{\tilde{J}}{2} \, \frac{\alpha_0^2}{\ln \Lambda}
    \frac{\bar{\phi}(1\!+\!r)}{(1\!+\!r) (t^{\star})^{1+r}}
    \left( \frac{k_B T}{\alpha D} \right)^r }           \nonumber \\
&\;\;\;-& 2 \tilde{t}_1 \, \frac{\alpha_0 \alpha_1}{\ln \Lambda}
    \frac{\bar{\phi}(1\!+\!r)}{(t^{\star})^{2+r}}
    \left( \frac{k_B T}{\alpha D} \right)^{1+r}         \nonumber \\
&\;\;\;+& 4 \tilde{U}_0 \left[ \frac{\alpha_0^2}{\ln \Lambda}
    \frac{\bar{\phi}(1\!+\!r)}{(1+r) (t^{\star})^{1+r}} \right]^2
    \left( \frac{k_B T}{\alpha D} \right)^{1+2r}        \label{chi_LM:int} \\
&\;\;\;+& \tilde{V}^2 \left( \frac{\alpha_0^2}{\ln\Lambda} \right)^2 \!\!
    \frac{1}{(t^{\star})^{2+2r}} \left[ \ln\Lambda \, r \bar{\phi}(2r)
    \left( \frac{k_B T}{\alpha D} \right)^{2r} \right. \nonumber \\
& & \left. \qquad + \, 2 \psi(r) \bar{\phi}(r_1\!+\!r)
    \left( \frac{k_B T}{\alpha D} \right)^{r_1+r} \right] .     \nonumber
\end{eqnarray}

The final step is to extrapolate the expressions for $F_{\text{imp}}$ and
$\chi_{\text{imp}}$ to the continuum limit.
For $\ln\Lambda\ll 1$,  Eqs.~(\ref{alpha}), (\ref{t*}),
and (\ref{alphas}) reduce to
\begin{equation}
\alpha, t^{\star} \approx 1 + {\cal O}(\ln\Lambda), \quad
\alpha_n^2 \approx \case{1}{2}(2n\!+\!1\!+\!r) \ln\Lambda.
                                                        \label{lim_Lambda=1}
\end{equation}
Substituting these values into Eqs.~(\ref{lambda:r>=1}), (\ref{F_LM:int})
and~(\ref{chi_LM:int}), and then letting $\ln\Lambda\rightarrow 0$, one
obtains Eqs.~(\ref{F_LM:cont}) and~(\ref{chi_LM:cont}) with
\begin{equation}
\lim_{\Lambda\rightarrow 1} \psi(r) = (r-1)^{-1}, \quad r > 1.
\end{equation}
In the special case $r=1$, application of Eqs.~(\ref{alpha}) and~(\ref{beta_N})
leads to the result
\begin{equation}
\lim_{N\rightarrow\infty,\Lambda\rightarrow 1} N\ln\Lambda =
-2\ln\left(\frac{\bar{\beta} k_B T}{D}\right),
\end{equation}
where, as stated above, $\bar{\beta}$ is the limiting value of $\beta_N$.
This in turn leads to the replacement in Eq.~(\ref{lambda:rep}).

\subsection{Symmetric strong-coupling regime}

The sums entering Eqs.~(\ref{F_SSC:sum}) and~(\ref{chi_SSC:sum}) can also be
transformed into integrals using the methods described in the previous
subsection.
The leading deviations from the fixed-point free-energy and susceptibility
arise from first-order terms in perturbation theory, so there are no double
summations to contribute logarithmic corrections to the simple power laws
in temperature.
For small but finite $\ln\Lambda$, one obtains
\begin{equation}
-\frac{F_{\text{imp}}}{k_B T} = r_1 \ln 4
    - 8 \tilde{t}_2 \frac{\beta_1 \beta_2}{\ln \Lambda}
    \frac{\phi(1\!-\!r)}{(t^{\star} \Lambda^{-r/2})^{2-r}}
    \left( \frac{k_B T}{\alpha D} \right)^{1-r}
\end{equation}
and
\begin{eqnarray}
\lefteqn{\frac{k_B T \chi_{\text{imp}}}{(g \mu_B)^2}
= \frac{r_1}{8} - 2 \tilde{t}_2 \frac{\beta_1 \beta_2}{\ln \Lambda}
    \frac{\bar{\phi}(1\!-\!r)}{(t^{\star} \Lambda^{-r/2})^{2-r}}
    \left( \frac{k_B T}{\alpha D} \right)^{1-r}}                \nonumber \\
& & \quad - 4 \tilde{U}_1
    \left( \frac{\beta_1^2}{\ln \Lambda} \right)^2
    \left[ \frac{\bar{\phi}(1\!-\!r)}%
                {(1\!-\!r)(t^{\star}\Lambda^{r/2})^{1-r}}\right]^2
    \left( \frac{k_B T}{\alpha D} \right)^{1-2r},               \nonumber \\
\end{eqnarray}
where $\phi(x)$ and $\bar{\phi}(x)$ are defined
in Eqs.~(\ref{phi's}).
Extrapolation to the continuum limit yields the final expressions
contained in Eqs.~(\ref{F_SSC:cont}) and~(\ref{chi_SSC:cont}).


\begin{thebibliography}{99}
\bibitem{xi_note}
    In Refs.~\onlinecite{Wilson} and~\onlinecite{KWW},
    $e_n = 0$ for all $n$ due to the symmetry of the
    conduction band about the Fermi energy; furthermore, $t_n$
    is instead called $\xi_{n-1}$.
\bibitem{N_note}
    For notational compactness, we replace all powers of $\Lambda^{N-1}$
    that appear in the conventional formulation of the numerical~RG method
    with powers of $\Lambda^N$.
    This change necessitates redefinition of the prefactors $\alpha$,
    $\alpha_n$, and $\beta_n$ [see Eqs.~(\ref{alpha}), (\ref{alphas}),
    and~(\ref{beta_nj}), respectively].
    However, all formulae for the eigenvalues, eigenvectors, and thermodynamic
    properties of the truncated Hamiltonian $H_N$ reduce --- for
    $r=0$ and, where necessary, $z=1$ --- to the corresponding results
    presented in Refs.~\onlinecite{Wilson}, \onlinecite{KWW},
    and~\onlinecite{Yoshida}.
\bibitem{all_r}
    This result arises solely from symmetry considerations, and holds for
    all~$r$.
    It is demonstrated for $r\!=0\!$ in Ref.~\onlinecite{Wilson}.
\bibitem{rescaling}
    The critical exchange $J_c$ of Ref.~\onlinecite{Ingersent:NRG}
    must be multiplied by $\rho_0$ given in Eq.~(\ref{rho0_pure}) to
    obtain the quantity plotted in Fig.~\ref{fig:Jc_A}.
\end{thebibliography}
\end{document}